\documentclass[a4paper]{aa}

\usepackage{graphicx,natbib}
\usepackage[varg]{txfonts}
\usepackage{color}

\newcommand{\kms}{km\,s$^{-1}$}

\newcommand{\bs}{$\langle B \rangle$}
\newcommand{\bzdi}{$\langle B_{\rm V} \rangle$}
\newcommand{\fe}{Fe~{\sc i}}
\newcommand{\vs}{$v_{\rm e}\sin i$}

\newcommand{\fifps}[2]{\centering\resizebox{#1}{!}{\includegraphics{#2}}}

\begin{document}

\title{Hidden magnetic fields of young suns}

\author{
O.~Kochukhov\inst{1}
\and T.~Hackman\inst{2}
\and J.J.~Lehtinen\inst{3,4}
\and A.~Wehrhahn\inst{1}
}

\institute{
Department of Physics and Astronomy, Uppsala University, Box 516, SE-75120 Uppsala, Sweden\\\email{oleg.kochukhov@physics.uu.se}
\and
Department of Physics, P.O. Box 64, FI-00014 University of Helsinki, Finland
\and
Max Planck Institute for Solar System Research, Justus-von-Liebig-Weg 3, D-37077 G\"ottingen, Germany
\and
ReSoLVE Centre of Excellence, Department of Computer Science, Aalto University, PO Box 15400, FI-00076 Aalto, Finland
}

\date{Received 00 November 2019 / Accepted 00 November 2019}

\titlerunning{Magnetic fields of young suns}
\authorrunning{O.~Kochukhov et al.}

\abstract{
Global magnetic fields of active solar-like stars are nowadays routinely detected with spectropolarimetric measurements and are mapped with Zeeman-Doppler imaging (ZDI). However, due to the cancellation of opposite field polarities, polarimetry captures only a tiny fraction of the magnetic flux and cannot assess the overall stellar surface magnetic field if it is dominated by a small-scale component. Analysis of Zeeman broadening in high-resolution intensity spectra can reveal these hidden complex magnetic fields. Historically, there were very few attempts to obtain such measurements for G dwarf stars due to the difficulty of disentangling Zeeman effect from other broadening mechanisms affecting spectral lines. Here we developed a new magnetic field diagnostic method based on relative Zeeman intensification of optical atomic lines with different magnetic sensitivity. Using this technique we obtained 78 field strength measurements for 15 Sun-like stars, including some of the best-studied young solar twins. We find that the average magnetic field strength $Bf$ drops from 1.3--2.0~kG in stars younger than about 120~Myr to 0.2--0.8~kG in older stars. The mean field strength shows a clear correlation with the Rossby number and with the coronal and chromospheric emission indicators. Our results suggest that magnetic regions have roughly the same local field strength $B\approx3.2$ kG in all stars, with the filling factor $f$ of these regions systematically increasing with stellar activity. Comparing our results with the spectropolarimetric analyses of global magnetic fields in the same stars, we find that ZDI recovers about 1\% of the total magnetic field energy in the most active stars. This figure drops to just 0.01\% for the least active targets. 
}


\maketitle

\section{Introduction}
\label{intro}

Magnetic fields play a central role in the surface activity of cool stars. It is now well established that magnetism is responsible for such phenomena as dark spots, flares, coronal mass ejections, enhanced chromospheric and X-ray emission. Magnetic fields directly affect stellar evolution by altering the mass loss and governing redistribution of angular momentum between different parts of stellar interiors. Planets orbiting cool stars are influenced by the stellar magnetic activity in many different ways. This makes understanding of stellar magnetism essential for studying evolution, atmospheres, and habitability of terrestrial exoplanets.

Despite availability of a massive body of circumstantial observations of stellar magnetic activity, direct detections and measurements of magnetic fields on stellar surfaces is still very challenging. This type of research relies on exploiting the signatures of the Zeeman effect in stellar spectra, which requires high-resolution, high signal-to-noise ratio spectroscopic and spectropolarimetric observational data. Two complementary approaches to studying Zeeman effect are commonly used to infer the presence of a magnetic field and derive its characteristics. The first method relies on line polarisation measurements with high-resolution spectropolarimetry. The second technique extracts information on the magnetic broadening and splitting of spectral lines from the usual intensity spectra.

Analyses of weak circular polarisation signals in spectral lines, often enhanced with a multi-line technique, have been very successful in studying cool-star magnetic fields \citep{donati:1997}. The polarimetric method yielded magnetic field detections for hundreds of stars using snapshot circular polarisation observations \citep[e.g.][]{fossati:2013,marsden:2014,moutou:2017}. It has also enabled reconstruction of detailed magnetic field maps for dozens of objects \citep[e.g.][]{donati:2003,petit:2008,morin:2008,rosen:2016,folsom:2016} with the Zeeman Doppler imaging \citep[ZDI,][]{kochukhov:2016} inversion technique applied to time-resolved spectropolarimetry. The success of polarimetric diagnostic methods stem from an unambiguous nature of the magnetic field detection and a relative simplicity of the theoretical modelling of weak stellar circular polarisation signatures, which can be carried out relying on a very basic line formation treatment \citep{folsom:2018}. 

However, it is also understood that polarimetry is able to recover only a small fraction of the magnetic field energy, vastly underestimating the true strength of magnetic structures on the surfaces of cool stars \citep{vidotto:2016,kochukhov:2017b,lehmann:2019}. This polarimetric bias is caused by the topological complexity of a typical cool-star magnetic field geometry, which comprises many unresolved magnetic features with opposite field polarities. Circular polarisation signals corresponding to these regions have opposite signs and mostly cancel out in any disk-integrated polarimetric observable. Thus, polarimetry is sensitive only to a large-scale magnetic field component, particularly for slowly rotating stars which exhibit no significant rotational Doppler broadening of their line profiles. At the moment, it is unclear how this large-scale field component is related to fields at smaller spatial scales where the bulk of the magnetic energy is concentrated.

On the other hand, the Zeeman splitting and broadening observed in stellar intensity spectra is proportional to the absolute value of magnetic field strength and thus includes contributions from magnetic structures at all spatial scales. In this way the Zeeman broadening diagnostic provides an unbiased estimate of the total surface magnetic field strength. The relative intensities of Zeeman components are only weakly sensitive to the field orientation, making it impossible to infer the vector field maps from intensity spectra considering typical field strengths of $\sim$\,1~kG encountered in cool stars. Moreover, a challenging aspect of this type of magnetic field analysis is that magnetic broadening has to be separated from many other broadening mechanisms affecting spectral lines. This makes the field detections from intensity spectra more ambiguous compared to the polarimetric method and often requires observational data of exceptional quality. Additionally, the Zeeman response of the intensity profiles of spectral lines is more complex and diverse than the circular polarisation in the same lines, impeding application of multi-line techniques and requiring the use of sophisticated polarised radiative transfer codes. 

These problems have been overcome, with varying degree of success, by a number of pioneering studies which inferred the presence of magnetic fields in different types of cool active stars \citep{robinson:1980,saar:1986b,basri:1988,valenti:1995}. The Zeeman broadening method was particularly successful in application to T~Tauri stars \citep{johns-krull:1999,johns-krull:2007,yang:2011,lavail:2017,lavail:2019} and active M dwarfs \citep{johns-krull:1996,shulyak:2014,shulyak:2017,shulyak:2019,kochukhov:2017c,kochukhov:2019a}, which typically have rather strong fields and narrow lines. At the same time, relatively little progress has been made for Sun-like G dwarfs \citep[see review by][]{reiners:2012}. The majority of Zeeman broadening field detections and measurements for these stars come from historic publications by S.~Saar and collaborators \citep{saar:1987,saar:1996,saar:2001,saar:1986,saar:1986a,saar:1992}, with a few measurements contributed by other studies \citep{basri:1988,rueedi:1997,anderson:2010}. To summarise, despite numerous recent polarimetric investigations of global magnetic fields of Sun-like stars \citep[e.g.][and references therein]{see:2019}, the properties of their overall magnetic fields, including typical surface field strengths, their rotational modulation and cyclic variation, relationship to large-scale fields and different indirect magnetic activity indicators, remain largely unexplored. 

This unsatisfactory situation is largely due to the absence of an efficient Zeeman broadening diagnostic technique that can be applied to moderate quality high-resolution optical spectra of solar-type stars. In this paper we develop such a technique and present its application to a sample of G dwarf stars with different activity levels. The rest of this paper is organised as follows. Section~\ref{methods} details various methodological aspects of our study, including a review of different manifestations of the Zeeman effect in stellar intensity spectra, description of our line profile modelling codes, motivation of the choice of key diagnostic lines, target selection and discussion of observational data. This is followed in Sect.~\ref{results} by the presentation of magnetic field measurement results for each target star. We discuss our results in Sect.~\ref{discussion}, where we establish correlations between magnetic field characteristics, stellar parameters and magnetic activity proxies. Finally, Sect.~\ref{conclusions} summarises main conclusions of our investigation.

\section{Methods}
\label{methods}

\subsection{Zeeman broadening and intensification of spectral lines}
\label{zbzi}

The presence of a magnetic field in stellar atmosphere leads to splitting of each spectral line into three groups of differently polarised Zeeman components. The linearly polarised $\pi$ components are distributed symmetrically around the line centre. Elliptically polarised $\sigma$ components form two groups, one shifted bluewards, and another redwards of the line centre. The consequence of this Zeeman effect on spectral lines observed in the usual intensity spectrum is twofold. First, lines are broadened (if the magnetic splitting is less than the non-magnetic line width) or split (if the field strength is large enough) due to a wavelength separation of the $\pi$ and $\sigma$ components. The magnitude of magnetic broadening increases linearly with the field strength and can be expressed in \kms\ units as
\begin{equation}
\Delta v_B = 1.4\times 10^{-4} g_{\rm eff} \lambda B
\end{equation}
for the field strength in kG and wavelength in \AA. The effective Land\'e factor, $g_{\rm eff}$, expresses the relative span of Zeeman splitting for lines with different magnetic sensitivity. Considering that the most magnetically sensitive lines one can find in stellar spectra have $g_{\rm eff}\approx3$, we get $\Delta v_B \la 2$~\kms\ for a 1~kG field and a line at $\lambda=5000$~\AA. This magnetic broadening is comparable to the intrinsic line width, dominated by the $\sim$\,2--3~\kms\ turbulent broadening, and is smaller than the instrumental broadening (3--6~\kms\ for the resolving power of $R=\lambda/\Delta\lambda=$\,0.5--1\,$\times10^5$) of most of the actively used night-time spectrographs. Furthermore, any significant rotational broadening, with \vs\ exceeding a few \kms, effectively renders Zeeman broadening unobservable in cool stars. All these factors limit practical applications of the Zeeman broadening diagnostic to active, very slowly rotating stars observed at $R\ga10^5$ with high signal-to-noise ratio spectra \citep{robinson:1980,marcy:1984,saar:1986b,basri:1988,johns-krull:1996,rueedi:1997,anderson:2010}. These restrictions can be partly relaxed with the help of observations at near-infrared wavelengths \citep{saar:1985,valenti:1995,johns-krull:2007,yang:2011,lavail:2017,lavail:2019,flores:2019} thanks to a faster increase of Zeeman splitting with wavelength (grows as $\lambda^2$) compared to other broadening mechanisms (grow as $\lambda$). For instance, $\Delta v_B\approx10$~\kms\ for a 1~kG field and a $g_{\rm eff}=3$ line at $\lambda=2.3$~$\mu$m, enabling magnetic measurements of moderately fast rotators using lower quality data.

Another, less commonly discussed, consequence of Zeeman effect is the overall strengthening of absorption lines. This Zeeman intensification effect occurs due to a separation of Zeeman components and the resulting desaturation of strong spectral lines. This effect is well known from studies of much stronger magnetic fields encountered in the early-type chemically peculiar stars \citep{babcock:1949,hensberge:1974,mathys:1990,takeda:1991,mathys:1992,kupka:1996,stift:2003,kochukhov:2004e,kochukhov:2013a}. It was first studied in cool stars by \citet{basri:1992} and \citet{basri:1994}. They demonstrated that Zeeman intensification is a complex function of both magnetic field parameters (field intensity and orientation) and spectral line characteristics (line strength and Zeeman splitting pattern). Unlike Zeeman broadening, magnetic intensification is most effective for strong spectral lines with a large number of uniformly spaced Zeeman components but not necessarily for lines with the largest Land\'e factors. A major advantage of Zeeman intensification analysis over Zeeman broadening is that the former does not require observational data of exceptional quality and can be applied to rapid rotators provided that diagnostic lines remain free from blends. On the other hand, quantitative interpretation of spectral line strengths in terms of Zeeman intensification requires detailed modelling of the magnetic desaturation process with a realistic polarised radiative transfer code and is more sensitive to errors of atomic parameters, particularly transition probabilities, which determine relative line intensities in the absence of a magnetic field. 

A series of recent studies of magnetic fields in active M dwarf stars using the \ion{Ti}{i} multiplet at $\lambda$ 9647--9788~\AA\ implemented a combined Zeeman broadening and intensification approach \citep{shulyak:2017,shulyak:2019,kochukhov:2017c,kochukhov:2019a}. These investigations demonstrated that a detection and analysis of $\ga2$~kG fields in stars rotating as fast as \vs\,=\,30--40~\kms\ is within reach. However, this methodology cannot be directly applied to Sun-like stars since this particular group of \ion{Ti}{i} lines becomes too weak at $T_{\rm eff}\ga4500$~K. Here we aim to develop an equivalent magnetic field measurement procedure for hotter stars based on a different set of diagnostic lines.

\subsection{Magnetic spectrum synthesis}
\label{magspec}

We use the polarised spectrum synthesis code {\sc Synmast} \citep{kochukhov:2007d,kochukhov:2010a} to model magnetic field effects on absorption lines in the spectra of active stars. This code solves the polarised radiative transfer equation with an efficient numerical algorithm \citep{de-la-cruz-rodriguez:2013} using realistic stellar model atmospheres. It treats ionisation of atomic species and dissociation of molecules with an up-to-date equation of state package shared with the {\sc SME} code \citep{piskunov:2017}\footnote{\url{http://www.stsci.edu/~valenti/sme.html}}. Theoretical spectra in four Stokes parameters are computed by {\sc Synmast} for a given limb angle and a depth-independent magnetic field vector. Information on atomic and molecular line parameters is obtained from the VALD database \citep{ryabchikova:2015}\footnote{\url{http://vald.astro.uu.se}}. In addition to the usual set of line parameters required for spectrum synthesis (central wavelength, excitation potential, oscillator strength and damping constants), VALD supplies Land\'e factors and $J$ quantum numbers of the upper and lower atomic levels necessary for calculation of Zeeman splitting patterns.

A magnetic broadening and intensification analysis of cool active stars does not require a detailed geometrical model of the surface magnetic field distribution. We follow previous Zeeman broadening studies \citep[e.g.][]{valenti:1995,johns-krull:2007,yang:2011,shulyak:2017,kochukhov:2017c,lavail:2019} in assuming that the field is uniform and oriented normally with respect to the stellar surface. The assumption of a radial field orientation is justified by the analogy with solar flux tubes \citep[e.g.][]{valenti:1995,johns-krull:1999,johns-krull:2004} and represents an intermediate case, in terms of the impact on line profiles, between (clearly unrealistic in the stellar case) extremes of the magnetic field vectors strictly parallel and strictly perpendicular to the observer's line of sight. In this way, a uniform radial field geometry provides a good mixture of field lines with a range of orientations to the line of sight \citep{yang:2011}. 
%This is consistent with ZDI results, which typically find significantly weaker fields than indicated by the Zeeman broadening studies (\citealt{reiners:2009,kochukhov:2019a,see:2019}, Sect.~\ref{zdicomparison} below).

Using this simple field geometry model, the local intensity spectra are computed with {\sc Synmast} at seven limb angles and then convolved with appropriate kernels to take into account the radial-tangential macroturbulence, rotational Doppler broadening, and instrumental broadening. The final theoretical stellar flux spectrum is obtained by adding these local calculations with the weights corresponding to relative areas of the seven annular regions and normalising by the continuum flux integrated over the visible stellar hemisphere in the same way. Further details on this disk integration procedure can be found in \citet{valenti:1996}.

We treat the field strength distribution with a standard two-component model adopted by many previous Zeeman broadening studies of active FGK stars \citep{basri:1988,rueedi:1997,saar:2001,anderson:2010,kochukhov:2017c}. This approach is inspired by the notion that small-scale magnetic fields are concentrated in distinct surface elements, referred to as flux tubes in solar physics \citep[e.g.][]{stenflo:1973,solanki:1984,stenflo:1994,solanki:2006}. In this case, the total observed stellar spectrum is given by the weighted superposition of magnetic and non-magnetic contributions
\begin{equation}
S (\lambda) = (1-f) \cdot S_0 (\lambda) + f\cdot S(\lambda,B),
\end{equation}
where $S_0(\lambda) $ and $S(\lambda,B)$ is the non-magnetic, continuum-normalised stellar flux spectrum and the spectrum calculated with the field strength $B$, respectively, and $f$ is the filling factor of magnetic regions. This is undoubtedly a highly simplified approximation of the actual continuous distribution of magnetic field strengths. Nevertheless, it has proven to be a successful practical approach to the problems of diagnosing small-scale solar magnetic fields (\citealt{stenflo:1994} and references therein) and inferring mean magnetic field strength from the stellar intensity spectra. Information on the magnetic filling factors determined within this framework (provided that $f$ can be reliably separated from $B$) is important for understanding a range of processes taking place in active cool stars \citep[e.g.][]{montesinos:1993,cranmer:2011,cranmer:2017,see:2019}.

The solar small-scale strong-field regions, such as flux tubes, are known to have a distinctly different thermodynamic structure. However, there are no reliable analytical models or numerical simulations that quantify this difference for active stars with much stronger mean magnetic fields than those observed in the quiet Sun. In the absence of suitable grids of one-dimensional models of magnetised cool star atmospheres we resort to employing the same normal model atmosphere for computing both $S_0(\lambda)$ and $S(\lambda,B)$. This approach is also obligatory to enable a meaningful comparison with the results of previous studies, most of which did not take a difference between the structures of magnetic and non-magnetic atmospheres into account. Nevertheless, expecting that this difference is primarily reflected in unequal temperatures of the regions with different magnetic fields, we will explore the impact of adopting different normal model atmospheres for calculation of $S_0(\lambda)$ and $S(\lambda,B)$. Since our approach relies on the differential magnetic intensification of atomic lines from the same multiplet, the main effect of a temperature difference is unequal continuum brightness of the spectra corresponding to magnetic and non-magnetic regions. This difference is subsumed, to first order, by the magnetic filling factor $f$. This parameter thus represents continuum-intensity-weighted fraction of the stellar surface covered by a magnetic field.

\subsection{Zeeman-sensitive lines in the optical solar spectrum}
\label{wide}

\begin{figure*}[!t]
\fifps{\hsize}{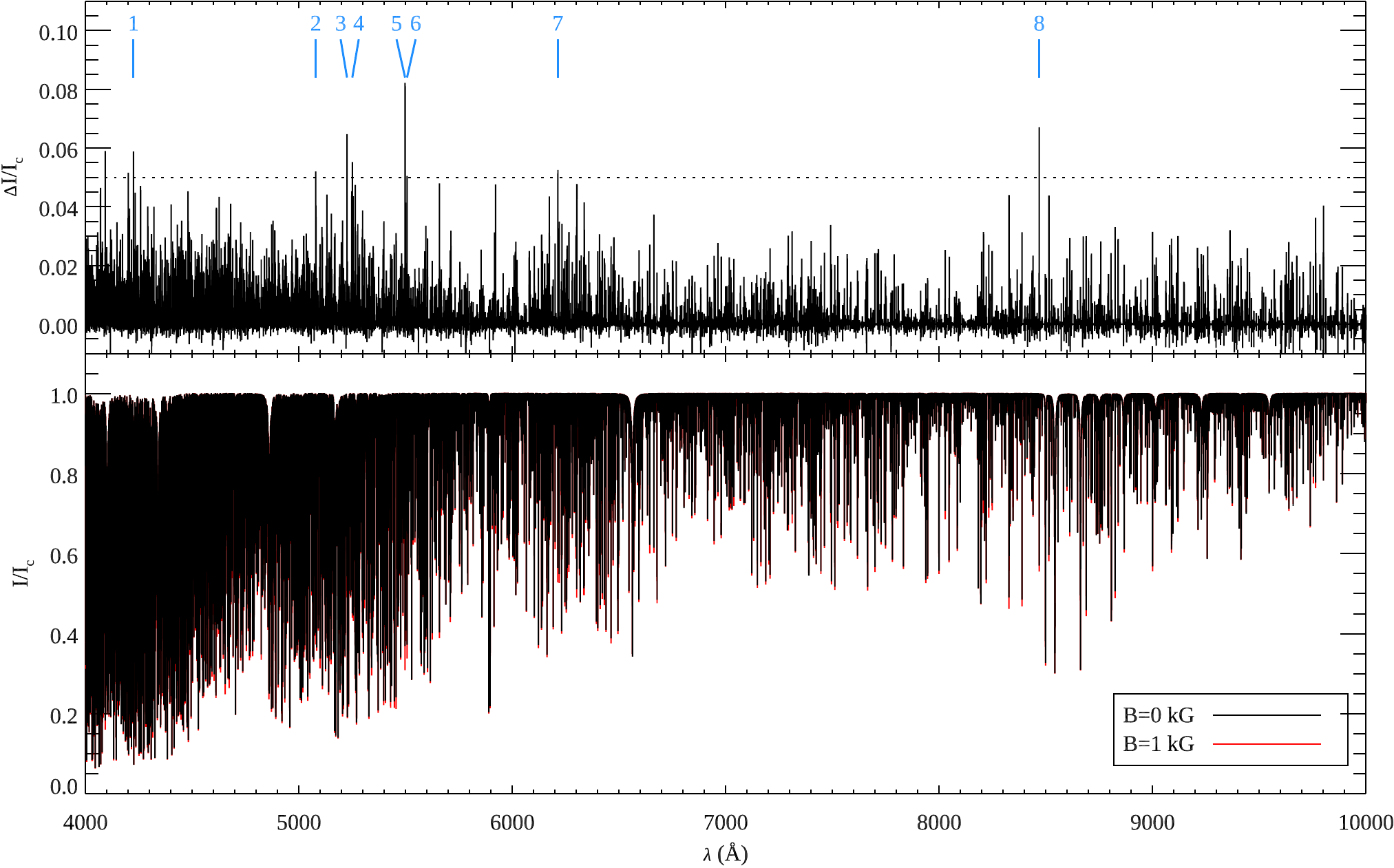}
\caption{
Response of the continuum normalised spectrum in the 4000--10000~\AA\ wavelength range to a 1 kG magnetic field covering the entire stellar surface. \textit{Lower panel}: synthetic spectra calculated for the solar atmospheric parameters and abundances with (light curve) and without (dark curve) magnetic field. \textit{Upper panel}: the difference between these two spectra. The lines showing the strongest Zeeman response are identified according to the numbering adopted in Table~\ref{tbl:lines_best}.
}
\label{fig:wide}
\end{figure*}

Several neutral Fe lines (\fe\ 5250, 6173, 6302, 8468 \AA) are commonly used for the analysis of polarisation and Zeeman broadening in the atmosphere of the Sun and solar-type stars. These lines are distinguished by large values of effective Land\'e factors and thus exhibit the largest profile shape modification when a magnetic field is present. However, as discussed in Sect.~\ref{zbzi}, these lines are not necessarily the most useful diagnostics when Zeeman intensification is considered. To assess the latter in a comprehensive and systematic manner, we have carried out magnetic spectrum synthesis calculations with {\sc Synmast} for the entire optical spectrum (400 nm to 1~$\mu$m) covered by modern echelle spectropolarimeters. These calculations were based on the solar model atmosphere from the {\sc MARCS} grid \citep{gustafsson:2008}, employed a line list retrieved from VALD and adopted the solar chemical abundances \citep{grevesse:2007}, microturblent velocity $v_{\rm mic}=0.85$~\kms, macrotubulent velocity $v_{\rm mac}=3$~\kms, projected rotational velocity $v_{\rm e}\sin i=5$~\kms, and instrumental resolution $R=\lambda/\Delta\lambda=10^5$. Two theoretical calculations were produced: one without a magnetic field and another one with a $B=1$~kG radial field covering the entire stellar surface. This choice of \vs\ and $B$ corresponds to a moderately active star rotating significantly faster than the Sun. This parameter combination is in the regime where a Zeeman broadening analysis is already quite challenging since the signatures of magnetic line broadening are largely washed out by the stellar rotation.

The resulting magnetic and non-magnetic synthetic spectra and their difference are shown in Fig.~\ref{fig:wide}. The spikes in the difference plot identify spectral features exhibiting the strongest intensification for the 1~kG magnetic field considered in this calculation. Setting an arbitrary threshold of 5\% for the change of the central line depth, we compiled the list of 8 unblended lines potentially useful for a Zeeman intensification analysis. These lines, all belonging to \fe, are listed in Table~\ref{tbl:lines_best}. The well-known magnetic diagnostic lines \fe\ 5250 and 8468~\AA\ are among the lines with the largest magnetic intensification. However, by far the strongest Zeeman response is found for the \fe\ 5497.5~\AA\ line. It has a moderately large, though not exceptional, effective Land\'e factor of 2.25. Interestingly, the nearby \fe\ 5506.8~\AA\ ($g_{\rm eff}=2.00$) is also present in Table~\ref{tbl:lines_best} and another line from the same multiplet, \fe\ 5501.5~\AA\ ($g_{\rm eff}=1.87$), exhibits a weaker but still significant Zeeman intensification response. 

\begin{table}[!t]
\centering
\caption{Spectral lines with the strongest Zeeman intensification in the solar optical spectrum. \label{tbl:lines_best}}
%{\normalsize
\begin{tabular}{cccccc}
\hline
\hline
No. & Ion & $\lambda$ (\AA) & $E_{\rm lo}$ (eV) & $g_{\rm eff}$ & $\Delta I/I_{\rm c}$ (\%) \\
\hline
1 & \fe & 4224.513 &  3.4302 & 2.780 & 5.88 \\
2 & \fe & 5078.974 &  4.3013 & 1.870 & 5.20 \\
3 & \fe & 5225.526 &  0.1101 & 2.250 & 6.47 \\
4 & \fe & 5250.209 &  0.1213 & 3.000 & 5.52 \\
5 & \fe & 5497.516 &  1.0111 & 2.255 & 8.22 \\
6 & \fe & 5506.778 &  0.9901 & 2.000 & 5.05 \\
7 & \fe & 6213.429 &  2.2227 & 1.995 & 5.25 \\
8 & \fe & 8468.406 &  2.2227 & 2.495 & 6.71 \\
\hline
\end{tabular}
%}
\tablefoot{The last column indicates change of the residual line depth due to a 1~kG radial magnetic field assuming $v_{\rm e} \sin i=5$~\kms, $v_{\rm mac}=3$~\kms, and $R=10^5$.}
\end{table}

\citet{stenflo:1984} discussed the solar Stokes $I$ and $V$ spectra of the \fe\ 5497.5--5506.8~\AA\ lines in a strong plage, noting that these lines are substantially more polarised than \fe\ 5250~\AA. \citet{rosen:2012} used these three \fe\ lines for numerical tests of ZDI, finding that the Stokes $I$ Zeeman intensification signal in the 5497.5~\AA\ line aids reconstruction of stellar magnetic field geometries provided that the intensity and circular polarisation spectra are modelled self-consistently. \citet{morgenthaler:2012} correlated the widths of the 5497.5 and 5506.8~\AA\ lines with chromospheric emission indicators for the active Sun-like star $\xi$~Boo~A (HD\,131156A). Apart from these few studies, to the best of our knowledge, the \fe\ 5497.5 and 5506.8~\AA\ lines have not been systematically utilised for either solar or stellar magnetic field diagnostic.

\subsection{Determination of magnetic field parameters}
\label{method}

The magnetic intensification of the three \fe\ lines 5497.5, 5501.6, and 5506.8~\AA\ represents a promising tool for measuring surface magnetic field strength in Sun-like stars, provided one can find a suitable reference spectral feature with weak or no magnetic field sensitivity. According to \citet{nave:1994}, these three lines belong to the \fe\ multiplet 88, also known as multiplet 15 in the older tables by \citet{moore:1959}, formed by transitions between the $a\,^5F$ and $z\,^5F^o$ energy levels in a neutral iron atom. This multiplet includes a handful of other strong unblended \fe\ lines, one of which, 5434.5~\AA, has a very small effective Land\'e factor and is therefore essentially insensitive to a magnetic field. Thus, a suitable method of extracting information on stellar magnetic fields from the \fe\ 5497.5--5506.8~\AA\ lines is to compare them with \fe\ 5434.5~\AA. The latter line can be employed to establish the stellar projected rotational velocity and Fe abundance. Then, the three magnetically sensitive lines can be used to determine magnetic field parameters. This two-step procedure comprises the new magnetic diagnostic method advanced in this paper.

Parameters of the four \fe\ lines are summarised in Table~\ref{tbl:lines_chosen}. The oscillator strengths listed in this table come from high-precision laboratory measurements \citep{fuhr:1988,obrian:1991}. Other sources of atomic data may provide oscillator strengths with a different overall scale, yet the relative strengths of the four lines are going to be identical since all these transitions come from the same multiplet. This alleviates the problem of oscillator strength uncertainties which complicated many previous attempts to measure stellar magnetic fields with Zeeman intensification and broadening \citep{basri:1992,valenti:1995,rueedi:1997,anderson:2010}. 

\begin{figure}[!t]
\fifps{\hsize}{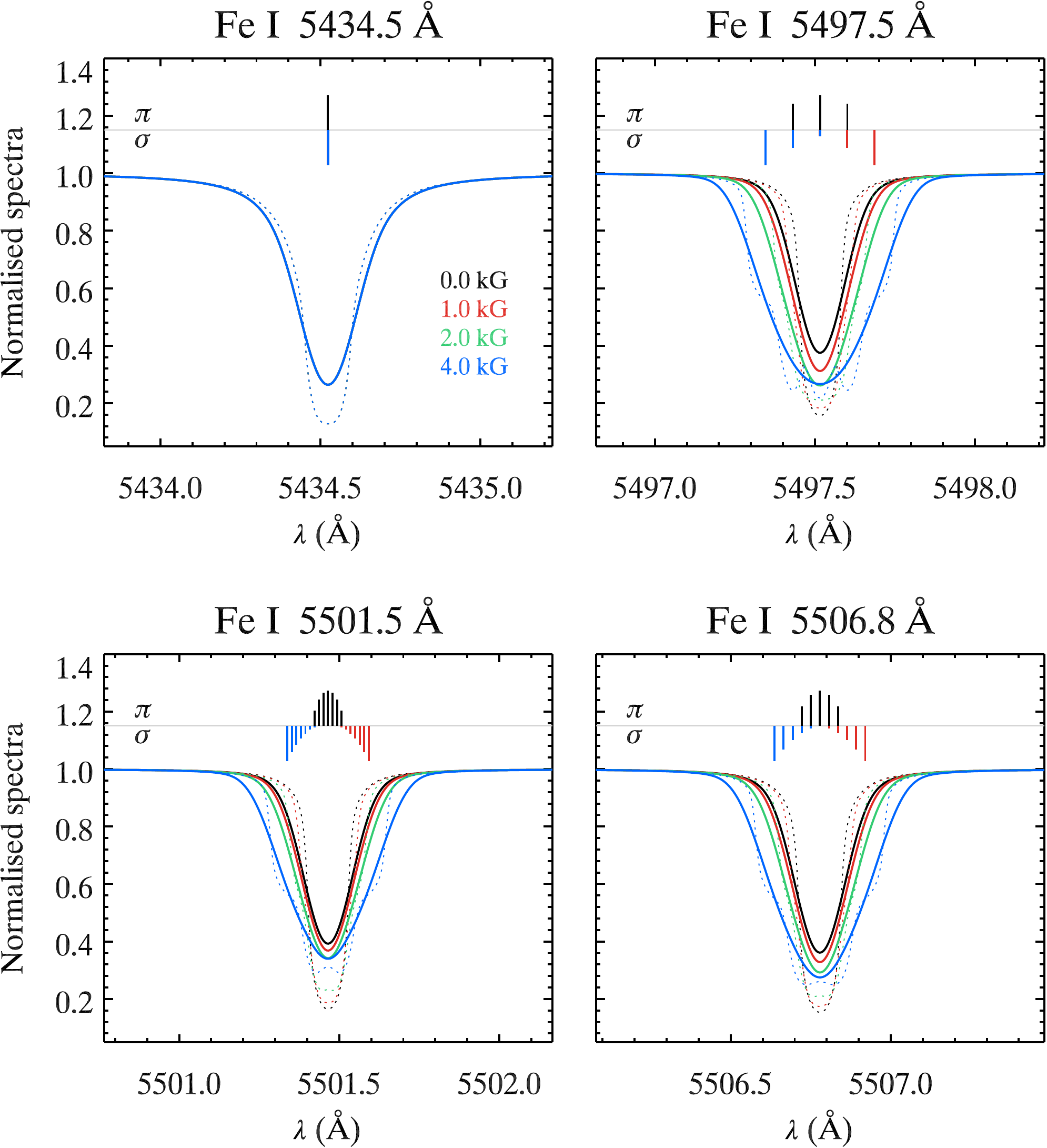}
\caption{
Theoretical profiles of the four \fe\ spectral lines studied in this paper. Calculations for the solar parameters and magnetic field strengths ranging from 0 to 4 kG are shown for representative rotational, macroturbulent and instrumental broadening (solid lines) and without broadening (dotted lines). The bar plots above line profiles schematically show the Zeeman splitting patterns for a field strength of 4~kG.
}
\label{fig:lines}
\end{figure}

\begin{figure}[!t]
\fifps{7.2cm}{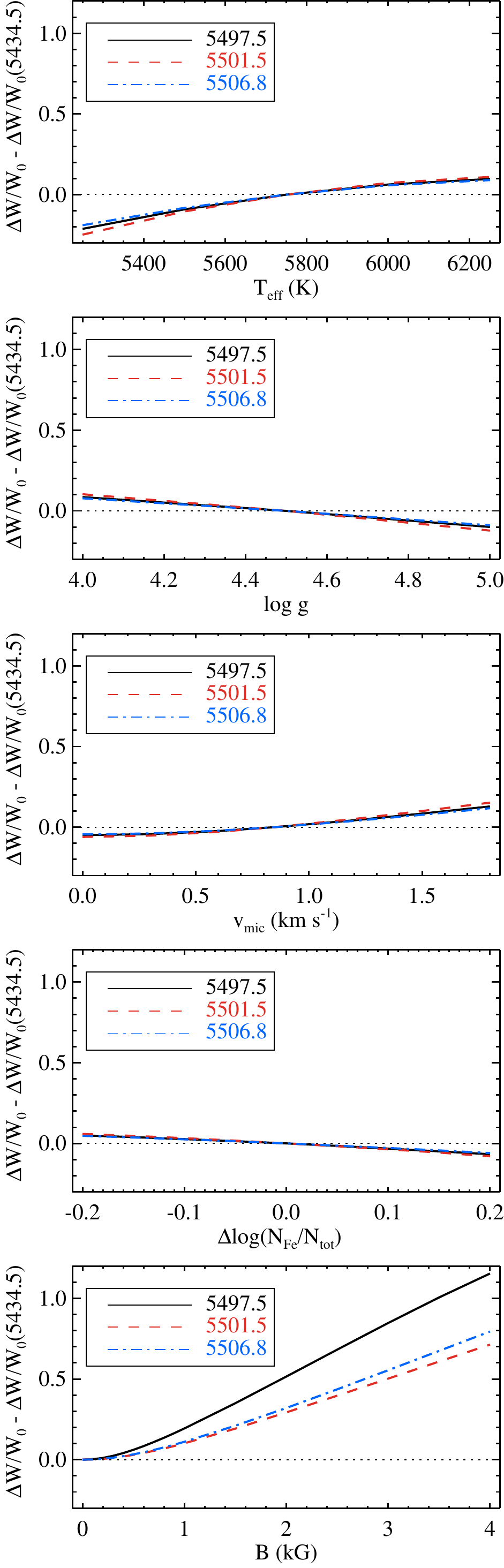}
\caption{
Equivalent width change of the \fe\ lines $\lambda$ 5497.5, 5501.5, and 5506.8~\AA\ relative to the \fe\ $\lambda$ 5435.5~\AA\ line in response to the variation (from top to bottom) of the effective temperature, surface gravity, microturbulent velocity, Fe abundance, and magnetic field strength.
}
\label{fig:ew}
\end{figure}

Figure~\ref{fig:lines} illustrates response of the four \fe\ diagnostic lines to a uniform radial magnetic field increasing in strength from 0 to 4 kG. These calculations were carried out with {\sc Synmast} for the same set of stellar parameters as used in Sect.~\ref{wide} and treating each of the four lines in isolation, ignoring possible blends. Two sets of calculations are shown: one without any broadening applied to the theoretical spectra and another one using representative values of $v_{\rm mac}$, \vs\ and $R$ adopted in Sect.~\ref{wide}. The Zeeman splitting patterns are schematically shown in Fig.~\ref{fig:lines} for the 4~kG field. It is evident that the 5497.5--5506.8~\AA\ lines are strongly influenced by the magnetic field compared to \fe\ 5434.5~\AA. The largest magnetic broadening and intensification effect is shown by \fe\ 5497.5~\AA\ and is due to its wide Zeeman splitting pattern composed of 5 groups of overlapping $\pi$ and $\sigma$ components.

Similarity of the oscillator strengths, wavelengths and excitation potentials of the four \fe\ lines studied here translates into their similar formation physics in a non-magnetic cool-star atmosphere. These lines exhibit little differential response to variation of thermodynamic structure, allowing one to disentangle magnetic intensification from other effects. A series of spectrum synthesis calculations documented by Fig.~\ref{fig:ew} compares the relative change of equivalent widths of the 5497.5--5506.8~\AA\ lines with respect to the 5434.5~\AA\ line caused by variation of magnetic field and stellar parameters. The reference model parameters adopted for these calculations were $T_{\rm eff}=5750$~K, $\log g=4.5$, $v_{\rm mic}=0.85$~\kms, and $B=0$~kG. The relative normalised equivalent width change was computed for each magnetically sensitive line by dividing its equivalent width change $\Delta W$ by the initial equivalent width $W_0$ and subtracting the same ratio for the \fe\ 5434.5~\AA\ line. The purpose of these line formation calculations was to explore to what extent inevitable uncertainties of stellar parameters can interfere with the determination of magnetic field strength using our method. The three upper panels in Fig.~\ref{fig:ew} suggest that an uncertainty of $\sim$\,50~K for $T_{\rm eff}$, $\sim$\,0.05~dex for $\log g$, and $\sim$\,0.1~\kms\ for $v_{\rm mic}$, typical of modern spectroscopic analyses of solar-type stars \citep{valenti:2005}, have little impact ($\Delta W/W_0 - \Delta W/W_0(5434.5) \la 0.01$) on the magnetic measurements unless $B$ is much smaller than $\sim$\,200~G. The studied lines show no significant differential response to a moderate variation of Fe abundance. One can also note that the differential equivalent width change due to temperature variation (top panel in Fig.~\ref{fig:ew}) is small and has a positive slope. This means that large cool spots, known to be present at the surfaces of the most active stars in our sample, should yield a relative equivalent width change opposite to that of the magnetic intensification effect.

\begin{table}[!t]
\centering
\caption{Parameters of the \fe\ spectral lines studied in this paper. \label{tbl:lines_chosen}}
%{\normalsize
\begin{tabular}{ccccc}
\hline
\hline
$\lambda$ (\AA) & $E_{\rm lo}$ (eV) & $E_{\rm up}$ (eV) & $\log gf$ & $g_{\rm eff}$ \\
\hline
5434.523 & 1.0111 & 3.2918 & $-2.122$ & $-0.010$ \\
5497.516 & 1.0111 & 3.2657 & $-2.849$ & 2.255 \\
5501.465 & 0.9582 & 3.2112 & $-3.047$ & 1.875 \\
5506.778 & 0.9901 & 3.2410 & $-2.797$ & 2.000 \\
\hline
\end{tabular}
%}
\end{table}

Among the three nuisance spectroscopic parameters considered here, the microturbulent velocity is characterised by the largest relative error. The choice of $v_{\rm mic}$ can influence the strengths of the 5497.5--5506.8~\AA\ lines relative to the 5434.5~\AA\ line because the latter is about 40\% stronger than any of the former and is more sensitive to $v_{\rm mic}$. Throughout this paper we follow \citet{valenti:2005} and \citet{brewer:2016} in using the same fixed microturbulent velocity, $v_{\rm mic}=0.85$~\kms, for all G-type targets. This is a reasonable assumption for weakly or moderately active stars within a narrow parameter range around the solar values. One can suspect that this assumption does not hold for stars significantly more active than the Sun due to a modification of their convective turbulent spectrum by a magnetic field. However, such targets also exhibit a larger Zeeman intensification due to stronger fields, making the $v_{\rm mic}$ uncertainty less of a concern.

The bottom panel in Fig.~\ref{fig:ew} demonstrates that the magnetic intensification curves of the three \fe\ lines follow theoretically expected $\propto$\,$B^2$ dependence \citep{polarization:2004} up to 200--300~G and then behave linearly with $B$. It may be more problematic to recognise the presence of the field using these \fe\ lines in the quadratic regime, corresponding to the hypothetical situation of \bs\,$\la$\,300~G and $f\approx1$. However, we did not encounter such situations in our analysis.

The increase in the equivalent width of the \fe\ 5497.5~\AA\ line is steeper than for the other two magnetically sensitive lines. This difference of the equivalent width responses, coupled with the Zeeman broadening of the 5497.5~\AA\ line detectable in slower rotators, enables disentangling, to some extent, the field strength $B$ from the magnetic filling factor $f$. Nevertheless, as we will show below, the filling factor and the field strength are still partially degenerate in our approach, yielding considerably larger individual errors of $B$ and $f$ compared to the uncertainty of their product, the mean field strength \bs\,=\,$B\cdot f$.

Drawing from the forward theoretical spectrum synthesis calculations described above we proceed to the analysis of the four \fe\ lines in the spectrum of the Sun. For this purpose we consider the HARPS solar flux spectrum calibrated in wavelength with the help of a laser frequency comb \citep{molaro:2013}. The atomic and molecular data were extracted from VALD, now taking into account all known absorption features around the four \fe\ lines studied here.

\begin{figure}[!t]
\fifps{\hsize}{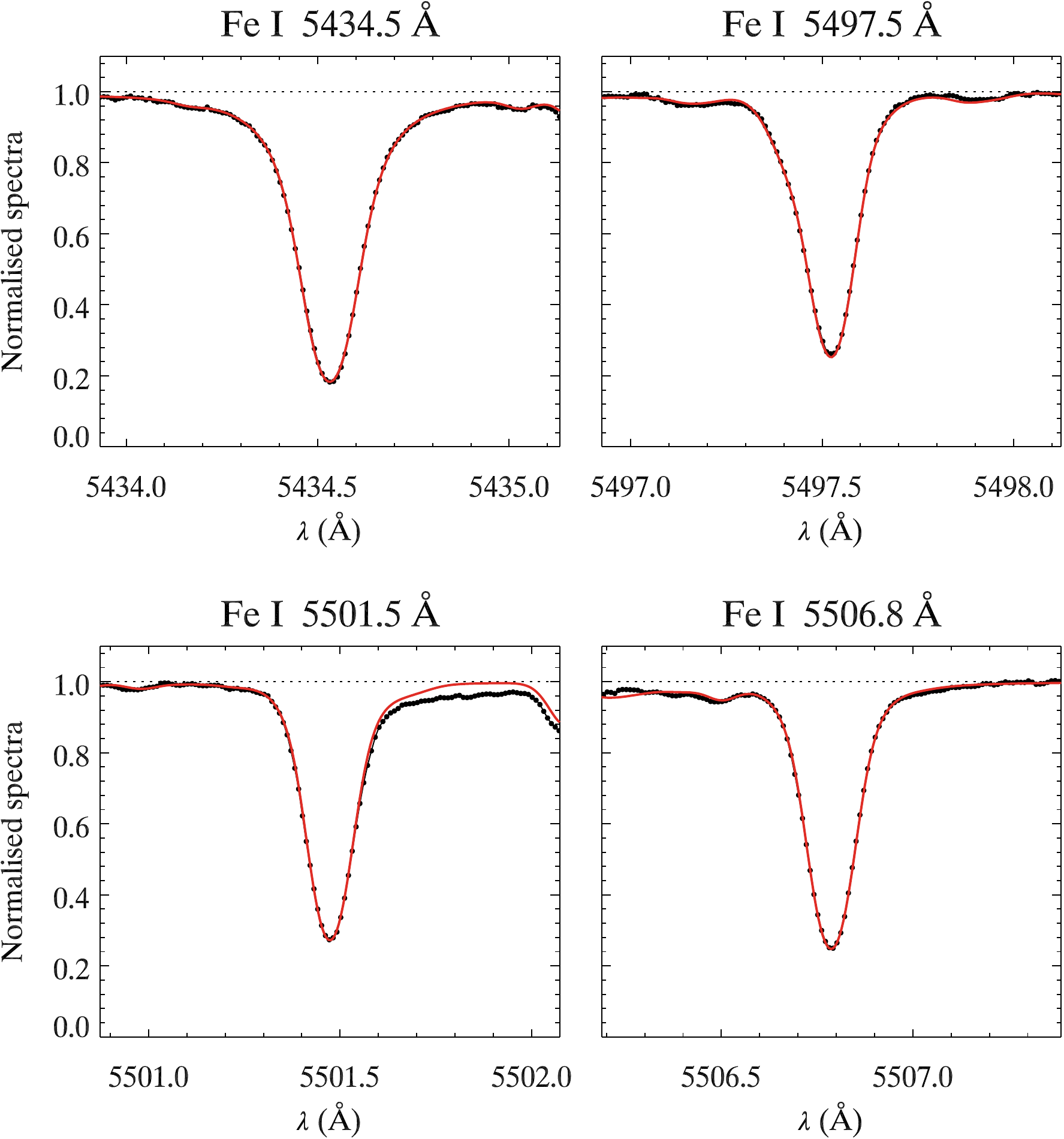}
\caption{
Solar flux spectrum obtained with the HARPS spectrograph (symbols connected by thin solid lines) compared to best fitting synthetic profiles (solid lines) for the four diagnostic \fe\ lines studied in this paper.
}
\label{fig:prfSun}
\end{figure}

A comparison between the observed solar spectrum and the best fit model calculation is shown in Fig.~\ref{fig:prfSun}. For this spectrum synthesis we adopted $v_{\rm mic}=0.85$~\kms, $v_{\rm e}\sin i=1.63$~\kms\ \citep{valenti:2005} and determined $v_{\rm mac}=2.83$~\kms\ and $\log (N_{\rm Fe}/N_{\rm tot})=-4.58$ from the 5434.5~\AA\ line. Then, the fit to the three magnetically sensitive \fe\ lines was optimised by changing both $B$ and $f$. This was accomplished with a straightforward grid search approach, varying $f$ from 0 to 1 with a step of 0.01 and $B$ from 0 to 7~kG with a 0.1~kG step and looking for a $B$, $f$ pair yielding the lowest chi-square. The chi-square probability statistics was employed to establish 68.3\% confidence limits for $B$, $f$, and their product $B\cdot f$. We found $B=2.6\substack{+3.2\\-1.2}$~kG, $f=0.07\substack{+0.07\\-0.03}$, and \bs\,=\,$0.18\substack{+0.11\\-0.05}$~kG. The 99.7\% confidence limits for \bs\ are 0.12 to 0.32 kG. This mean field strength is comparable to the average field strength of 130--220~G reported by some Zeeman and Hanle studies of the quiet Sun magnetism \citep{trujillo-bueno:2004,shchukina:2011,danilovic:2010,orozco-suarez:2012}. The solution with $f=1$ (one-component model) increases the chi-square by a factor of 1.9 relative to the best fitting two-component model and is well outside the 99.7\% confidence range.

As part of the analysis of the solar spectrum we adjusted oscillator strengths of several weak lines adjacent to the four \fe\ diagnostic features. Only one of these lines, \ion{Y}{ii} 5497.4~\AA, for which the oscillator strength had to be reduced by 0.12~dex, directly affects the wing of one of the studied \fe\ lines in the solar spectrum. Some other blending features will contribute to the wings of these \fe\ lines in the spectra of faster rotators. As can be seen from Fig.~\ref{fig:prfSun}, we have not succeeded in reproducing the shallow diffuse absorption in the far red wing of \fe\ 5501.6~\AA. Several weak C$_2$ lines contribute to the solar spectrum at those wavelengths. It is possible that the VALD C$_2$ line data are incomplete or inaccurate, explaining the discrepancy between the observed solar spectrum and the model calculation. This disagreement is unimportant for the Sun due to its low \vs. But this issue becomes progressively more problematic as \vs\ increases. Consequently, we systematically excluded the far red wing of the \fe\ 5501.6~\AA\ line from the set of wavelength intervals employed for chi-square calculations.

Here we also present a detailed account of the magnetic parameter inference for the active star HD\,129333 (EK Dra). For this series of calculations we considered the average spectrum of this star corresponding to the epoch 2009.02. This spectrum was obtained by co-adding 4 individual observations obtained over 9 nights (see Sect.~\ref{obs} for further details on the observational data employed in this study). The \fe\ 5434.5~\AA\ line profile was used to determine $v_{\rm e}\sin i=17.0$~\kms\ and $\log (N_{\rm Fe}/N_{\rm tot})=-4.46$ adopting $v_{\rm mac}=3.0$~\kms. This macroturbulent broadening parameter follows from the $v_{\rm mac}(T_{\rm eff}, \log g)$ calibration by \citet{doyle:2014} with the solar value replaced by $v_{\rm mac}^\odot=2.83$~\kms\ determined above. The macroturbulent velocity was calculated in this manner for all G-type stars in this study. However, the exact choice of the macrotubulent velocity value is unimportant in the context of our analysis. For narrow-line stars considered here a change of $v_{\rm mac}$ can be compensated by modification of \vs, resulting in the same chi-square of the fit to observations and identical magnetic field parameters.

\begin{figure}[!t]
\fifps{\hsize}{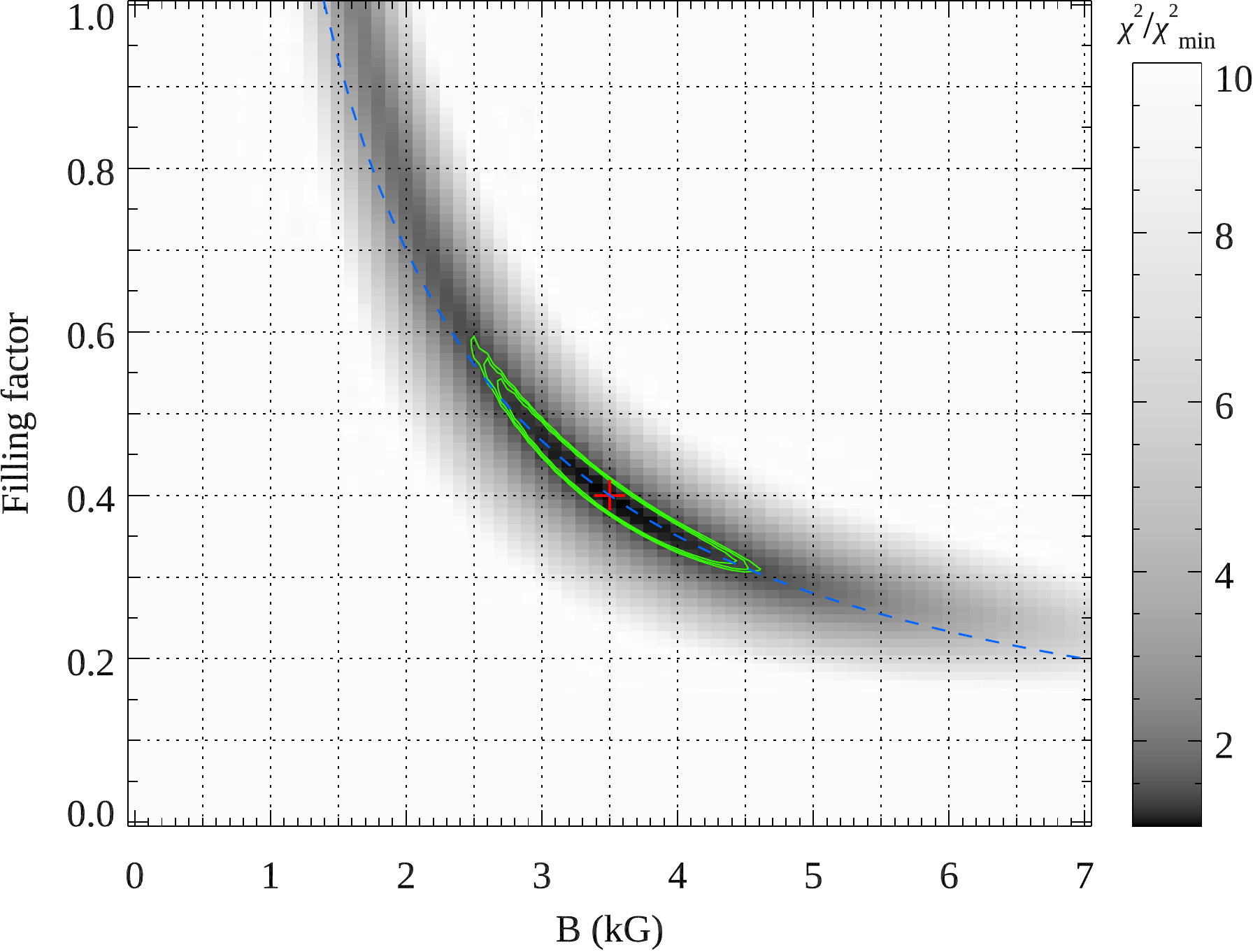}
\caption{
Illustration of the determination of magnetic field strength $B$ and filling factor $f$ for HD\,129333 (2009.02 epoch). The greyscale corresponds to the $\chi^2$ of the fit to the three magnetically sensitive \fe\ lines, as quantified by the side bar. The best fitting parameters, $B=3.5$~kG and $f=0.4$, are indicated with the cross. Solid lines show 68.3, 95.5, and 99.7\% confidence limits. The dashed line corresponds to the $B\cdot f=\mathrm{const}$ curve.
}
\label{fig:bf129333}
\end{figure}

After constraining the Fe abundance and \vs\ of HD\,129333 with the 5434.5~\AA\ line we proceeded to determination of $B$ and $f$ using the same grid search technique as was employed for modelling the solar spectrum. The resulting chi-square surface is shown in Fig.~\ref{fig:bf129333}. The best fitting parameters are $B=3.5$~kG, $f=0.40$, and $B\cdot f=1.40$~kG. The corresponding 68.3\% confidence limits are 2.7--4.4~kG, 0.32--0.54, and 1.33--1.47~kG respectively. As expected, there is a significant anti-correlation between $B$ and $f$ along the $B\cdot f=\mathrm{const}$ line. The error determination procedure adopted here fully takes this anti-correlation into account, allowing to derive realistic constraints on $B$ and $f$. In this particular case, the best one-component model ($f=1$, $B$\,=\,\bs\,=\,1.6~kG) is clearly excluded as it yields a chi-square increase by a factor of 1.8. We emphasise that the 68.3\% confidence limits discussed in this section and elsewhere in the paper should not be treated as one-$\sigma$ error bars of the normal distribution. For example, for the 2009.02 observation of HD\,129333 considered here, the 95.5\% confidence limits (corresponding to the second contour in Fig.~\ref{fig:bf129333}) are 2.6--4.5~kG, 0.31--0.56, and 1.33--1.48~kG for $B$, $f$, and \bs\ respectively. 

The two-step grid search procedure described above, with the initial determination of \vs\ and the Fe abundance using the 5434.5~\AA\ line followed by the measurement of $B$, $f$, and \bs\ from magnetically sensitive lines, is largely equivalent to a general chi-square optimisation using all four lines simultaneously. However, our approach is computationally faster, more straightforward, reproducible and less prone to degeneracies thanks to a clear separation of the information content of lines with different magnetic sensitivity. Anyway, we have verified that application of a general least-squares fitting algorithm to the observation of HD\,129333 discussed above yields the same set of magnetic field parameters, \vs, and Fe abundance as was obtained with our two-step grid search procedure.

\begin{figure}[!t]
\fifps{\hsize}{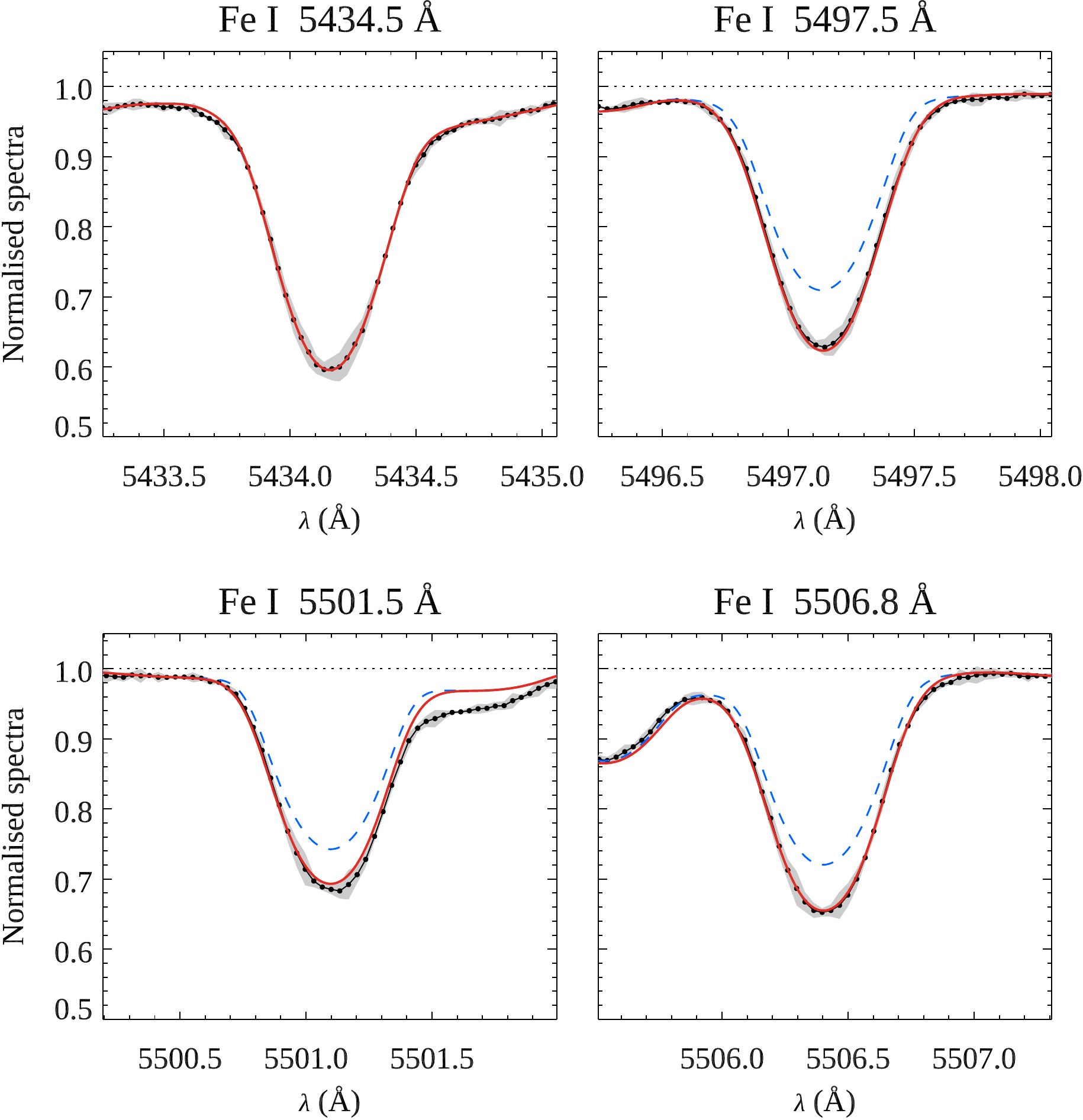}
\caption{
Magnetic field measurement for HD\,129333 (2009.02 epoch). The average observed profiles of the four \fe\ lines are shown with black symbols connected by thin solid lines. The grey curve in the background corresponds to $\pm$ twice the standard deviation for each pixel of the observed spectra. The thick red solid line shows theoretical spectrum for the magnetic model yielding the lowest $\chi^2$ according to Fig.~\ref{fig:bf129333}. The dashed blue line shows the corresponding non-magnetic spectrum.
}
\label{fig:prf129333}
\end{figure}

The observed spectrum of HD\,129333 and the best fitting magnetic model spectrum are displayed in Fig.~\ref{fig:prf129333}. This plot also shows the non-magnetic theoretical calculation for the same set of stellar parameters. The magnetic intensification of the 5497.5--5506.8~\AA\ lines is readily apparent. The equivalent width and the residual central depth of the \fe\ 5497.5~\AA\ line increase by 28\% and 9\%, respectively. Such an effect can be easily detected for this very active star even using moderate quality spectra. Fig.~\ref{fig:prf129333} also demonstrates that the rotational variability of the \fe\ lines in HD\,129333 induced by cool spots is much smaller than the Zeeman intensification signature. 

We conclude the assessment of the new magnetic field measurement methodology with investigation of the sensitivity of the analysis results to variation of stellar parameters. For this purpose we use the same observation of HD\,129333 as was discussed above. Determination of \vs, $\log (N_{\rm Fe}/N_{\rm tot})$ from the 5434.5~\AA\ line, followed by measurement of the magnetic field parameters from the 5497.5--5506.8~\AA\ lines, was repeated varying $T_{\rm eff}$ by $\pm100$~K, $\log g$ by $\pm0.1$ dex, and $v_{\rm mic}$ by $\pm0.2$~\kms. These uncertainties are about a factor of two larger than the typical errors of $T_{\rm eff}$, $\log g$, and $v_{\rm mic}$ reported by spectroscopic studies of Sun-like stars and, in this respect, represent a conservative estimate of possible systematic errors. The outcome of this error analysis is summarised in Table~\ref{tbl:errors}. We found that $B$ changes by up to 0.3~kG, $f$ by up to 0.03, and \bs\ by up to 0.11~kG in response to the variation of stellar  parameters. In all cases these changes are compatible with the formal error bars of the magnetic field parameters obtained using the reference set of $T_{\rm eff}$, $\log g$, and $v_{\rm mic}$. This suggests that our magnetic field measurements are not strongly affected by the uncertainties of stellar parameters adopted from the literature.

Finally, we study the impact of neglecting the multi-component nature of active star atmospheres on our magnetic field analysis. The work by \citet{jarvinen:2018} demonstrated that HD\,129333 has several cool spots with 200--1000~K temperature contrast occupying 14\% of the stellar surface. We therefore repeated determination of the magnetic field parameters for the 2009.02 observation of HD\,129333 assuming that 20\% of the star is 500~K cooler than the rest of the surface. In this test we assumed that both hot and cool atmospheric components have the same distribution of small-scale magnetic field strengths, i.e. both have the same $B$ and $f$. Table~\ref{tbl:errors} shows that adopting this multi-component model has a very minor impact. Both $B$ and $f$ remain within 68\% confidence limits of the reference determination whereas \bs\ is altered by just 0.02~kG.

A systematic temperature difference between magnetic and non-magnetic regions causes a more significant modification of our spectrum fitting results. To test  this effect, we repeated analysis of the 2009.02 spectrum of HD\,129333 assuming that (a) magnetic regions are 100~K cooler relative to the mean stellar effective temperature $T_{\rm eff}=5845$~K and non-magnetic regions are hotter by the same amount ($T_{\rm mag}=5745$~K, $T_0=5945$~K) and (b) the temperature difference is reversed ($T_{\rm mag}=5945$~K, $T_0=5745$~K). In both of these situations the 5434.5~\AA\ line is affected by the choice of the filling factor $f$, so our two-step procedure was iterated until convergence was achieved for $f$ and Fe abundance. The results of these tests are reported in the last two rows of Table~\ref{tbl:errors}. A 200~K temperature contrast modifies $B$, $f$, and \bs\ systematically by the amount comparable to formal error bars. This change is explained primarily by the difference of continuum brightness of the two spectral contributions, as predicted in Sect.~\ref{magspec}. If magnetic regions are cooler, their contribution to the total spectrum is diminished (case a) and a larger \bs\ is required to fit the observations. The opposite is happening in the case (b). In fact, one can analytically predict \bs\ correction factors of 1.08 and 0.91 for the cases (a) and (b), respectively, considering the continuum brightness at $\lambda=5500$~\AA\ of the spectra corresponding to $T_{\rm eff}=5745$ and 5945~K. The actual change of the mean field strength according to Table~\ref{tbl:errors} is a factor of 1.07 and 0.92 for the scenarios with cooler and hotter magnetic regions, respectively.

\begin{table*}[!t]
\centering
\caption{Sensitivity of the magnetic field analysis results for HD\,129333 (2009.02 epoch) to the variation of stellar parameters.\label{tbl:errors}}
\renewcommand{\arraystretch}{1.4}
\begin{tabular}{lcccccc}
\hline
\hline
Changed parameter & $\log (N_{\rm Fe}/N_{\rm tot})$ & $v_{\rm e}\sin i$ (\kms) & $B$ (kG) & $f$ & \bs\ (kG) & $\Delta$\bs\ (kG) \\
\hline
reference $T_{\rm eff}$, $\log g$, $v_{\rm mic}$ & $-4.46$ & 17.0 & $3.5\substack{+0.9 \\ -0.8}$ & $0.40\substack{+0.14 \\ -0.08}$ & $1.40\substack{+0.07 \\ -0.07}$ &         \\
$T_{\rm eff}+100$~K                              & $-4.35$ & 16.9 & $3.8\substack{+0.8 \\ -0.9}$ & $0.37\substack{+0.13 \\ -0.06}$ & $1.41\substack{+0.07 \\ -0.06}$ & $+0.01$ \\
$T_{\rm eff}-100$~K                              & $-4.56$ & 17.1 & $3.2\substack{+1.0 \\ -0.7}$ & $0.42\substack{+0.14 \\ -0.10}$ & $1.34\substack{+0.08 \\ -0.06}$ & $-0.06$ \\ 
$\log g +0.1$                                    & $-4.51$ & 16.9 & $3.6\substack{+0.8 \\ -0.9}$ & $0.40\substack{+0.15 \\ -0.07}$ & $1.44\substack{+0.06 \\ -0.07}$ & $+0.04$ \\
$\log g -0.1$                                    & $-4.42$ & 17.0 & $3.5\substack{+1.0 \\ -0.9}$ & $0.38\substack{+0.15 \\ -0.08}$ & $1.33\substack{+0.06 \\ -0.08}$ & $-0.07$ \\
$v_{\rm mic}+0.2$~\kms\                          & $-4.49$ & 17.1 & $3.4\substack{+0.9 \\ -1.0}$ & $0.38\substack{+0.20 \\ -0.08}$ & $1.29\substack{+0.10 \\ -0.07}$ & $-0.11$ \\
$v_{\rm mic}-0.2$~\kms\                          & $-4.44$ & 16.9 & $3.7\substack{+0.8 \\ -0.8}$ & $0.40\substack{+0.12 \\ -0.07}$ & $1.48\substack{+0.05 \\ -0.08}$ & $+0.08$ \\
$\Delta T_{\rm spot}=500$~K, $f_{\rm spot}=0.2$ & $-4.56$ & 16.9 & $3.2\substack{+1.0 \\ -0.7}$ & $0.43\substack{+0.15 \\ -0.10}$ & $1.38\substack{+0.07 \\ -0.07}$ & $-0.02$ \\
$T_0 - T_{\rm mag}=+200$~K  & $-4.45$ & 17.0 & $3.0\substack{+1.2 \\ -1.2}$ & $0.50\substack{+0.29 \\ -0.12}$ & $1.50\substack{+0.10 \\ -0.11}$ & $+0.10$ \\
$T_0 - T_{\rm mag}=-200$~K & $-4.49$ & 17.0 & $3.2\substack{+1.0 \\ -0.7}$ & $0.34\substack{+0.07 \\ -0.06}$ & $1.29\substack{+0.11 \\ -0.09}$ & $-0.11$ \\
\hline
\end{tabular}
\end{table*}

\subsection{Target selection and stellar parameters}

The targets for our study were selected according to the following criteria (i) fundamental stellar parameters close to solar values, (ii) information on the global magnetic field is available from previous ZDI studies, and (iii) high-quality optical spectra are available. The first of these constraints is motivated by our general goal of expanding the number of early-G dwarfs with reliable magnetic field measurements and certain limitations of our magnetic diagnostic method (see Sect.~\ref{method}), which relies on solar calibration of the line list and turbulent velocities. Starting from the summaries of ZDI studies published by \citet{vidotto:2014} and \citet{see:2019}, we identified 14 dwarfs in the spectral type range from G0 to G7 which satisfy these criteria. This list includes 6 stars (HD\,1835, HD\,20630, HD\,39587, HD\,72905, HD\,129333, HD\,206860) from the well-studied ``Sun in Time'' reference sample \citep{ribas:2005,gudel:2007,rosen:2016,fichtinger:2017,pognan:2018} composed of solar twins at different evolutionary stages. To this G dwarf sample we added one additional very active cooler star, LQ~Hya (HD\,82558), to test applicability of our method to faster rotators. Except this early-K dwarf, believed to have a mass of about 0.8$M_\odot$, all our targets have masses within $\pm10$\% of the solar value \citep[see Table~1 in][]{see:2019}.

Table~\ref{tbl:targets} summarises relevant parameters of our targets. The first three columns list the HD number, the commonly used name, and the spectral type adopted from the Hipparcos input catalogue \citep{turon:1993}. This is followed by the effective temperature $T_{\rm eff}$ and surface gravity $\log g$ taken mainly from \citet{valenti:2005}. The stellar ages and rotational periods reported in columns 6 and 7 are adopted primarily from \citet{vidotto:2014} and \citet{see:2019}, respectively. The 8th column in Table~\ref{tbl:targets} provides Rossby numbers calculated by dividing the rotational period by the convective turnover time. The latter was computed according to the prescription given by \citet{cranmer:2011}. The 9th column reports \vs\ determined for each star as part of our spectrum synthesis analysis described in Sect.~\ref{method}. The last two columns in Table~\ref{tbl:targets} list two widely used proxies of the stellar magnetic activity. The ratio of the X-ray to bolometric luminosity $\log L_{\rm X}/L_{\rm bol}$ was adopted from \citet{vidotto:2014} and \citet{wright:2011} or calculated using the X-ray fluxes from \citet{boller:2016}. The \ion{Ca}{ii} H\&K chromospheric emission indicator $\log R^\prime_{\rm HK}$ is the median value of the measurements found in the catalogue by \citet{boro-saikia:2018}.

As discussed above, we used $v_{\rm mac}$ calculated with the modified calibration of \citet{doyle:2014} and employed the same $v_{\rm mic}=0.85$~\kms\ for all G dwarfs. For HD\,82558 we used $v_{\rm mac}=1.5$~\kms\ and $v_{\rm mic}=0.5$~\kms\ \citep{cole:2015,flores-soriano:2017}. All model atmospheres required for the line profile synthesis were extracted from the {\sc MARCS} \citep{gustafsson:2008}\footnote{\url{http://marcs.astro.uu.se}} model atmosphere grid. The logarithm of surface gravity is close to 4.5 for the majority of stars in our sample. Therefore, we used a pair of MARCS models with $\log g=4.5$ and $T_{\rm eff}$ values bracketing the stellar effective temperature. Theoretical spectra were obtained using linear interpolation between {\sc Synmast} calculations with these two atmospheric models. For HD\,131156A ($\xi$~Boo~A), which has $\log g =4.65$, we used a bilinear interpolation between theoretical spectra calculated with four model atmospheres. For HD\,82558 (LQ~Hya) we used a single model atmosphere with $T_{\rm eff}=5000$~K and $\log g=4.0$ \citep{cole:2015}.

\begin{table*}[!t]
\centering
\caption{Parameters of target stars. \label{tbl:targets}}
%{\footnotesize
\begin{tabular}{lllllllllll}
\hline
\hline
%HD       & Name           & Sp. type$^1$ & $T_{\rm eff}$ (K)$^2$ & $\log g$$^2$ & Age (Myr)$^7$ & $P_{\rm rot}$ (d)$^{10}$ & Ro & $v_{\rm e}\sin i$ (\kms) &  $\log L_{\rm X}/L_{\rm bol}$$^{7}$   & $\log R^\prime_{\rm HK}$$^{14}$ \\
HD       & Name           & Sp. type$^1$ & $T_{\rm eff}$$^2$ & $\log g$$^2$ & Age$^7$ & $P_{\rm rot}$$^{10}$ & Ro & $v_{\rm e}\sin i$ & $\log L_{\rm X}/L_{\rm bol}$$^{7}$  & $\log R^\prime_{\rm HK}$$^{14}$ \\
& & & (K) & & (Myr) & (d) & & (\kms) & & \\
\hline
   1835  & BE~Cet         & G3\,V   & 5837     & 4.47     & 600$^8$ & 7.78$^{11}$  & 0.659 &  6.3 & $-4.43^{11}$ & $-4.43$ \\
  20630  & $\kappa^1$~Cet & G5\,V   & 5742     & 4.49     & 600     & 9.3          & 0.696 &  4.7 & $-4.71$      & $-4.40$ \\
  29615  &                & G3\,V   & 5866$^3$ & 4.41$^6$ &  27     & 2.34         & 0.207 & 20.1 & $-3.62^{13}$ &         \\
  39587  & $\chi^1$~Ori   & G0\,V   & 5882     & 4.34     & 500     & 4.83         & 0.437 &  9.4 & $-4.64$      & $-4.37$ \\
  56124  &                & G0\,V   & 5848     & 4.46     & 4500    & 18           & 1.549 &  0.6 & $-5.23^{13}$ & $-4.78$ \\
  72905  & $\pi^1$~UMa    & G1.5\,V & 5873$^4$ & 4.44     & 500     & 4.9          & 0.437 &  9.6 & $-4.64$      & $-4.33$ \\
  73350  & V401~Hya       & G5\,V   & 5802     & 4.48     & 510     & 12.3         & 0.993 &  3.2 & $-4.80$      & $-4.50$ \\
  76151  &                & G3\,V   & 5790     & 4.55     & 3600$^2$& 20.5         & 1.629 &  0.0 & $-5.12^{13}$ & $-4.66$ \\
  82558  & LQ~Hya         & K1\,V   & 5000$^5$ & 4.00$^5$ &  50     & 1.601$^5$    & 0.067 & 28.3 & $-3.06$      & $-3.97$ \\
 129333  & EK~Dra         & G1.5\,V & 5845     & 4.47     & 120     & 2.606$^{12}$ & 0.223 & 17.0 & $-3.60$      & $-4.09$ \\
 131156A & $\xi$~Boo~A    & G7\,V   & 5570     & 4.65     & 200$^9$ & 6.4          & 0.400 &  4.9 & $-4.44$      & $-4.32$ \\
 166435  &                & G1\,IV  & 5843     & 4.44     & 3800    & 3.43         & 0.293 &  7.6 & $-4.08$      & $-4.26$ \\
 175726  &                & G0\,V   & 5998     & 4.41     & 500     & 3.92         & 0.434 & 12.4 & $-4.58$      & $-4.38$ \\
 190771  &                & G2\,V   & 5834     & 4.44     & 2700    & 8.8          & 0.742 &  3.4 & $-4.45$      & $-4.39$ \\
 206860  & HN~Peg         & G0\,V   & 5974     & 4.47     & 260     & 4.55         & 0.481 & 10.1 & $-4.65$      & $-4.37$ \\
\hline
\end{tabular}
%}
\tablebib{(1) \citealt{turon:1993}; (2) \citealt{valenti:2005}; (3) \citealt{mcdonald:2012}; (4) \citealt{gonzalez:2010}; (5) \citealt{cole:2015}; (6) \citealt{allende-prieto:1999a}; (7) \citealt{vidotto:2014}; (8) \citealt{rosen:2016}; (9) \citealt{olah:2016}; (10) \citealt{see:2019}; (11) \citealt{wright:2011}; (12) \citealt{jarvinen:2018}; (13) \citealt{boller:2016}; (14) \citealt{boro-saikia:2018}.}
\end{table*}

\subsection{Observational data}
\label{obs}

High-resolution archival spectra of target stars were collected from the two sources. We used the PolarBase archive \citep{petit:2014}\footnote{\url{http://polarbase.irap.omp.eu}} to retrieve 755 observations of 15 stars obtained with the twin spectropolarimeters ESPaDOnS and Narval, installed at the 3.6m Canada-France-Hawaii Telescope (CFHT) and the 2m Telescope Bernard Lyot (TBL) respectively. These data were processed with an automatic reduction pipeline \citep{donati:1997} running at the telescopes and are available from PolarBase in fully reduced format. Each observation covers the 3700--10500~\AA\ spectral window at the resolving power of $R=65000$. We have also used 107 spectropolarimetric observations of 5 stars acquired with the HARPSpol instrument mounted at the 3.6m ESO telescope. These spectra cover the 3780--6910~\AA\ wavelength region at the resolution of $R=110000$. The HARPSpol spectra is a mixture of public data available from the ESO archive\footnote{\url{http://archive.eso.org}} and observations which we have collected during several recent visitor observing runs described elsewhere \citep{hackman:2016,rosen:2016,lehtinen:2020}. All HARPSpol observations were reduced with the {\sc REDUCE} package \citep{piskunov:2002} following the procedure detailed in \citet{makaganiuk:2012} and \citet{rusomarov:2013}. Observations from all three instruments were normalised to the continuum with the method described by \citet{rosen:2018}. 

Most of the datasets considered here were obtained for the purpose of monitoring rotational variation and reconstructing global magnetic field topologies with ZDI. Consequently, each observing epoch is represented by anywhere between 2 and 44 individual observations obtained over the time span from one night to a few months. On average, there are 10 observations taken over 25 days. Our first-look analysis did not reveal any differential variability of the magnetically sensitive \fe\ lines relative to \fe\ 5434~\AA. This indicates that the small-scale magnetic fields investigated in this paper are  distributed approximately uniformly over stellar surfaces. Moreover, in all cases the amplitude of rotational modulation due to cool spots was found to be significantly smaller than the differential magnetic intensification signature (e.g. see Fig.~\ref{fig:prf129333}). This justifies co-adding all spectra obtained within the same observing run to enhance the signal-to-noise ratio and average out profile distortions caused by cool spots. This approach yielded 78 high-quality average spectra for 15 stars, with the largest number of epochs (10) available for HD\,206860 and the smallest number (2) for HD\,29615 and HD\,175726. Information on individual average spectra employed in our study is given in Table~\ref{tbl:results}. The first three columns of this table list the observing epoch, the number of individual observations used to calculate the average spectrum, and the facility where these data were acquired.

\section{Magnetic fields of active Sun-like stars}
\label{results}

\subsection{HD\,1835 (BE Cet)}

This star is part of the group of young solar analogues included in the ``Sun in Time'' sample \citep{ribas:2005,gudel:2007}. However, its magnetic activity was studied relatively infrequently in the past compared to other stars in that sample. Here we report three field strength measurements based on the two epochs of HARPSpol data and a pair of spectra obtained with ESPaDOnS. We infer \bs\,=\,0.61--0.75~kG, with the difference between the extreme values being statistically significant. Comparison between one of the observations and the model spectra, shown in Fig.~\ref{fig:prf1835}, reveals a clear evidence of both Zeeman broadening and intensification effects. A single \bs\,=\,0.45~kG determination can be found for this star in the literature \citep{saar:1987}. The global magnetic field of HD\,1835 was investigated by \citet{rosen:2016} based on the earliest data set analysed here.

\subsection{HD\,20630 ($\kappa^1$~Cet)}

This object is considered to be one of the best young Sun proxies \citep[e.g.][]{ribas:2010,fichtinger:2017,lynch:2019} and is intensely studied in this role. We derived 4 field strength measurements in the range from 0.45 to 0.55~kG using observations spanning from 2012 to 2017. The difference between the extreme \bs\ determinations is not statistically significant. Figure~\ref{fig:prf20630} shows an example of the fit to the four \fe\ lines with and without magnetic field. Multiple previous estimates of \bs\ are available for HD\,20630, with values ranging between 0.32 and 0.52~kG \citep{saar:1987,saar:1992}. These determinations are generally consistent with our results. The global field topology of HD\,20630 was independently mapped by \citet{rosen:2016} and \citet{do-nascimento:2016} using the circular polarisation spectra from the two earlier data sets analysed here.

\subsection{HD\,29615}

Magnetic activity of this very young rapidly rotating (\vs\,=\,20~\kms, $P_{\rm rot}$\,=\,2.34~d) star was studied by \citet{waite:2015} and \citet{hackman:2016}. These authors focused on mapping distribution of cool spots using high-resolution spectra and reconstructed global magnetic field topology with ZDI. Despite a significant rotational broadening, we detect an unambiguous signature of magnetic intensification (Fig.~\ref{fig:prf29615}). The strengthening of magnetically sensitive lines does not depend on rotational phase and is much stronger than the line profile variability caused by cool spots. We derive a mean field strength of 1.30--1.38~kG, with about 50\% of the stellar surface covered by $\approx$\,2.7~kG field. Our analysis was based on the two sets of HARPSpol spectra collected in 2013 and 2017.

\subsection{HD\,39587 ($\chi^1$~Ori)}

This star is another frequently studied young solar analogue. We were able to obtain 8 individual \bs\ measurements using spectra recorded in the time interval from 2007 to 2017. An example of the field detection based on the earliest data set is presented in Fig.~\ref{fig:prf39587}. Most of our field strength determinations cluster around \bs\,=\,0.46~kG. The difference between the extreme values (0.38 and 0.50~kG) is marginally significant, hinting at a long-term trend of decreasing average photospheric magnetic field strength. All our field strength measurements appear to be weaker than the single estimate \bs\,=\,0.60~kG available in the literature \citep{saar:1987}. The global field topology of HD\,39587 was studied by \citet{rosen:2016} using 4 out of 8 observing epochs considered here.

\subsection{HD\,56124}

This star is frequently included in studies based on ZDI analyses of cool stars \citep[e.g.][]{vidotto:2014,see:2019}, although a detailed account of its magnetic mapping is yet to published (Petit et al., in prep.). HD\,56124 is the least active object in our study in terms of its $\log L_{\rm X}/L_{\rm bol}$ and $\log R^\prime_{\rm HK}$ indices (Table~\ref{tbl:targets}). Magnetic field effects on its line profiles are very subtle (see Fig.~\ref{fig:prf56124}). Nevertheless, we were able to determine a consistent and formally significant \bs\ of about 0.22~kG from the Narval spectra corresponding to four different observing epochs in 2008--2012. The mean magnetic field of HD\,56124 is the weakest among the stars studied here and is formally compatible with the solar average field strength inferred with our method in Sec.~\ref{method}. HD\,56124 is also the only object other than the Sun for which our analysis yields a magnetic field filling factor below 10\%.

\subsection{HD\,72905 ($\pi^1$~UMa)}

This is another very frequently studied young solar twin with an age, rotation rate and activity indices very similar to HD\,39587. We derived five mean field strength measurements using Narval spectra taken in 2007--2016. A comparison of one of the observed spectrum with the best fitting magnetic model calculation is presented in Fig.~\ref{fig:prf72905}. All our \bs\ determinations fall in a narrow range around 0.59~kG, with an insignificant scatter. These results show that HD\,72905 possess a stronger average magnetic field than HD\,39587. The same difference was also found in the ZDI analysis of the global fields of the these two stars \citep{rosen:2016}. No previous Zeeman broadening estimates of magnetic field strength are available for HD\,72905.

\subsection{HD\,73350 (V401~Hya)}

This object was included in the ZDI study of four solar twin stars with different rotation rates \citep{petit:2008}. Here we analysed four epochs of Narval observations, including the data used by \citet{petit:2008}. We inferred \bs\ to be in the 0.43--0.52~kG interval and found no conclusive evidence of the field strength variation from one epoch to the next. One of our field strength measurements is illustrated in Fig.~\ref{fig:prf73350}.

\subsection{HD\,76151}

This star was also part of the ZDI study by \citet{petit:2008}. It is an old star with a relatively low activity level and the longest rotational period ($P_{\rm rot}=20.5$~d) among the stars studied here. We derived 5 \bs\ measurements based on the spectropolarimetric observations collected with Narval in 2007--2015. One of these measurements is shown in Fig.~\ref{fig:prf76151}. All field strength determinations obtained for HD\,76151 are consistent, within error bars, with \bs\,=\,0.41~kG. This indicates that HD\,76151 is a more active object compared to HD\,56124 despite having a slightly longer rotational period. The tomographic mapping of the global magnetic field topology with ZDI points to a similar disparity between magnetism of these two stars \citep{petit:2008,see:2019}.

\subsection{HD\,82558 (LQ~Hya)}

LQ~Hya is one of the most frequently studied very active, young, rapidly rotating late-type dwarf stars. It is a popular target for Doppler mapping of the surface distributions of cool spots and global magnetic field \citep{donati:1999b,donati:2003,kovari:2004,cole:2015,flores-soriano:2017}. This star is the fastest rotator (\vs\,=\,28~\kms, $P_{\rm rot}=1.60$~d) and shows the strongest X-ray and chromospheric emission in our sample. It is also somewhat cooler and less massive than the rest of the stars studied here. Due to its rapid rotation, we had to slightly change the field strength measurement methodology by excluding \fe\ 5506.78~\AA\ from the group of magnetically sensitive lines. This was motivated by the difficulty of modelling the blending of the blue wing of this line by \ion{Mn}{i} 5505.87~\AA\ and \fe\ 5505.68~\AA. Nevertheless, as illustrated by Fig.~\ref{fig:prf82558}, the evidence of a magnetic intensification in the two remaining magnetic diagnostic lines is unambiguous and cannot be confused with the rotational profile variations. Simultaneous fit of the three \fe\ lines requires \bs\ of about 2~kG. This mean field strength was consistently obtained from three epochs of HARPSpol observations and one data set obtained with ESPaDOnS. The same four epochs of spectropolarimetric data were analysed with ZDI by \citet{lehtinen:2020}, allowing us to make a direct comparison of the global and total magnetic fields for this star.

Considering previous Zeeman broadening studies of HD\,82558, \citet{saar:1996} reported a single $B\cdot f$ measurement of 2.45~kG using near-infrared spectra, in good agreement with the outcome of our study. On the other hand, \citet{saar:1992a} investigated the possibility of deriving a spatially resolved distribution of magnetic field strength for HD\,82558 by combining temperature DI with a magnetic intensification analysis in the optical. Their preliminary study, based on a small number of low-quality spectra, suggested a high-contrast map of $B\cdot f$ with the extremes at 0.1 and 2.5~kG and a surface-averaged field strength of only $\approx$\,1.0~kG. Such a large field strength variation across the stellar surface appears to contradict our observations of the magnetically sensitive lines in the spectrum of HD\,82558 because these features do not exhibit any noticeable additional variability compared to the reference magnetic null line (see Fig.~\ref{fig:prf82558}).

\subsection{HD\,129333 (EK~Dra)}

EK~Dra is the most active object among the well-established young solar twins. It is frequently studied with DI \citep{strassmeier:1998,jarvinen:2007,jarvinen:2018} and ZDI \citep{rosen:2016,waite:2017} inversion techniques, providing a key reference point for investigations of different activity-rotation-age relationships among young Suns. Here we analysed 6 epochs of observing data obtained with the Narval and ESPaDOnS spectropolarimeters. Four of these data sets were previously used to produce global magnetic field maps. A detailed discussion of the derivation of \bs, $B$, and $f$ for one of the average spectra of HD\,129333 was given in Sect.~\ref{method} and illustrated by Figs.~\ref{fig:bf129333} and \ref{fig:prf129333}. Considering results for all 6 epochs, we found \bs\ to be in the 1.36--1.48~kG range for this star. The difference between these extremes is not statistically significant. However, there is a trend of increasing \bs\ from the earliest observation in 2006 to the latest data set taken in 2016.

\subsection{HD\,131156A ($\xi$~Boo~A)}

This star is a moderately active late-G dwarf in a wide binary system with an active K dwarf. Both components of this system are often used for benchmarking relations between different magnetic activity indicators \citep[e.g.][]{wood:2010,finley:2019}. The global magnetic field of HD\,131156A was studied in detail by \citet{morgenthaler:2012}. Here we  used the same 6 epochs as were analysed in that paper as well as two more recent Narval data sets. These observations cover the time interval from 2005 to 2015. We determined the mean field strength of 0.78~kG, with a formally insignificant scatter around this value for individual epochs. There is, however, a hint of a long-term trend in the \bs\ data as the field strength is systematically increasing from 2008--2010 until the latest observing epochs. An example of the fit to observations corresponding to one of the recent epochs is shown in Fig.~\ref{fig:prf131156A}.

Separate determination of $B$ and $f$, albeit rather uncertain, yields anomalous results for HD\,131156A. All other stars in our sample exhibit $B$ around 3~kG with a magnetic filling factor gradually increasing with the stellar activity level up to $f$\,$\la$\,50\%. In contrast to this behaviour, HD\,131156A shows a weaker field covering $69\pm28$\% of the stellar surface. This is the largest magnetic filling factor derived in our study. There are no obvious reasons, besides a lower mass and thus a somewhat thicker convection zone, for HD\,131156A to show a different surface field strength distribution compared to earlier G dwarfs.

\subsection{HD\,166435}

This star is a moderately active solar analogue with a rotational period of 3.43~d. We derived \bs\,=\,0.69~kG with essentially no scatter in the mean field strength values corresponding to four observing epochs. One of the field strength determinations is illustrated in Fig.~\ref{fig:prf166435}. All observations of HD\,166435 used in our study were obtained with Narval in 2010--2016. The corresponding ZDI analysis (Petit et al., in prep.) is not published yet, but the summary of the global field mapping results is available in the literature \citep{vidotto:2014,see:2019}.

\subsection{HD\,175726}

This object is similar to HD\,166435 in terms of the rotational period and the fact that only a summary of ZDI results has been published. At the same time, HD\,175726 exhibits a systematically weaker magnetic activity according to the coronal and chromospheric emission indices. It is therefore not surprising that we obtained nearly a factor of two weaker mean field, \bs\,=\,0.37~kG, for this star compared to HD\,166435. These measurements relied on the average spectra corresponding to two Narval data sets collected in 2008 and 2012. Figure~\ref{fig:prf175726} shows a comparison between theoretical model and observations for the former epoch.

\subsection{HD\,190771}

This star was included in the ZDI study of the four solar twins by \citet{petit:2008}. Subsequently, \citet{petit:2009} reported a polarity reversal of the global magnetic field and published ZDI results for three separate epochs. Here we analysed observations obtained at 9 epochs, covering 10 years starting from 2007 and ending in 2016. This is the second largest data set (after HD\,206860) investigated in this study. Our field strength measurements indicate \bs\,=\,0.59~kG without any evidence of an epoch-to-epoch scatter or a long-term variation. An example of the field strength determination for this star is shown in Fig.~\ref{fig:prf190771}.

\subsection{HD\,206860 (HN~Peg)}

This active, young Sun-like star was studied with ZDI by \citet{boro-saikia:2015} and \citet{rosen:2016}. Combining the data analysed in these studies with newer observing material yields 10 epochs spread  between 2007 and 2016. This is the largest collection of spectra, in terms of the number of epochs, analysed in our study. The modelling of the \fe\ magnetically sensitive lines suggests \bs\ around 0.45~kG, with individual field strength determinations ranging from 0.40 to 0.55~kG. One of our field strength measurements is shown in Fig.~\ref{fig:prf206860}. Given the formal error bars, the \bs\ scatter appears to be significant. A long-term behaviour of the mean field strength suggests a quasi-periodic variation, with apparent magnetic minima in 2008 and 2014.

\section{Discussion}
\label{discussion}

\subsection{Correlation with stellar parameters}

\begin{table*}[!t]
\centering
\caption{Mean magnetic field characteristics derived in this study in comparison with the global magnetic field strength inferred with ZDI. \label{tbl:mean_results}}
\renewcommand{\arraystretch}{1.4}
\begin{tabular}{lcccll}
\hline
\hline
Star & $B$ (kG) & $f$ & \bs\ (kG) & \bzdi\ (G) & ZDI reference \\
\hline
HD\,1835                     & $3.1\substack{+0.6 \\ -0.6}$ & $0.22\substack{+0.06 \\ -0.04}$ & $0.68\substack{+0.05 \\ -0.05}$ & 19                         & \citet{rosen:2016} \\
HD\,20630                    & $2.7\substack{+0.6 \\ -0.7}$ & $0.19\substack{+0.06 \\ -0.04}$ & $0.50\substack{+0.05 \\ -0.06}$ & $24\substack{+3 \\ -3}$    & \citet{rosen:2016} \\
HD\,29615                    & $2.7\substack{+2.0 \\ -1.2}$ & $0.49\substack{+0.51 \\ -0.22}$ & $1.34\substack{+0.24 \\ -0.16}$ & 89                         & \citet{hackman:2016} \\
HD\,39587                    & $2.9\substack{+1.4 \\ -1.5}$ & $0.16\substack{+0.21 \\ -0.05}$ & $0.46\substack{+0.09 \\ -0.07}$ & $16\substack{+4 \\ -3}$    & \citet{rosen:2016} \\
HD\,56124                    & $3.2\substack{+2.2 \\ -1.1}$ & $0.07\substack{+0.04 \\ -0.02}$ & $0.22\substack{+0.09 \\ -0.05}$ & 2.2                        & \citet{see:2019} \\
HD\,72905                    & $3.2\substack{+1.1 \\ -1.2}$ & $0.19\substack{+0.14 \\ -0.04}$ & $0.59\substack{+0.06 \\ -0.07}$ & $28\substack{+4 \\ -4}$    & \citet{rosen:2016} \\
HD\,73350\tablefootmark{a}   & $2.9\substack{+1.0 \\ -0.7}$ & $0.17\substack{+0.07 \\ -0.04}$ & $0.49\substack{+0.07 \\ -0.07}$ & 11                         & \citet{petit:2008} \\
HD\,76151\tablefootmark{a}   & $2.0\substack{+0.9 \\ -0.7}$ & $0.21\substack{+0.15 \\ -0.07}$ & $0.41\substack{+0.07 \\ -0.07}$ & 3.0                        & \citet{petit:2008} \\
HD\,82558                    & $4.5\substack{+1.4 \\ -2.1}$ & $0.45\substack{+0.54 \\ -0.10}$ & $2.01\substack{+0.32 \\ -0.15}$ & $169\substack{+87 \\ -12}$ & \citet{lehtinen:2020} \\
HD\,129333                   & $3.7\substack{+0.9 \\ -1.2}$ & $0.38\substack{+0.17 \\ -0.08}$ & $1.40\substack{+0.09 \\ -0.07}$ & $78\substack{+15 \\ -21}$  & \citet{waite:2017} \\
HD\,131156A\tablefootmark{a} & $1.2\substack{+0.4 \\ -0.3}$ & $0.69\substack{+0.29 \\ -0.28}$ & $0.78\substack{+0.13 \\ -0.13}$ & $36\substack{+26 \\ -13}$  & \citet{morgenthaler:2012} \\
HD\,166435                   & $2.9\substack{+1.1 \\ -1.1}$ & $0.24\substack{+0.17 \\ -0.07}$ & $0.69\substack{+0.09 \\ -0.10}$ & 20                         & \citet{see:2019} \\
HD\,175726                   & $3.9\substack{+2.0 \\ -2.6}$ & $0.10\substack{+0.27 \\ -0.02}$ & $0.37\substack{+0.11 \\ -0.08}$ & 10                         & \citet{see:2019} \\
HD\,190771\tablefootmark{a}  & $3.1\substack{+0.8 \\ -0.7}$ & $0.19\substack{+0.07 \\ -0.04}$ & $0.59\substack{+0.08 \\ -0.07}$ & 14                         & \citet{petit:2008} \\
HD\,206860                   & $3.6\substack{+1.5 \\ -1.6}$ & $0.13\substack{+0.14 \\ -0.03}$ & $0.45\substack{+0.08 \\ -0.08}$ & $22\substack{+3 \\ -9}$    & \citet{rosen:2016} \\
\hline
\end{tabular}
\tablefoot{The 5th column lists the mean ZDI field strength, \bzdi, taken from the references listed in the last column. For stars with multiple ZDI maps, column 5 gives the median field strength and the error bars corresponding to extreme values.\\
\tablefoottext{a}{\bzdi\ is given according to the compilation by \citet{see:2019}.}
}
\end{table*}

\begin{figure*}[!t]
\fifps{6.0cm}{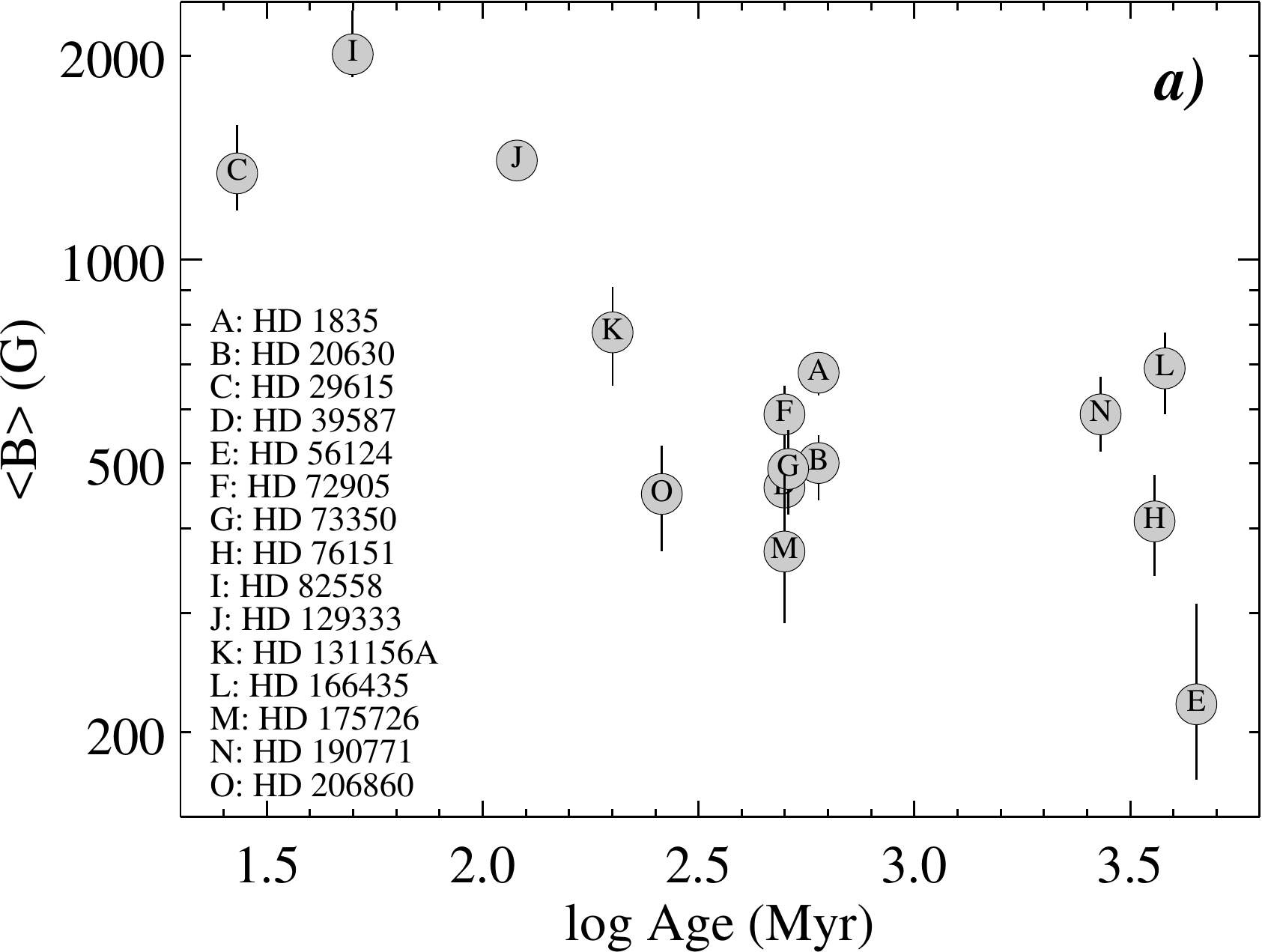}
\fifps{6.1cm}{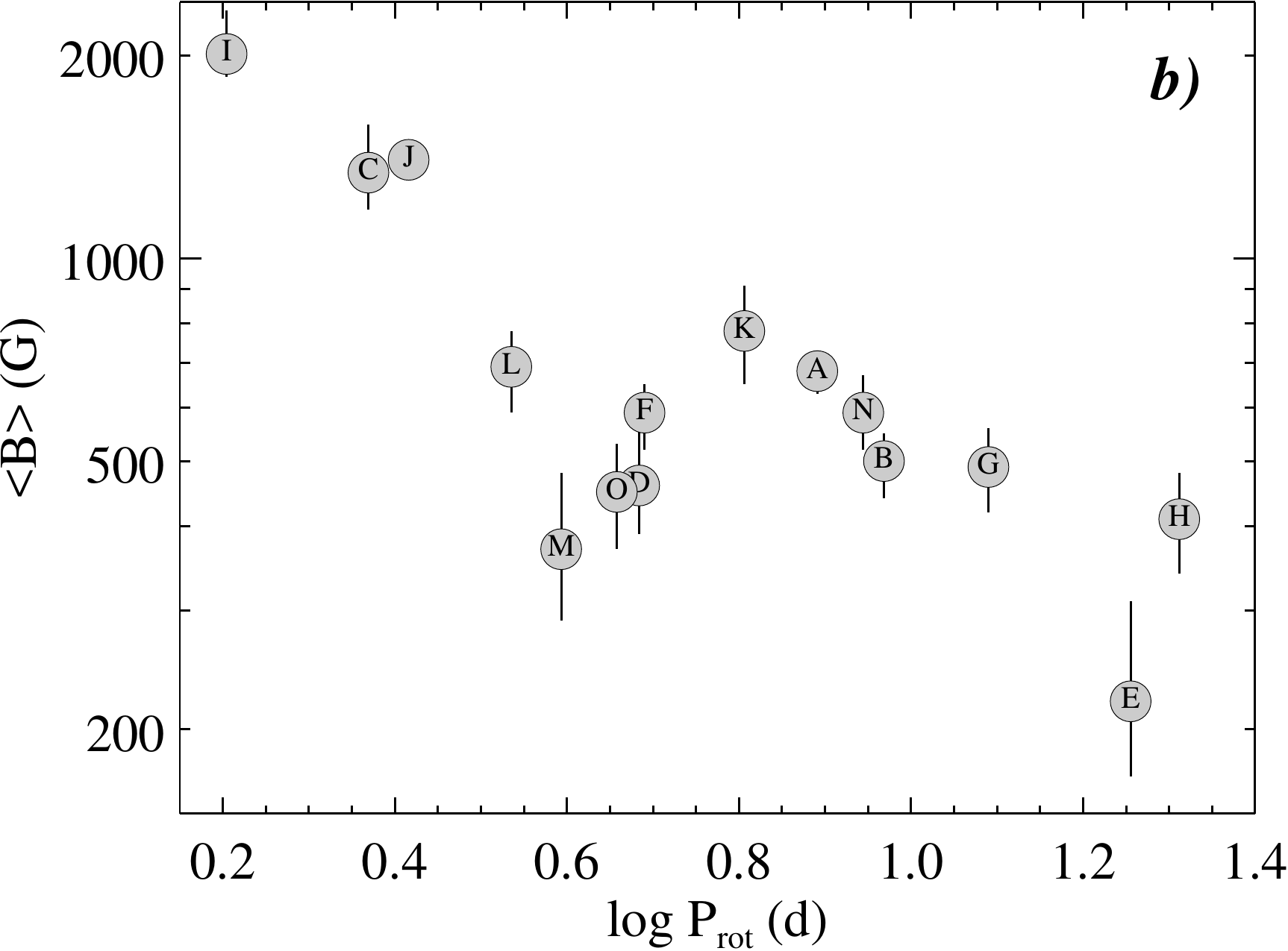}
\fifps{6.0cm}{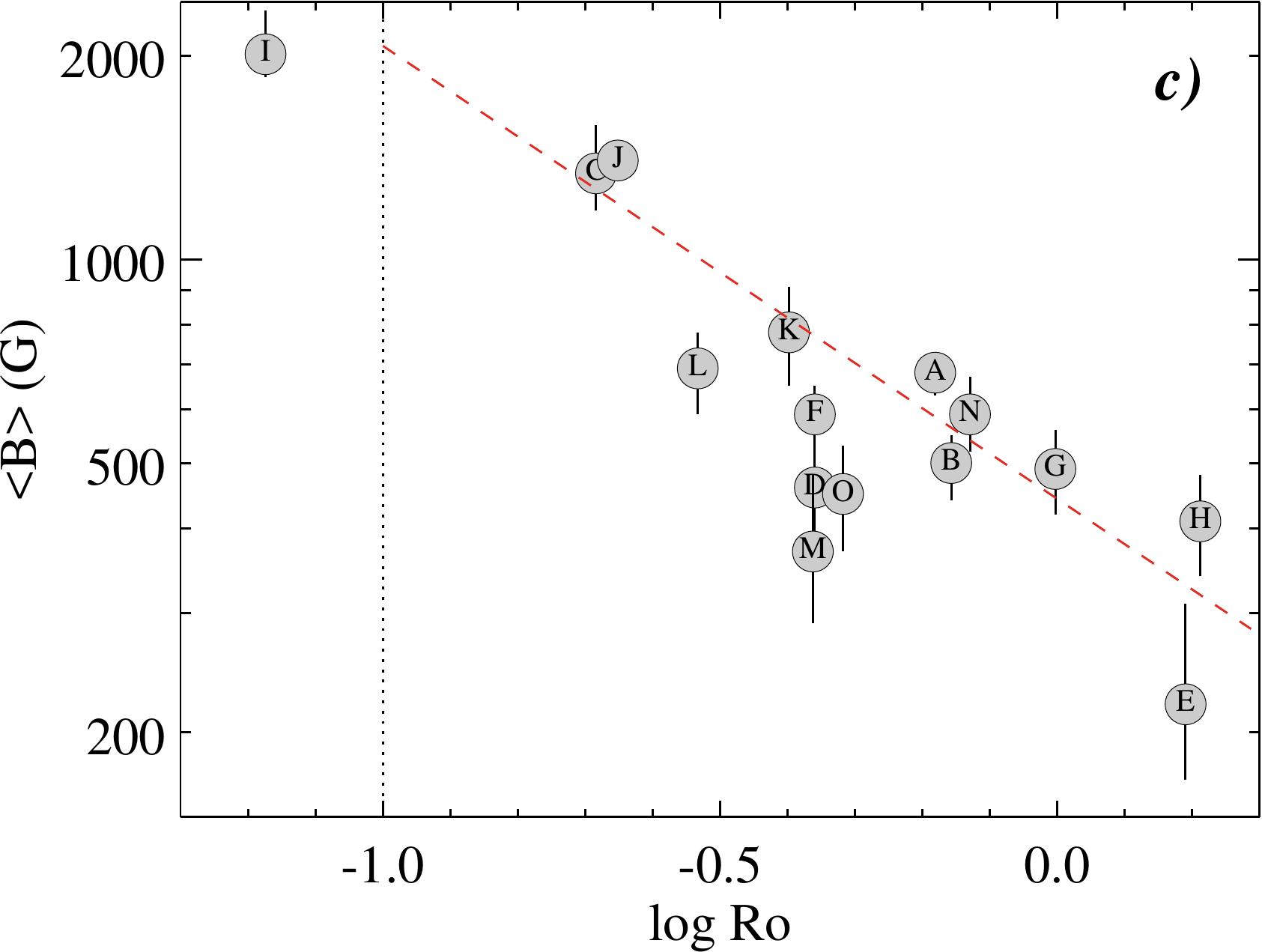}
\caption{
Correlation of the mean magnetic field strength with the stellar age (a), rotational period (b), and Rossby number (c). Individual targets are identified with Latin letters. The vertical dotted line at Ro\,=\,0.1 in panel (c) indicates the saturation limit. The dashed line shows the best fitting relation $\langle B \rangle \propto {\rm Ro}^{-0.58\pm0.07}$.
}
\label{fig:bi_age_rot}
\end{figure*}

The overall results of the Zeeman intensification magnetic field measurements carried out in this study are summarised in Table~\ref{tbl:mean_results}. Columns 2--4 of this table list the time-averaged values of the magnetic field strength $B$, the filling factor $f$, and the mean field strength \bs. These parameters, and the corresponding asymmetric error bars, were obtained by calculating median values of individual measurements in Table~\ref{tbl:results} for targets with two or more observing epochs and using mean values otherwise. In this section, we use these time-averaged magnetic field measurements to assess correlations between photospheric magnetic field characteristics and fundamental parameters, rotation, and magnetic activity indicators of the target stars.

Figure~\ref{fig:bi_age_rot}a illustrates dependence of the mean magnetic field strength on the stellar age. The latter was obtained from diverse literature sources and has different reliability, depending on the star. Typically, the age is known relatively well for young stars (age\,$\la$\,1~Gyr) thanks to memberships in open clusters and young moving groups. Conversely, the age of older objects in our sample (HD\,56124, HD\,76151, HD\,166435, HD\,190771) is constrained with far lesser precision by comparing stellar spectroscopic parameters with theoretical isochrones \citep[e.g.][]{valenti:2005}. Despite this caveat, Fig.~\ref{fig:bi_age_rot}a shows a clear overall decline of the mean field strength with age. The targets studied here can be broadly separated into three age groups with different magnetic field characteristics. The youngest group (age\,$\le$\,120~Myr), represented by HD\,29615, HD\,82558, and HD\,129333, has \bs\,=\,1.3--2.0~kG. The intermediate group (ages between 200 and 600~Myr), comprising 8 stars, shows fields in the 0.4--0.8~kG range. The four oldest stars (age\,$\ge$\,2.7~Gyr) seemingly exhibit \bs\ in a wide range of 0.2--0.7~kG and lack any correlation with age.

Next, we examine correlation between the mean field strength and stellar rotation. Figure~\ref{fig:bi_age_rot}b shows a general decrease of \bs\ with increasing $P_{\rm rot}$. This trend is arguably dominated by the three rapid rotators with the strongest fields (HD\,29615, HD\,82558, HD\,129333) and becomes less pronounced when considering the rest of the sample. The magnetic field-rotation relation can be recast in terms of the dependence on Rossby number (Fig.~\ref{fig:bi_age_rot}c), which is known to reduce the scatter. As established by many studies \citep[e.g.][]{noyes:1984,wright:2011,douglas:2014,vidotto:2014,folsom:2016}, both indirect magnetic activity indicators and direct field measurements correlate with Ro until stellar dynamo reaches a saturated state at Ro\,$\approx$\,0.1. All our targets except HD\,82558 (LQ~Hya) have Ro\,$>$\,0.1 and therefore are expected to be in the unsaturated dynamo regime. A weighted least-squares power law fit of \bs\ as a function of Ro, excluding HD\,82558, yields
\begin{equation}
\log \langle B \rangle = (2.65\pm0.05) - (0.67\pm0.11) \cdot \log \mathrm{Ro}.
\label{eq:ro}
\end{equation}

One can note that the same targets tend to deviate from the general trends in all three panels of Fig.~\ref{fig:bi_age_rot}. For example, HD\,56124 and HD\,175726 appear to have magnetic fields that are too weak for their age and rotation. It cannot be excluded that parameters other than those considered here (e.g. inclination of stellar rotational axis, phase in a long-term activity cycle, etc.) contribute to the scatter in magnetic field-rotation-age relationships.

Finally, we consider separate determinations of the local magnetic field strength $B$ and filling factor $f$. As discussed above, the relative precision with which these parameters can be constrained individually is much worse than that of their product \bs\,=\,$B\cdot f$. Nevertheless, our error determination procedure accounts for the anti-correlation between $B$ and $f$, allowing us to derive useful constraints on these parameters in many cases. For example, it can be established that a homogeneous one-component ($f=1$) model is excluded at more than 99.7\% confidence level for all stars except HD\,29615, HD\,82558, and HD\,131156A.

\begin{figure}[!t]
\fifps{7.0cm}{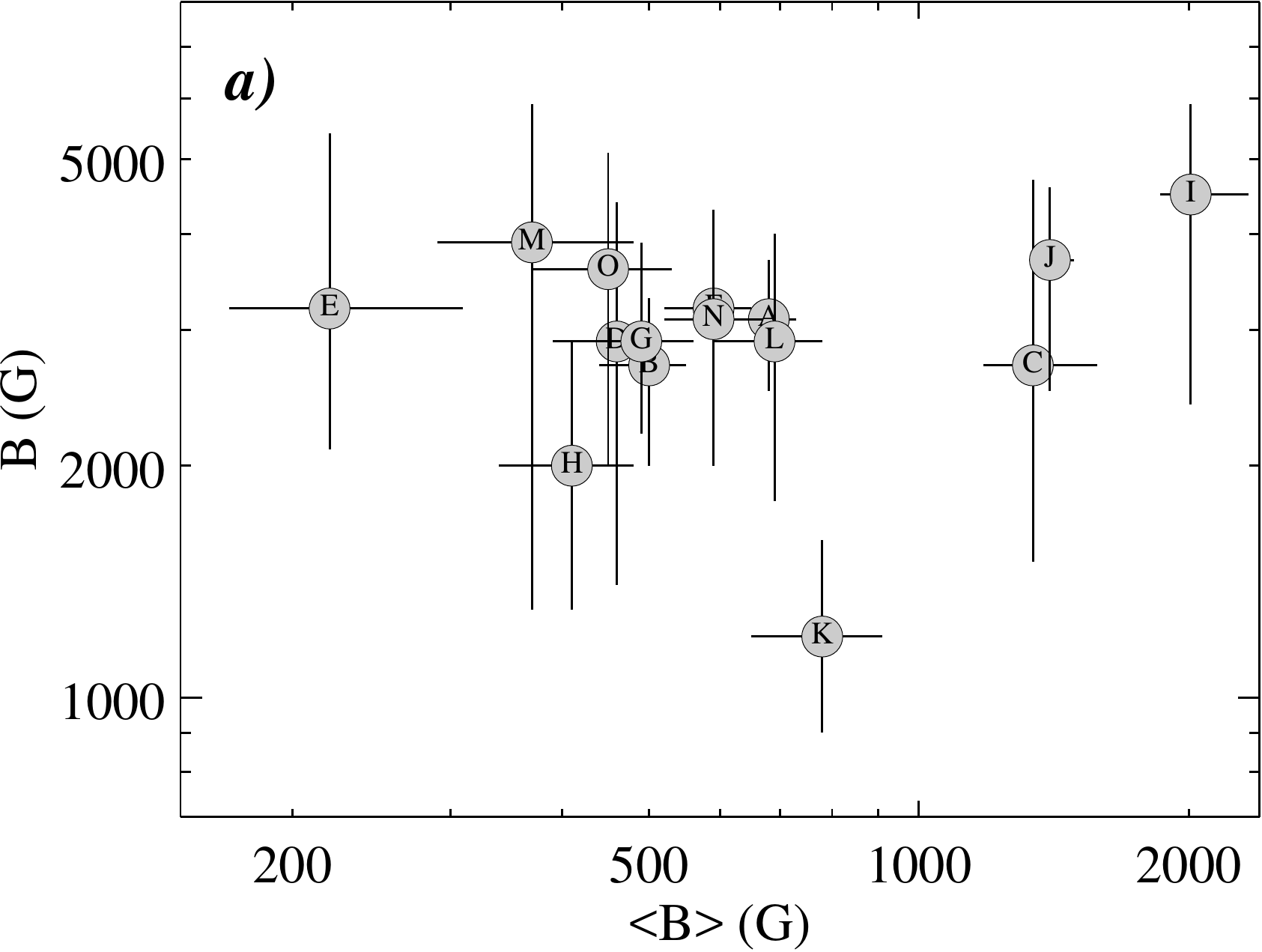}\vspace*{0.5cm}
\fifps{7.0cm}{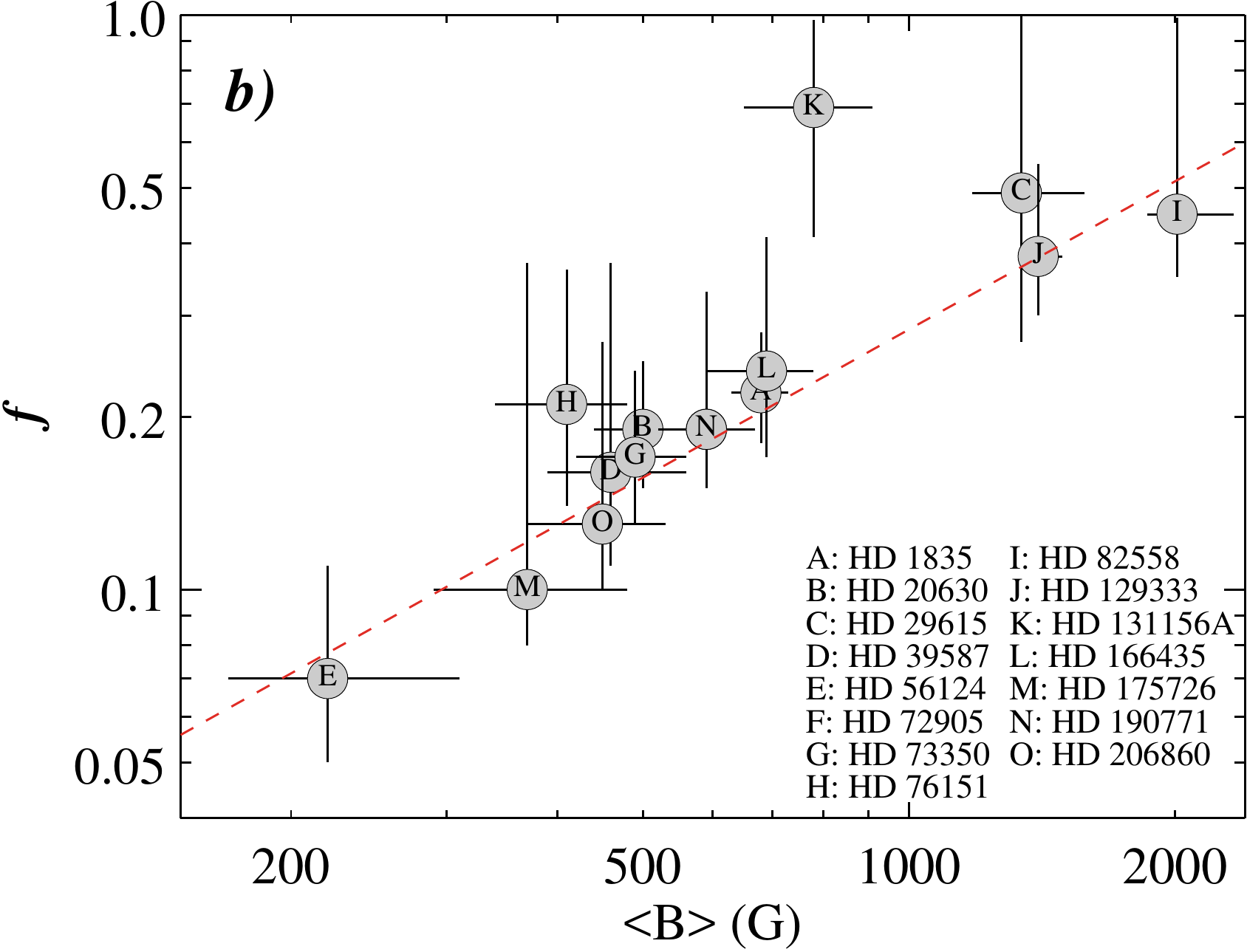}
\caption{
Correlations of the magnetic field strength $B$ (a) and magnetic filling factor $f$ (b) with the average field strength \bs\,=\,$B f$. Individual targets are identified with Latin letters. The dashed line in the lower panel shows the best fitting relation $\log f \propto \langle B \rangle^{0.86}$.
}
\label{fig:bi_bf}
\end{figure}

An interesting result emerges when we look at the plot of $B$ and $f$ as a function of \bs\ (Fig.~\ref{fig:bi_bf}). The upper panel of Fig.~\ref{fig:bi_bf} demonstrates that all stars except HD\,131156A ($\xi$~Boo~A) exhibit essentially the same local field intensity $B=3.2\pm0.6$~kG. This means that any variation of the mean field strength \bs\ is due to changes of the magnetic filling factor $f$, as conclusively demonstrated by Fig.~\ref{fig:bi_bf}b. Excluding HD\,131156A, the relation between $f$ and \bs\ in the unsaturated dynamo regime can be represented by 
\begin{equation}
\log f = -(3.12\pm0.20) + (0.86\pm0.07) \cdot \log \langle B \rangle.
\end{equation}

Thus, our results give further strong support, now based on a set of accurate field strength measurements for a sample of structurally similar G dwarf stars, that magnetic regions exhibit approximately the same local field intensity and that stellar magnetic activity is primarily modulated by the fraction of stellar surface occupied by these regions \citep{saar:1986,montesinos:1993,cranmer:2011}.

The common local field strength of 3.2~kG derived in our study noticeably exceeds the thermal equipartition field strength $B_{\rm eq}$\,=\,1.7--1.9~kG that can be calculated with $B_{\rm eq}=\sqrt{8\pi P_{\rm gas}}$ for the gas pressure at $\tau_{5000}=1$ in $T_{\rm eff}$\,=\,5750--6000~K, $\log g$\,=\,4.5 {\sc MARCS} model atmospheres. A marginal evidence of superequipartition fields was found for previous heterogenous samples of, mostly K, dwarf stars \citep[e.g.][]{cranmer:2011}. The presence of such fields is now definitively established for active M dwarfs \citep{shulyak:2017,kochukhov:2017c,kochukhov:2019a} and T Tauri stars \citep{sokal:2020}. This suggests that local field strength is not limited by the confinement of magnetic flux tubes by the photospheric gas pressure as was repeatedly assumed in the past \citep{saar:1986,cranmer:2017,see:2019}.

\subsection{Correlation with activity proxies}

\begin{figure}[!t]
\fifps{7.0cm}{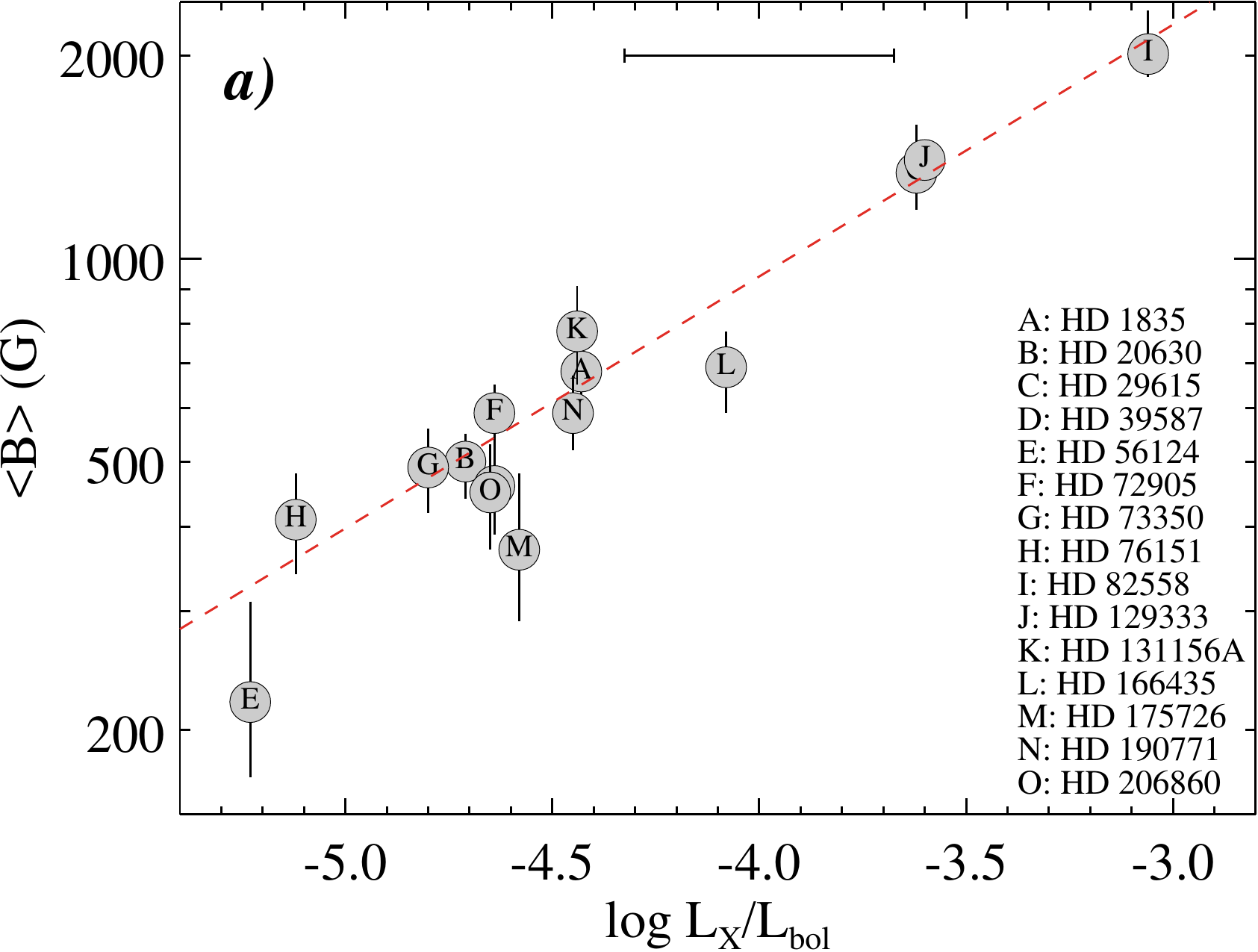}\vspace*{0.5cm}
\fifps{7.0cm}{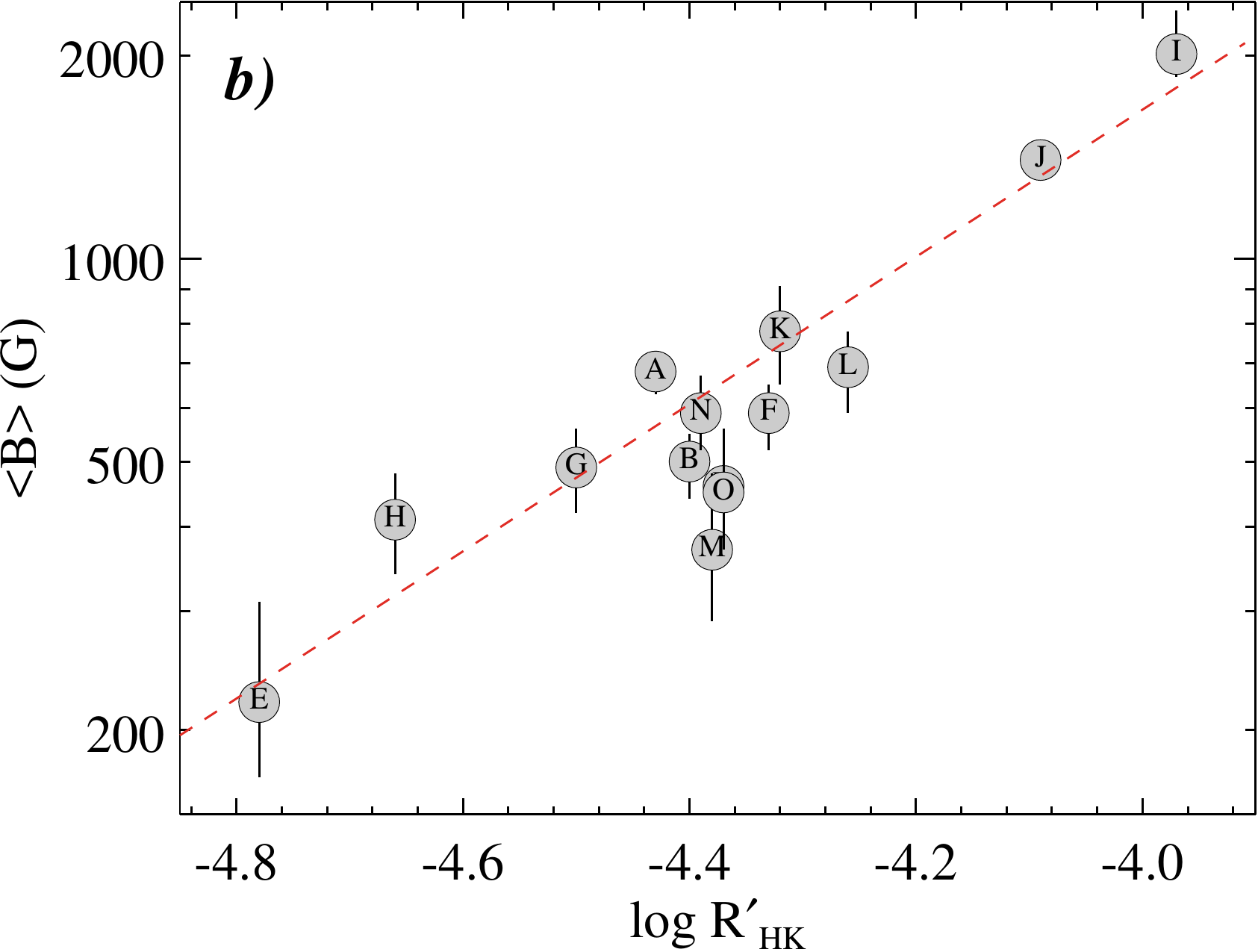}
\caption{
(a) Correlation between the total magnetic field strength and X-ray-to-bolometric luminosity ratio $\log L_{\rm X}/L_{\rm bol}$. Individual targets are identified with Latin letters. The dashed line shows the best fitting power-law relation $\langle B \rangle \propto (L_{\rm X}/L_{\rm bol})^{0.37}$. The horizontal bar at the top illustrates typical variation of $\log L_{\rm X}/L_{\rm bol}$ over stellar activity cycles. (b) Correlation between the total magnetic field strength and the Ca HK emission measure $\log R^\prime_{\rm HK}$. The dashed line corresponds to the best fitting relation $\langle B \rangle \propto {R^\prime_{\rm HK}}^{1.09}$.}
\label{fig:bi_proxy}
\end{figure}

The coronal (X-ray) emission is a frequently considered indirect proxy of magnetic activity \citep[e.g.][]{pevtsov:2003,vidotto:2014,wright:2011}. Using our new magnetic field measurements and the ratio of X-ray to bolometric luminosity reported for the target stars in Table~\ref{tbl:targets}, we are able to assess a correlation between these quantities for the first time for a sample of active solar analogues. The correlation between \bs\ and $\log L_{\rm X}/L_{\rm bol}$ is shown in Fig.~\ref{fig:bi_proxy}a. It should be noted that $\log L_{\rm X}/L_{\rm bol}$ may exhibit a large variation over stellar activity cycles. Based on the solar X-ray variation \citet{vidotto:2014} estimated the representative cyclic changes to be as large as 0.65~dex in $\log L_{\rm X}/L_{\rm bol}$. Despite this, Fig.~\ref{fig:bi_proxy}a suggests a relatively tight correlation between magnetic field strength and X-ray luminosity. We derive
\begin{equation}
\log \langle B \rangle = (4.47\pm0.09) + (0.37\pm0.02) \cdot \log L_{\rm X}/L_{\rm bol}
\label{eq:rx}
\end{equation}
by fitting a power law relation with \bs\ measured in G. 

Eq.~(\ref{eq:rx}) is equivalent to $L_{\rm X}/L_{\rm bol} \propto \langle B \rangle^{2.68\pm0.16}$. In comparison, \citet{vidotto:2014} deduced $L_{\rm X}/L_{\rm bol} \propto \langle B \rangle^{2.25}$ through indirect means, by combining correlations of \bs\ and $L_{\rm X}/L_{\rm bol}$ with Ro published by \citet{saar:2001} and \citet{wright:2011} respectively. Large uncertainties involved in this estimate of \bs\ vs. $L_{\rm X}/L_{\rm bol}$ dependence precluded \citet{vidotto:2014} from making a meaningful comparison with the corresponding relation for the global field \bzdi\ measured by ZDI studies, $L_{\rm X}/L_{\rm bol} \propto \langle B_{\rm V} \rangle^{1.61\pm0.15}$. Our analysis shows that the latter relation is significantly flatter than the dependence of X-ray to bolometric luminosity on the total field strength \bs. This may be taken as an indication that small-scale fields provide a dominant contribution to the coronal emission of active Sun-like stars.

\begin{figure*}[!t]
\fifps{6.06cm}{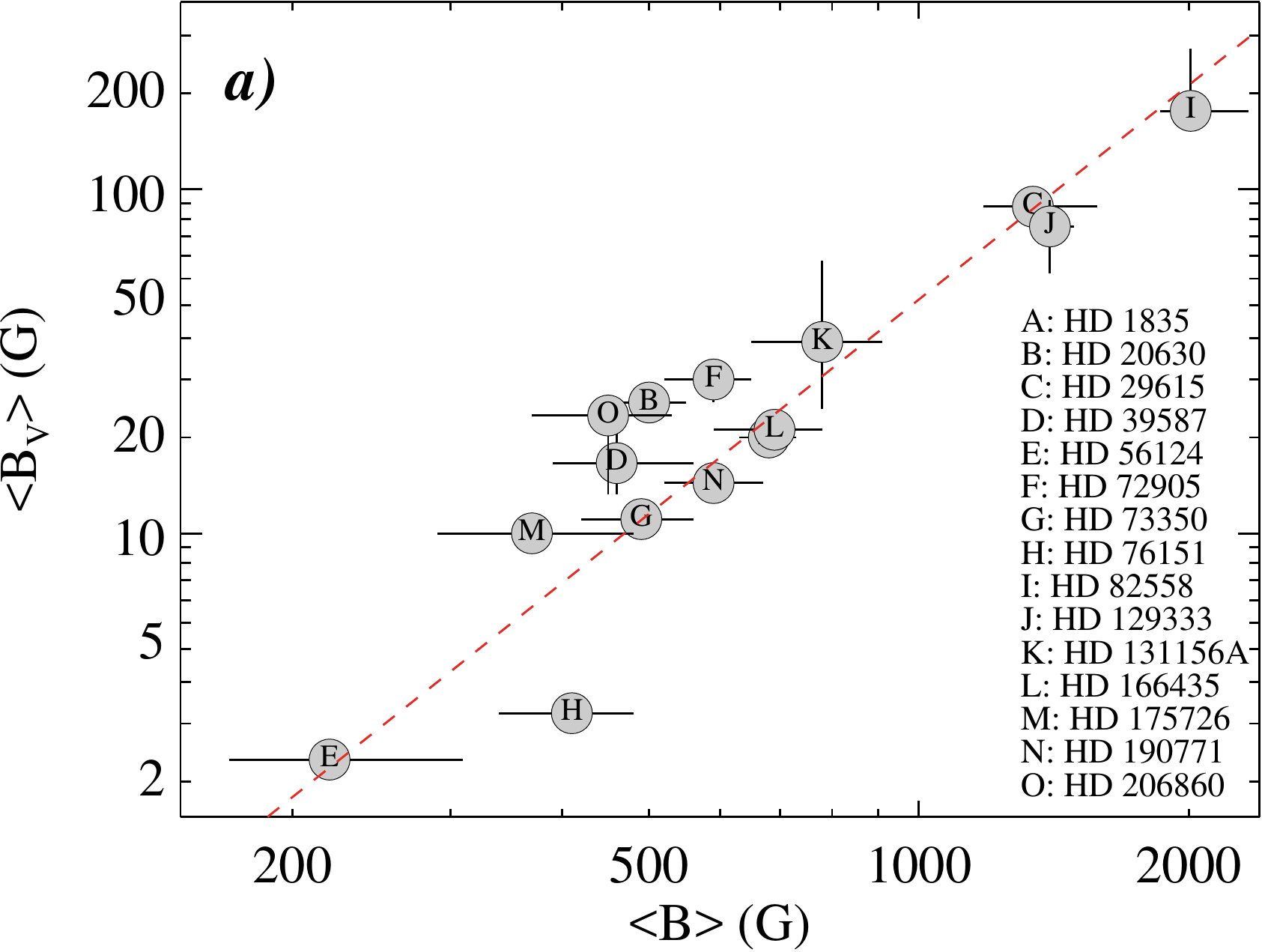}
\fifps{5.94cm}{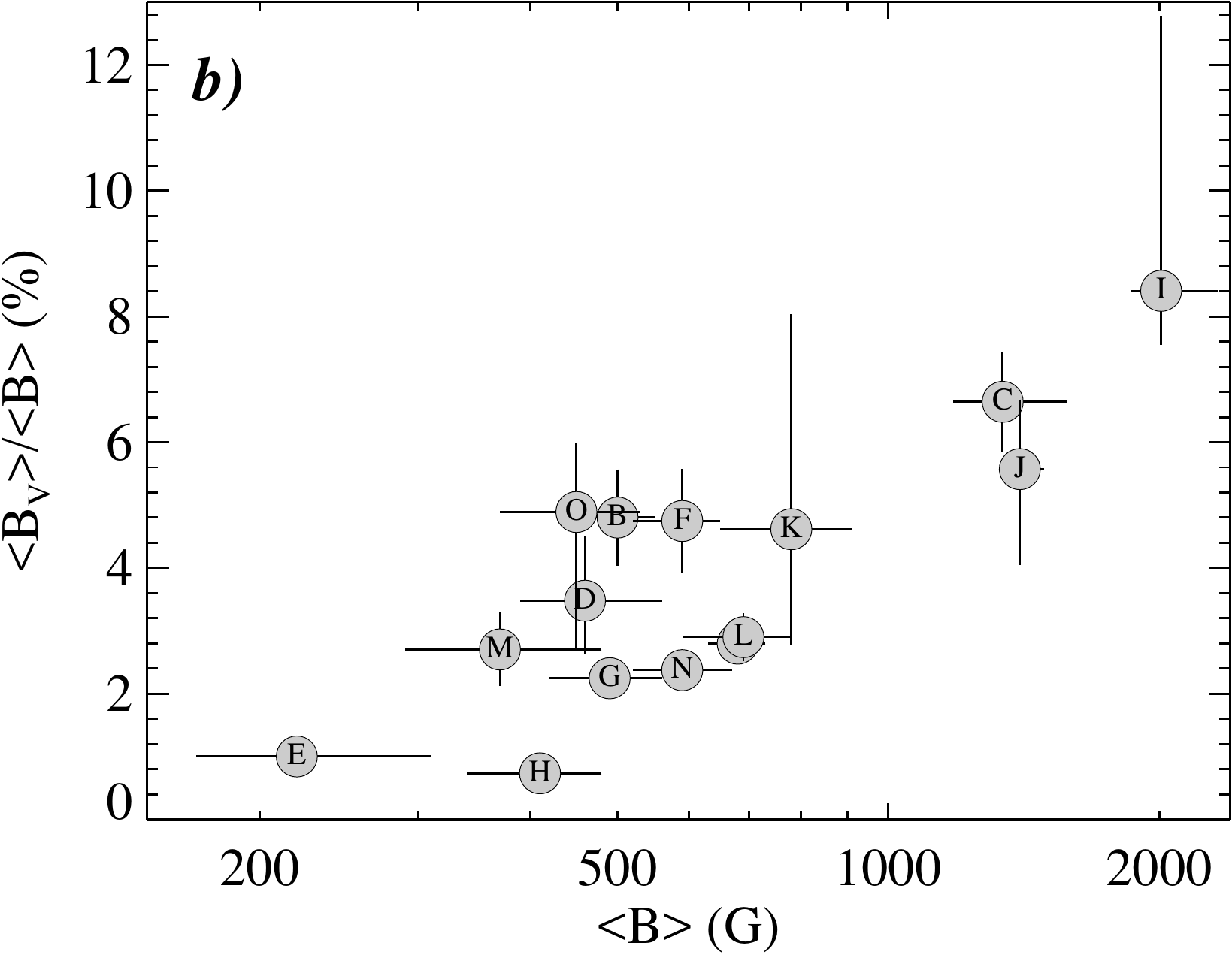}
\fifps{6.0cm}{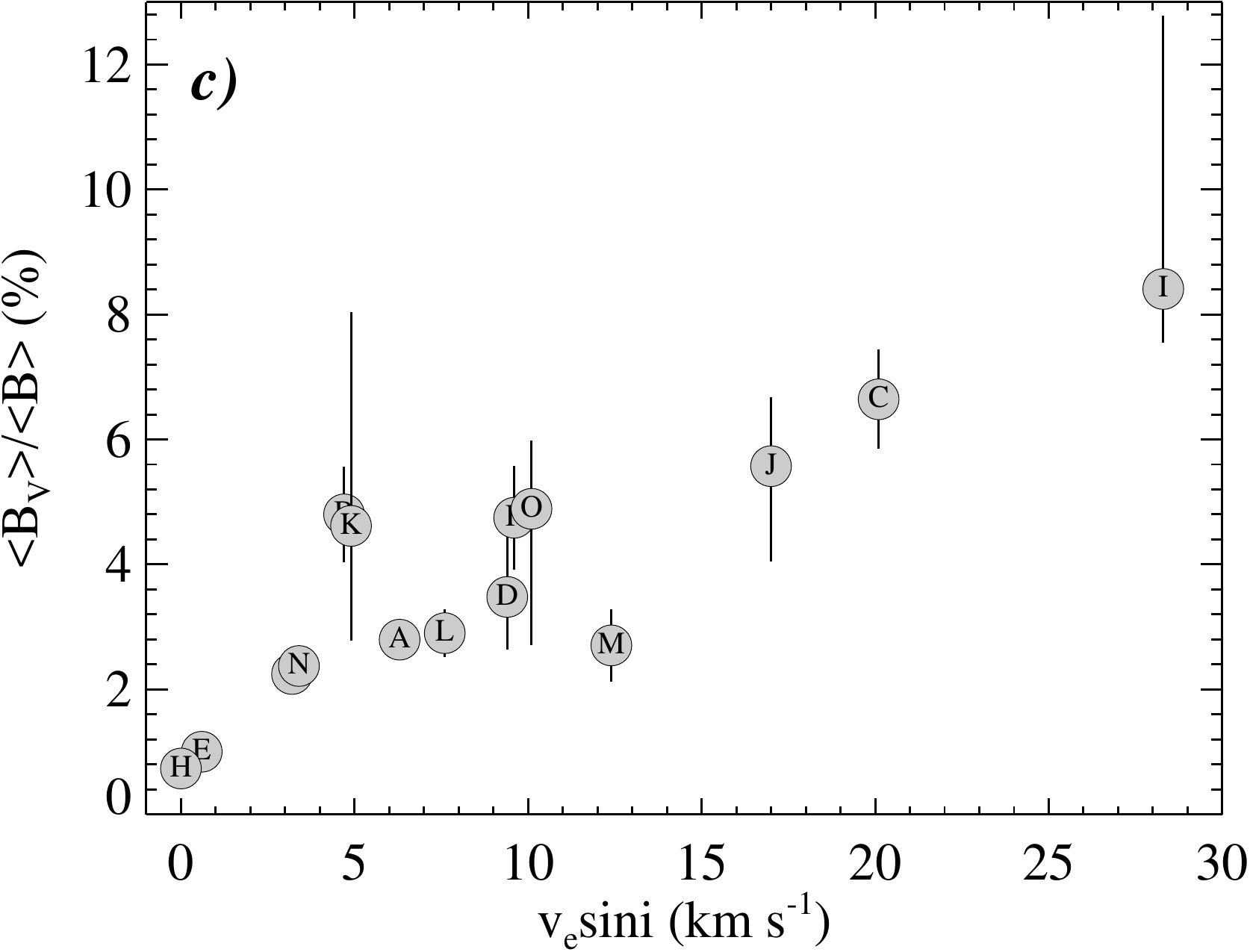}
\caption{
(a) Strength of the global magnetic field estimated with ZDI as a function of the total magnetic field strength derived in this study. Individual targets are identified with Latin letters. The dashed line shows the best fitting power-law relation $\langle B \rangle \propto \langle B_{\rm V} \rangle^{0.48}$. (b) The ratio $\langle B_{\rm V} \rangle /\langle B \rangle$ as a function of the total field strength. (c) The same ratio as a function of \vs.
}
\label{fig:bibv}
\end{figure*}

The Ca H\&K chromospheric emission is another popular magnetic activity proxy that loosely correlates with the global field strength \citep{marsden:2014} and is coupled to X-ray emission \citep{mittag:2018}. The relation between $\log \langle B \rangle$ and the chromospheric emission measure $\log R^\prime_{\rm HK}$ for our stellar sample is presented in Fig.~\ref{fig:bi_proxy}b. HD\,29615 is missing in this plot since no $\log R^\prime_{\rm HK}$ measurements can be found for this star in the catalogue by \citet{boro-saikia:2018} or elsewhere in the literature. A weighted least-squares power law fit yields 
\begin{equation}
\log \langle B \rangle = (7.59\pm0.54) + (1.09\pm0.12) \cdot \log R^\prime_{\rm HK}
\label{eq:rhk}
\end{equation}
with \bs\ in G. 

The steep ramping up of X-ray activity with increasing \bs, as given by Eq.~(\ref{eq:rx}), indicates that the coronal X-ray emission cannot itself be taken as a direct measure of the efficiency of magnetic field generation. This means that the canonically steep slope $L_{\rm X}/L_{\rm bol}\propto$\,Ro$^{-2}$ of the X-ray rotation-activity relation \citep[e.g.][]{pizzolato:2003} does not represent correctly the relation between dynamo efficiency and the stellar rotation rate, nor can observed deviations from this slope be directly interpreted in terms of predictions from dynamo models \citep{wright:2011}. In order to understand the correct scaling of the dynamo efficiency in stars, it is necessary to look at the scaling of the magnetic field itself (Eq.~\ref{eq:ro}), which indicates a rather shallow rotation dependence. Alternatively, the chromospheric Ca H\&K emission also provides a decent proxy for the magnetic field strength, since the two follow a nearly linear relation with each other (Eq.~\ref{eq:rhk}).

\subsection{Comparison with ZDI results}
\label{zdicomparison}

The sample of active Sun-like stars investigated here was chosen among the stars monitored with high-resolution spectropolarimetry and analysed with ZDI. This gives us a unique opportunity to directly compare characteristics of the global magnetic field reconstructed using ZDI with the true total magnetic field strength inferred by our Zeeman intensification analysis. Previously similar comparisons were done using large heterogeneous samples of active stars based on historic Zeeman broadening measurements collated from diverse literature sources spanning several decades \citep{vidotto:2014,see:2019}. In contrast, in this study we are able to intercompare parameters of the global and total magnetic fields obtained with modern techniques from the same observational data.

We use the mean field strength \bzdi\ as a measure of the strength of global magnetic field distributions recovered with ZDI. This is the same global field strength parameter as the one considered by \citet{see:2019}. On the other hand, it differs from the mean unsigned radial field $\langle |B_{\rm r}| \rangle$ analysed by \citet{vidotto:2014}. The main difference between these two characteristics, besides the fact that \bzdi\ is more readily available from original ZDI publications, is that \bzdi\ includes contributions from both poloidal and toroidal global field structures whereas $\langle |B_{\rm r}| \rangle$ is insensitive to toroidal field since the latter does not have a radial component.

Columns 5-6 in Table~\ref{tbl:mean_results} provide \bzdi\ for all stars in our sample along with the references to ZDI studies where magnetic maps were published. In the case of HD\,73350, HD\,76151, HD\,190771 \citep{petit:2008,petit:2009}, and HD 131156A \citep{morgenthaler:2012} the mean field strengths reported in the ZDI papers are incompatible with the actual magnetic maps presented in those studies. For these three stars we adopted \bzdi\ from the compilation by \citet{see:2019}. The same paper was used as a source of \bzdi\ values corresponding to the unpublished ZDI analyses of HD\,56124, HD\,166435, and HD\,175726 (Petit et al., in prep.). Whenever multiple ZDI maps were available for the same star, we calculated the median global field strength value and adopted the error bars corresponding to the full range covered by individual \bzdi\ measurements.

Figure~\ref{fig:bibv}a presents \bzdi\ as a function of \bs. A weighted least-squares fit with a power law function yields the following relation between these two quantities
\begin{equation}
\log \langle B \rangle = (2.19\pm0.08) + (0.48\pm0.05) \cdot \log \langle B_{\rm V} \rangle
\label{eq:bibv}
\end{equation}
for the field strengths measured in G. This relation is somewhat flatter and considerably more precise than $\langle B \rangle \propto \langle B_{\rm V} \rangle^{0.78\pm0.12}$ derived by \citet{see:2019} for the $M$\,$>$\,$0.5M_{\odot}$ part of their heterogeneous active star sample.

The ratio \bzdi$/$\bs\ is plotted as a function of \bs\ in Fig.~\ref{fig:bibv}b. As expected, the mean global field strength obtained with ZDI is vastly weaker than the total field strength diagnosed from Stokes $I$. The relation between \bzdi\ and \bs\ cannot be described by a constant scaling factor \citep{cranmer:2017}. Instead,  the global-to-total field strength ratio clearly grows with \bs. The highest \bzdi$/$\bs\ of $\sim$\,10\% is found for HD\,82558 (LQ~Hya). This ratio drops to about 6\% for HD\,29615 and HD\,129333 and lies in the 1--5\% range for the remaining targets. In other words, ZDI recovers merely 0.01--1\% of the total magnetic field energy (approximated by square of the mean field strength). This means that the total magnetic field measured in our study is dominated by a small-scale magnetic component invisible to ZDI. Nevertheless, the tightness of the relation described by Eq.~(\ref{eq:bibv}) suggests that the large- and small-scale fields are closely coupled and are likely to be produced by the same underlying dynamo mechanism.

The systematic increase of the fraction of the total magnetic field recovered by ZDI studies with \bs\ evident in Fig.~\ref{fig:bibv}b can be due to an intrinsic shift of the magnetic energy to larger spatial scales in more active stars. Alternatively, this effect can be a consequence of a better surface resolution achieved by ZDI studies of faster rotating stars due to a larger Doppler broadening of their line profiles. A correlation of \bzdi$/$\bs\ with \vs\ seen in Fig.~\ref{fig:bibv}c supports the latter hypothesis.

\subsection{Effect of magnetic field on abundance determination}

The strengthening of spectral lines due to magnetic intensification is usually ignored in stellar abundance analyses, especially in the context of modern spectroscopic studies dealing with large stellar samples. Based on the outcome of our in-depth investigation of magnetic fields of selected active solar analogues, we are able to perform a quantitative assessment of the likely errors incurred by neglecting magnetic field. To this end, we calculated a set of synthetic profiles of the three magnetically sensitive \fe\ line for fixed $B=3.2$~kG and changed the filling factor to get $B\cdot f$ in the interval from 0.05 to 2.5~kG. We then adjusted Fe abundance in the non-magnetic spectrum synthesis model until it matched the equivalent widths measured in magnetic spectra. All calculations adopted the same model atmosphere with $T_{\rm eff}=5750$~K, $\log g=4.5$, and assumed $v_{\rm mic}=0.85$~\kms.

\begin{figure}[!t]
\fifps{8cm}{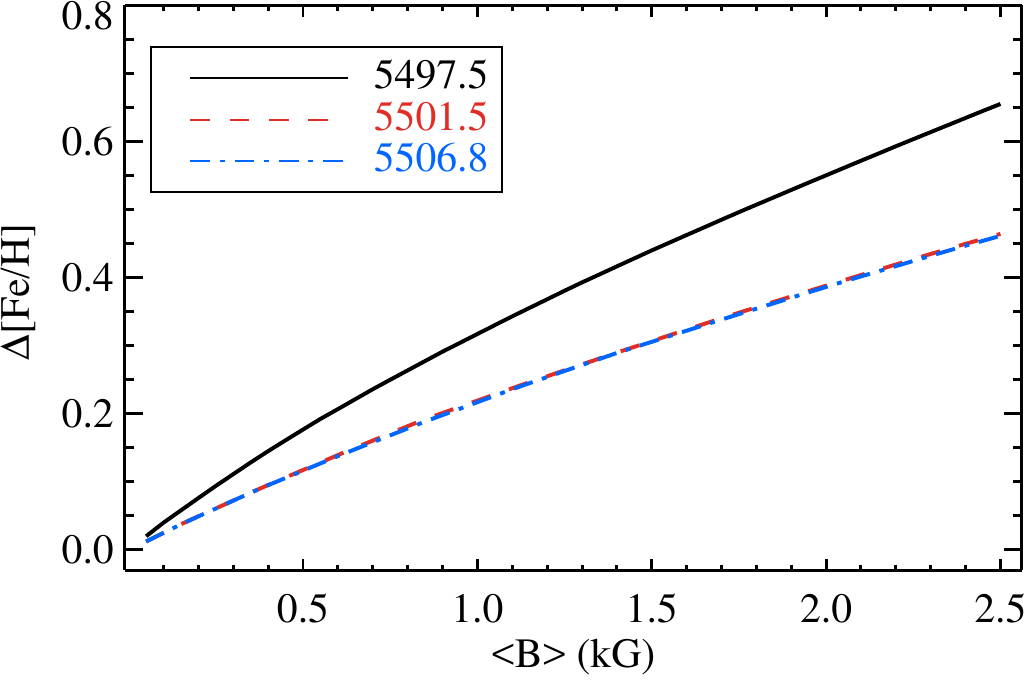}
\caption{
Error of Fe abundance determination incurred by ignoring magnetic intensification of \fe\ lines.
}
\label{fig:abn_error}
\end{figure}

The resulting error of the Fe abundance determination is illustrated in Fig.~\ref{fig:abn_error}. The curve corresponding to the \fe\ 5497.5~\AA\ line represents the upper limit of possible errors of an optical abundance analysis based on a single absorption feature. The curves corresponding to the other two lines are closer to average multi-line abundance determination bias. This plot indicates that an abundance error of up to 0.4--0.5~dex can be encountered for the most active stars ($\langle B \rangle=1.5$--2.0~kG), such as HD\,82558. The errors are still significant (0.15--0.30~dex) for stars with a more modest activity levels characterised by $\langle B \rangle =0.5$--1.0~kG. This analysis thus suggests that metallicities and individual abundances reported for active stars \citep[e.g.][]{valenti:2005,brewer:2016} may have been systematically overestimated. An unusually high line-to-line abundance scatter or a trend of abundance with wavelength might be a symptom of unaccounted Zeeman intensification.

\begin{figure*}[!t]
\fifps{13.0cm}{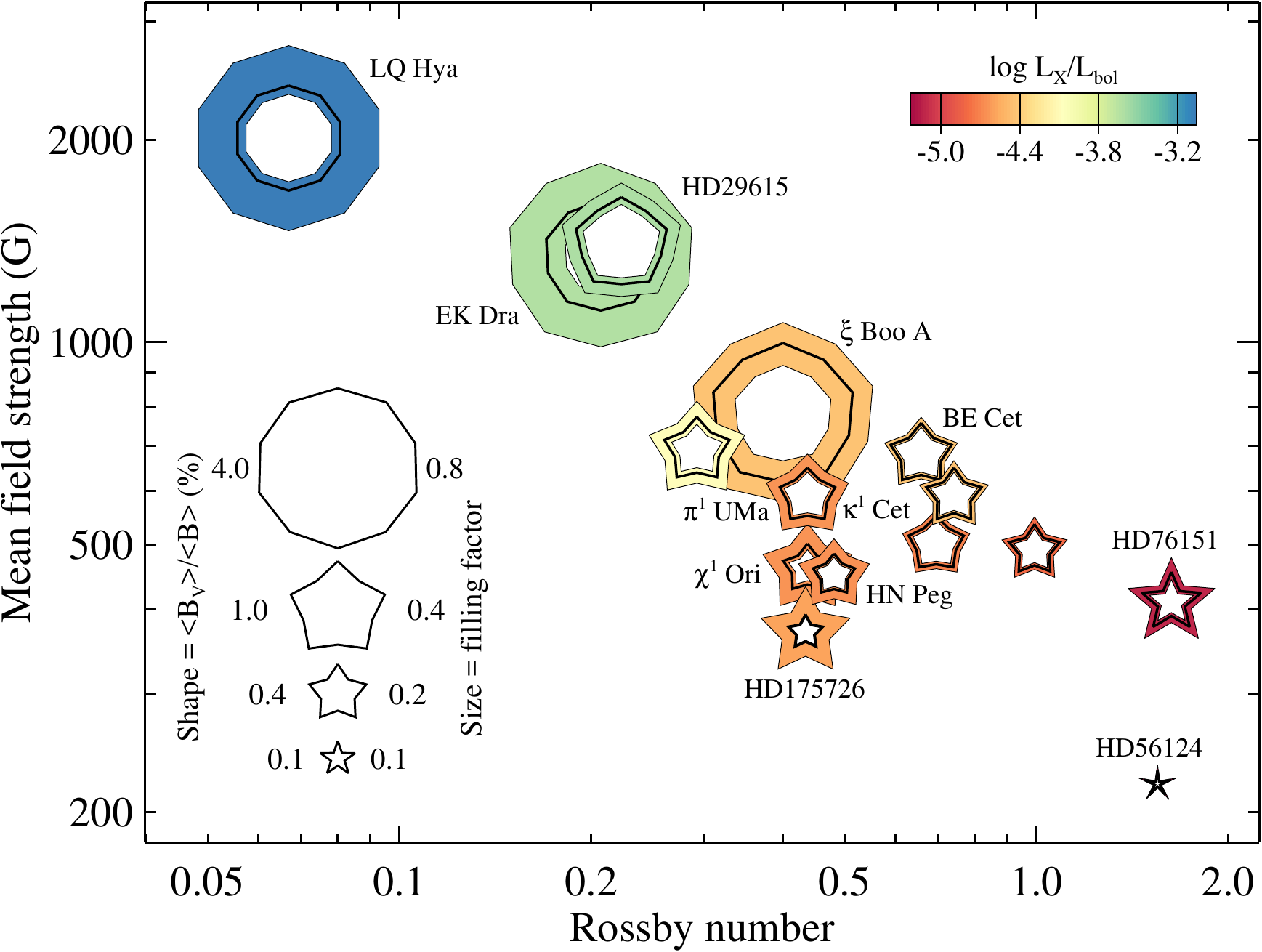}
\caption{Summary of the magnetic field analysis results. The average field strength is plotted as a function of Rossby number, similar to Fig.~\ref{fig:bi_age_rot}c. The symbol size increases with the magnetic filling factor $f$. The inner and outer outlines reflect uncertainty of the filling factor determination. The symbol shape corresponds to the ratio of the global magnetic field strength obtained with ZDI and the total field strength, $\langle B_{\rm V} \rangle /\langle B \rangle$. This ratio ranges from about 4\% to less than 0.1\%. The symbol colour corresponds to the X-ray-to-bolometric luminosity ratio, $\log L_{\rm X}/L_{\rm bol}$. 
}
\label{fig:summary}
\end{figure*}

\section{Conclusions}
\label{conclusions}

In this paper we have developed a new methodology for measuring average unsigned magnetic field strength in the atmospheres of Sun-like stars. This method takes advantage of the differential Zeeman intensification of \fe\ lines in the optical wavelength region. The new magnetic field diagnostic procedure is less restrictive compared to traditional Zeeman broadening analysis. In particular, it can be applied to moderately fast rotators and does not require spectroscopic data of exceptionally high quality. The technique can be implemented using typical high-resolution optical echelle spectra that are currently provided by many different night-time astronomical facilities around the world.

We have performed magnetic field measurements for a sample of 14 G dwarf stars and one early K dwarf, thereby considerably extending the number of solar analogues at different ages and activity levels with direct observational constraints on the total surface magnetic field. We derived 78 individual mean field strength ($\langle B \rangle = B\cdot f$) measurements and, in most cases, were able to provide meaningful constraints on the local field strength $B$ and the fraction $f$ of the stellar surface covered by a magnetic field. These data greatly improve our understanding of different evolutionary phases of the solar-like magnetic dynamo and can be used to estimate magnetic characteristics of the Sun during the first $\sim$\,1~Gyr of its life.

Main conclusions of our study can be summarised as follows.
\begin{itemize}
\item We have measured mean magnetic field strengths in the range from 0.2 to 2.0 kG and detected a systematic decline of the field intensity from $B\cdot f > 1$~kG in stars younger than $\sim$\,100~Myr to weaker fields in older stars.
\item A clear anti-correlation of the mean field strength and the stellar rotational period, or Rossby number, is established and a new empirical calibration of $B\cdot f$ as a function of Ro is obtained.
\item Separate analysis of the local field strength $B$ and magnetic filling factor $f$ suggests that the increase of $B\cdot f$ associated with the transition from less active to more active stars is mostly due to a systematic growth of $f$ from $\sim$\,10\% to $>$\,50\% of the stellar surface. The local field strength $B$ remains at the level of $\sim$\,3~kG in all stars.
\item The mean magnetic field strength determined in our study exhibits clear correlations with the chromospheric and coronal emission indicators. This allowed us to derive new calibrations of the widely used activity indices $\log L_{\rm X}/L_{\rm bol}$ and $\log R^\prime_{\rm HK}$ in terms of the surface magnetic field strength.
\item We have compared our magnetic field strength measurements with results of ZDI analyses of the global magnetic field topologies, which in most cases relied on the same observational data as used in our study. We found a tight correlation of the ratio of the global and total magnetic field strengths, \bzdi$/$\bs, with the stellar activity level. In terms of this ratio, the fraction of magnetic field detected by spectropolarimetric studies varies from $\sim$\,10\% in the most active stars to less than 1\% in the least active objects.
\item Ignoring magnetic intensification in an optical spectroscopic abundance analysis of active Sun-like stars may lead to overestimation of element concentrations by up to 0.4--0.5~dex if this analysis is based on lines with a large magnetic sensitivity.
\end{itemize}

The most important correlations derived in our study are schematically illustrated in Fig.~\ref{fig:summary}. This plot combines information from Figs.~\ref{fig:bi_age_rot}--\ref{fig:bibv} to show how the field strength, the filling factor and the ratio of the global to total field strength change with Rossby number and how these trends relate to X-ray emission of the target stars.

There are numerous further applications of the magnetic diagnostic procedure developed in our paper. For example, this technique can be systematically applied to large volumes of high-resolution spectroscopic data accumulated for Sun-like stars by the radial velocity exoplanet surveys \citep[e.g.][]{adibekyan:2012,delgado-mena:2017,hojjatpanah:2019}. This method can be also utilised to investigate the variation of magnetic field strength during stellar activity cycles for active stars with sufficiently long spectroscopic time-series \citep{alvarado-gomez:2018,galarza:2019}. One can also envisage incorporating Zeeman intensification of the \fe\ lines studied here into Doppler imaging inversions with the goal of simultaneously reconstructing both cool spot maps and distributions of small-scale magnetic field intensity \citep{saar:1992a}.

\begin{acknowledgements}
O.K. acknowledges support by the Knut and Alice Wallenberg Foundation (project grant ``The New Milky Way''), the Swedish Research Council (projects 621-2014-5720, 2019-03548), and the Swedish National Space Board (projects 185/14, 137/17).
J.J.L. acknowledges the Academy of Finland ``ReSoLVE'' Centre of Excellence (project number 307411) and the Max Planck Research Group ``SOLSTAR'' funding.
This work is based on observations made with the HARPSpol instrument on the ESO 3.6 m telescope at La Silla (Chile), under the programs 091.D-0836 and 0100.D-0176. Also based on archival observations obtained at the Bernard Lyot Telescope (Pic du Midi, France) of the Midi-Pyr\'en\'ees Observatory, which is operated by the Institut National des Sciences de l'Univers of the Centre National de la Recherche Scientifique of France, and on observations obtained at the Canada-France-Hawaii Telescope, which is operated by the National Research Council of Canada, the Institut National des Sciences de l'Univers of the Centre National de la Recherche Scientifique of France, and the University of Hawaii.
This study has made use of the VALD database, operated at Uppsala University, the Institute of Astronomy RAS in Moscow, and the University of Vienna.
\end{acknowledgements}

%\bibliographystyle{aa}
%\bibliography{astro_papers}

%\clearpage

\begin{appendix}

\section{Additional figures}

\begin{figure}[!h]
\fifps{\hsize}{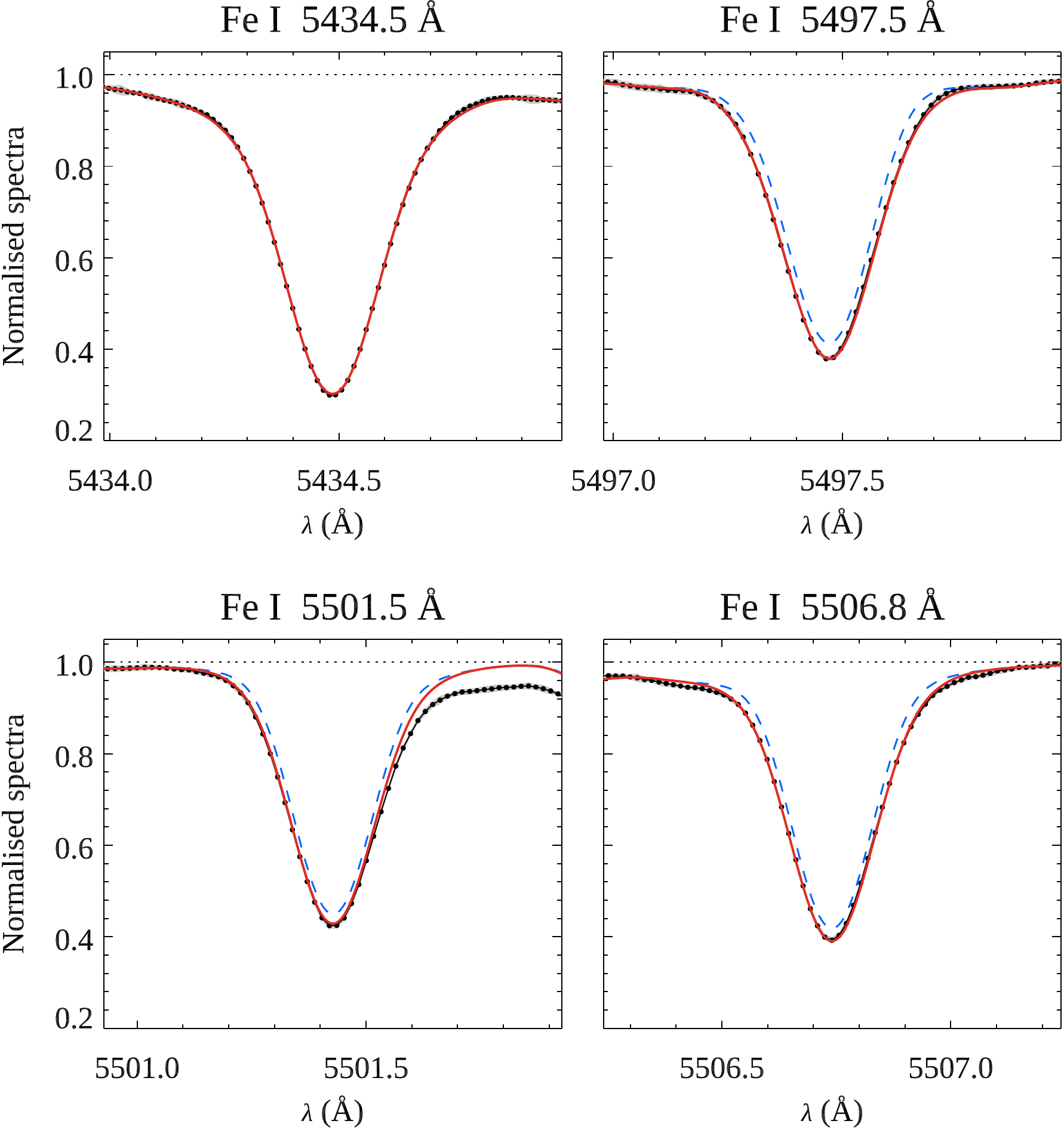}
\caption{Same as Fig.~\ref{fig:prf129333} for HD\,1835 (2013.70 epoch).}
\label{fig:prf1835}
\end{figure}

\begin{figure}[!h]
\fifps{\hsize}{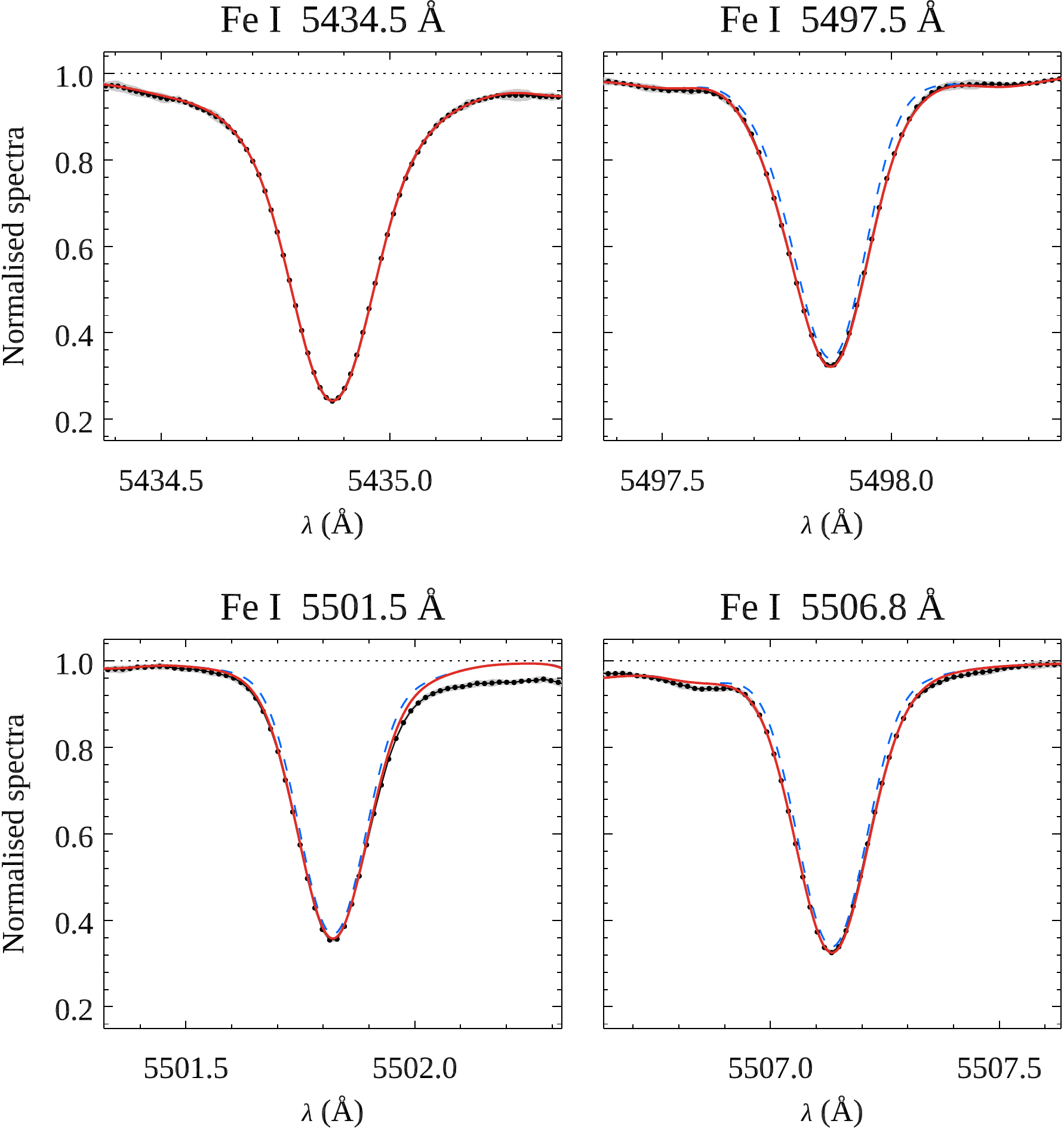}
\caption{Same as Fig.~\ref{fig:prf129333} for HD\,20630 (2013.70 epoch).}
\label{fig:prf20630}
\end{figure}

\begin{figure}[!h]
\fifps{\hsize}{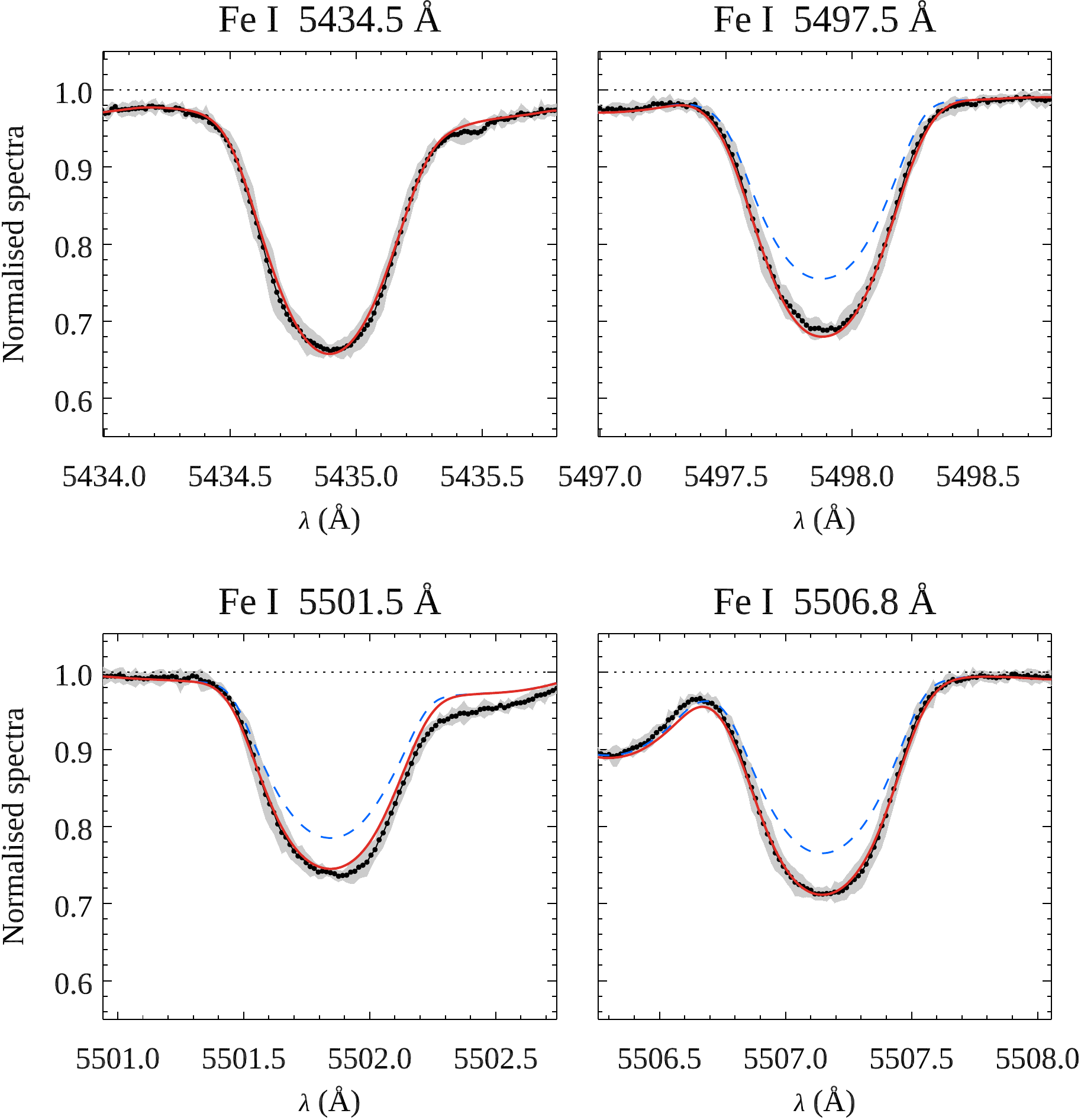}
\caption{Same as Fig.~\ref{fig:prf129333} for HD\,29615 (2017.95 epoch).}
\label{fig:prf29615}
\end{figure}

\begin{figure}[!h]
\fifps{\hsize}{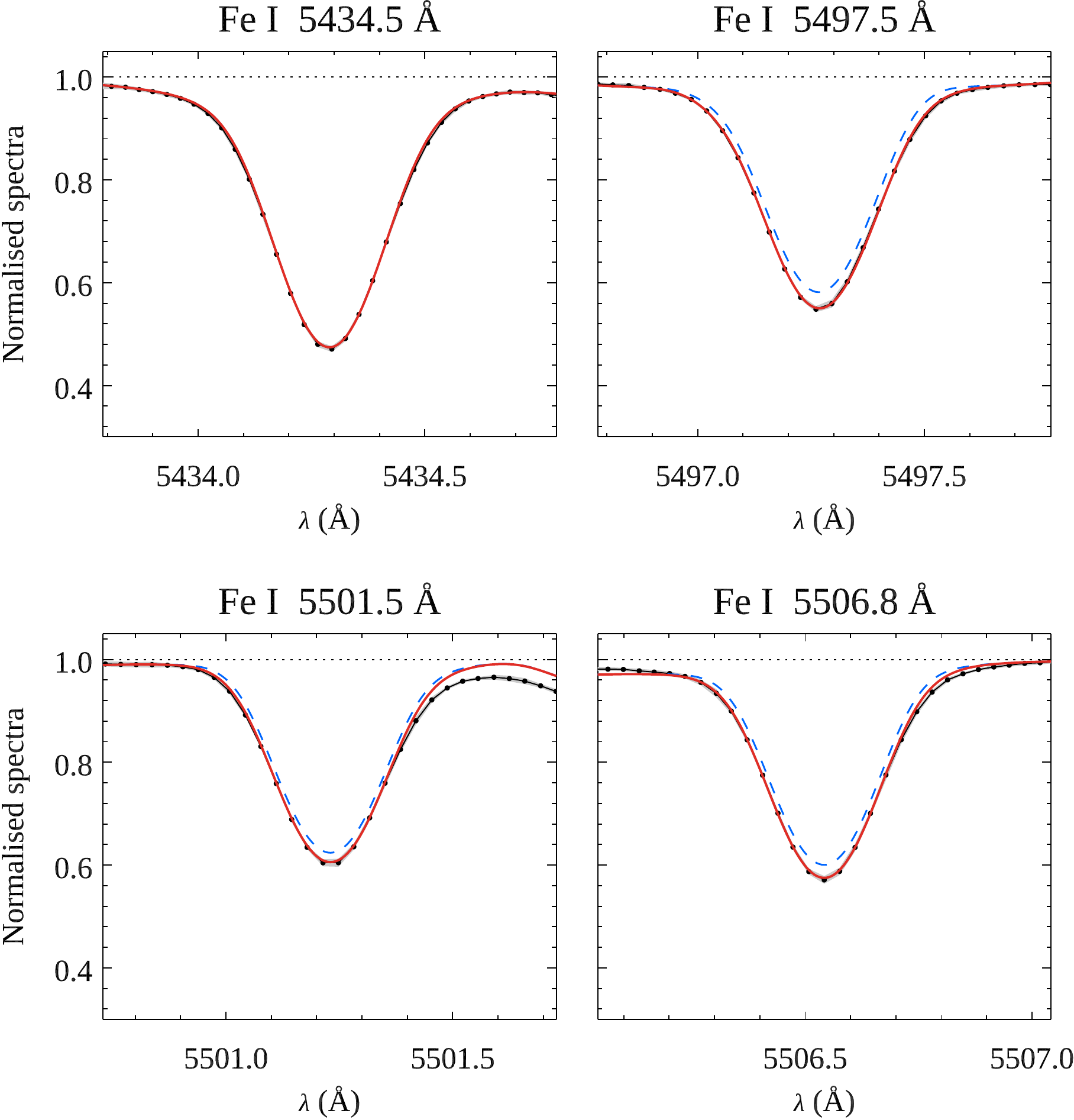}
\caption{Same as Fig.~\ref{fig:prf129333} for HD\,39587 (2007.09 epoch).}
\label{fig:prf39587}
\end{figure}

\begin{figure}[!h]
\fifps{\hsize}{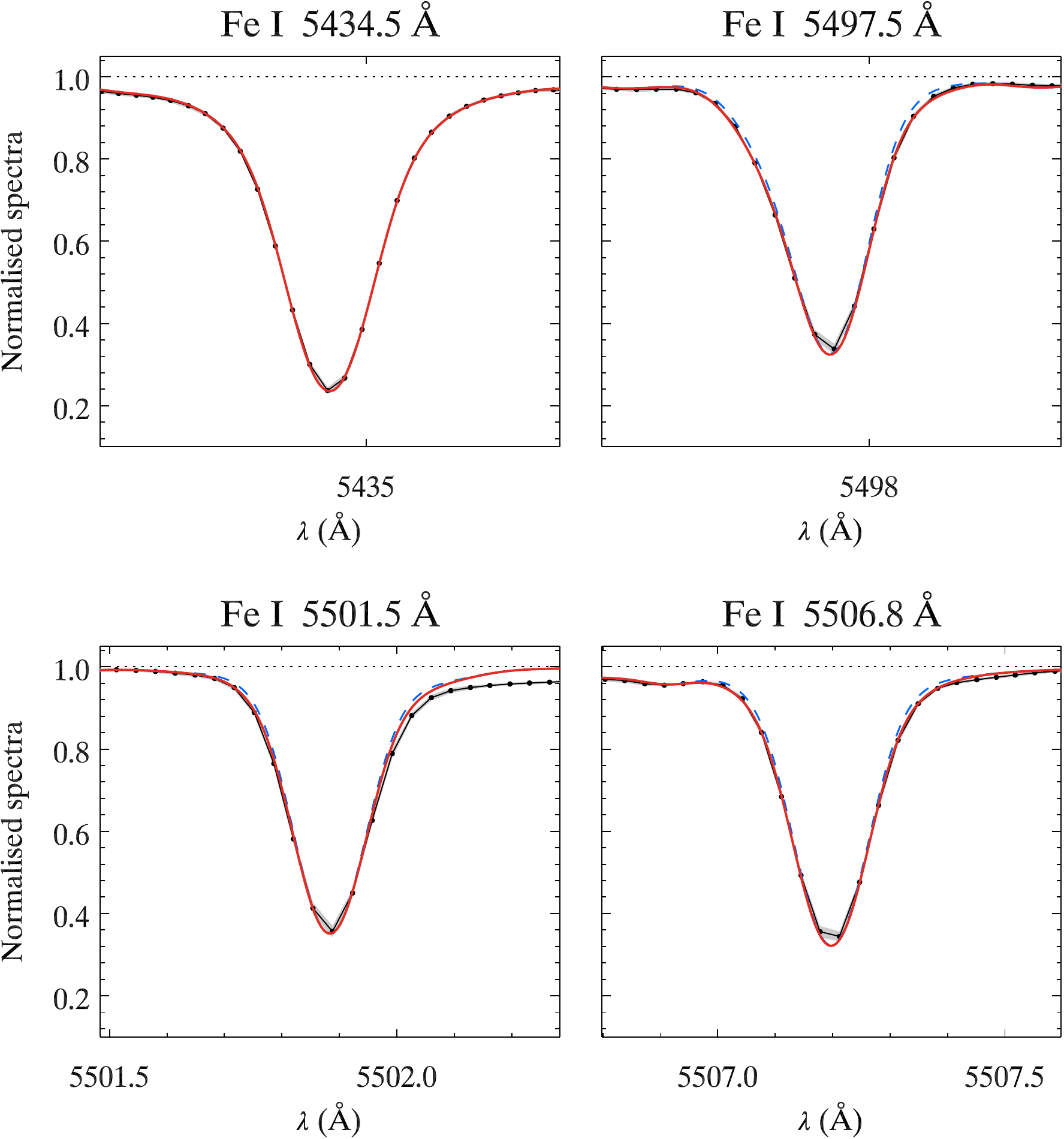}
\caption{Same as Fig.~\ref{fig:prf129333} for HD\,56124 (2008.09 epoch).}
\label{fig:prf56124}
\end{figure}

\begin{figure}[!h]
\fifps{\hsize}{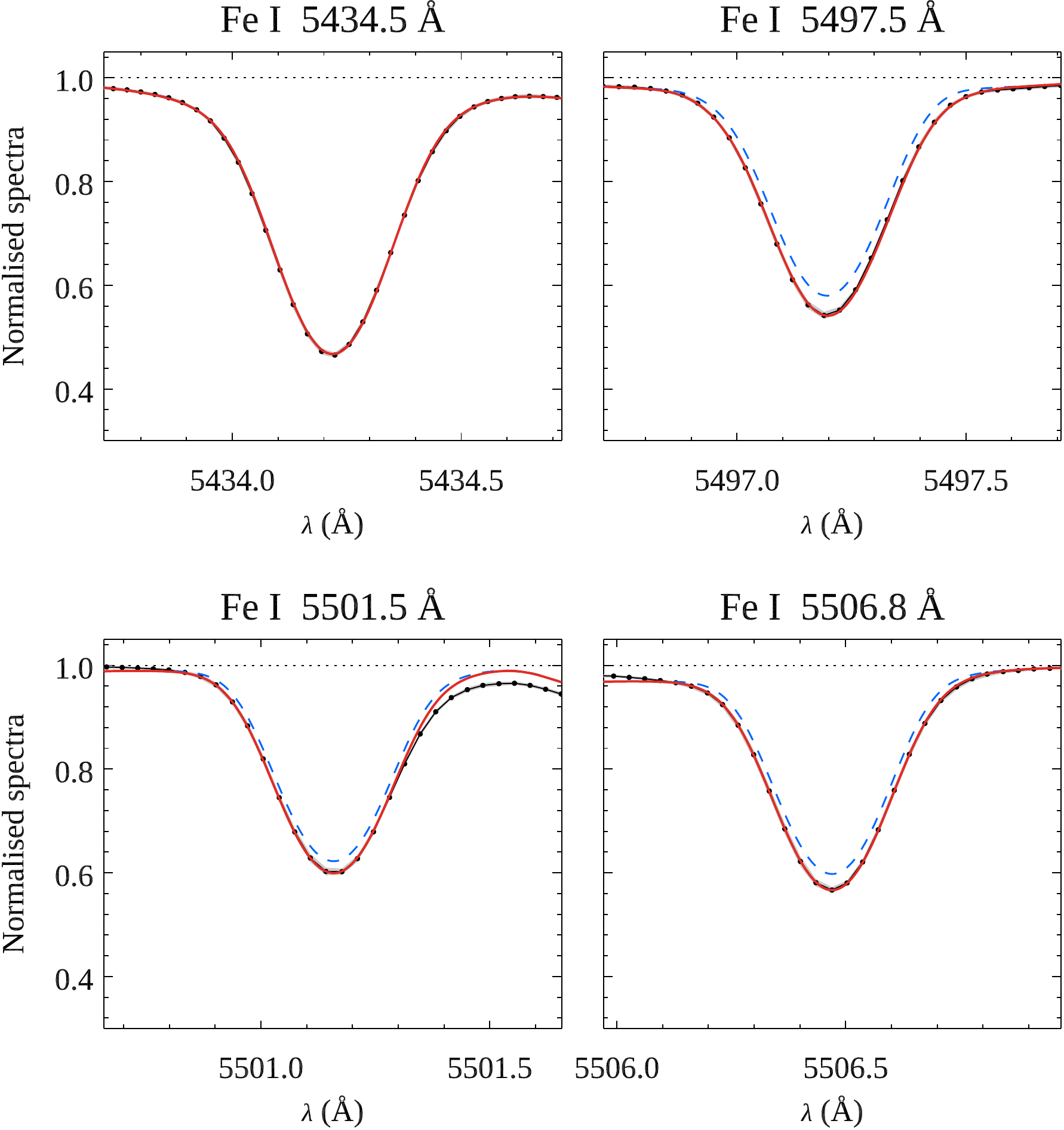}
\caption{Same as Fig.~\ref{fig:prf129333} for HD\,72905 (2014.32 epoch).}
\label{fig:prf72905}
\end{figure}

\begin{figure}[!h]
\fifps{\hsize}{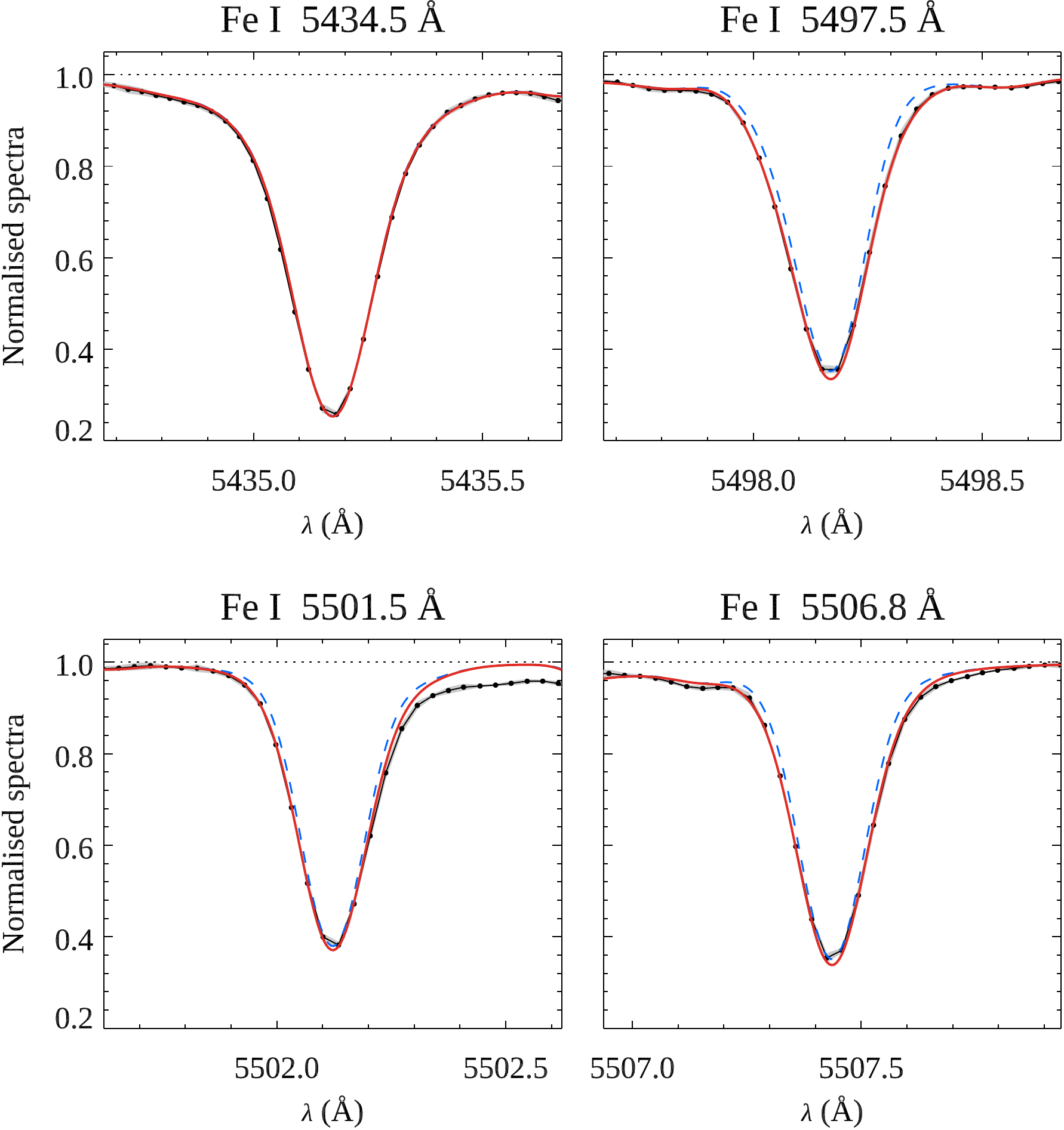}
\caption{Same as Fig.~\ref{fig:prf129333} for HD\,73350 (2011.05 epoch).}
\label{fig:prf73350}
\end{figure}

\begin{figure}[!h]
\fifps{\hsize}{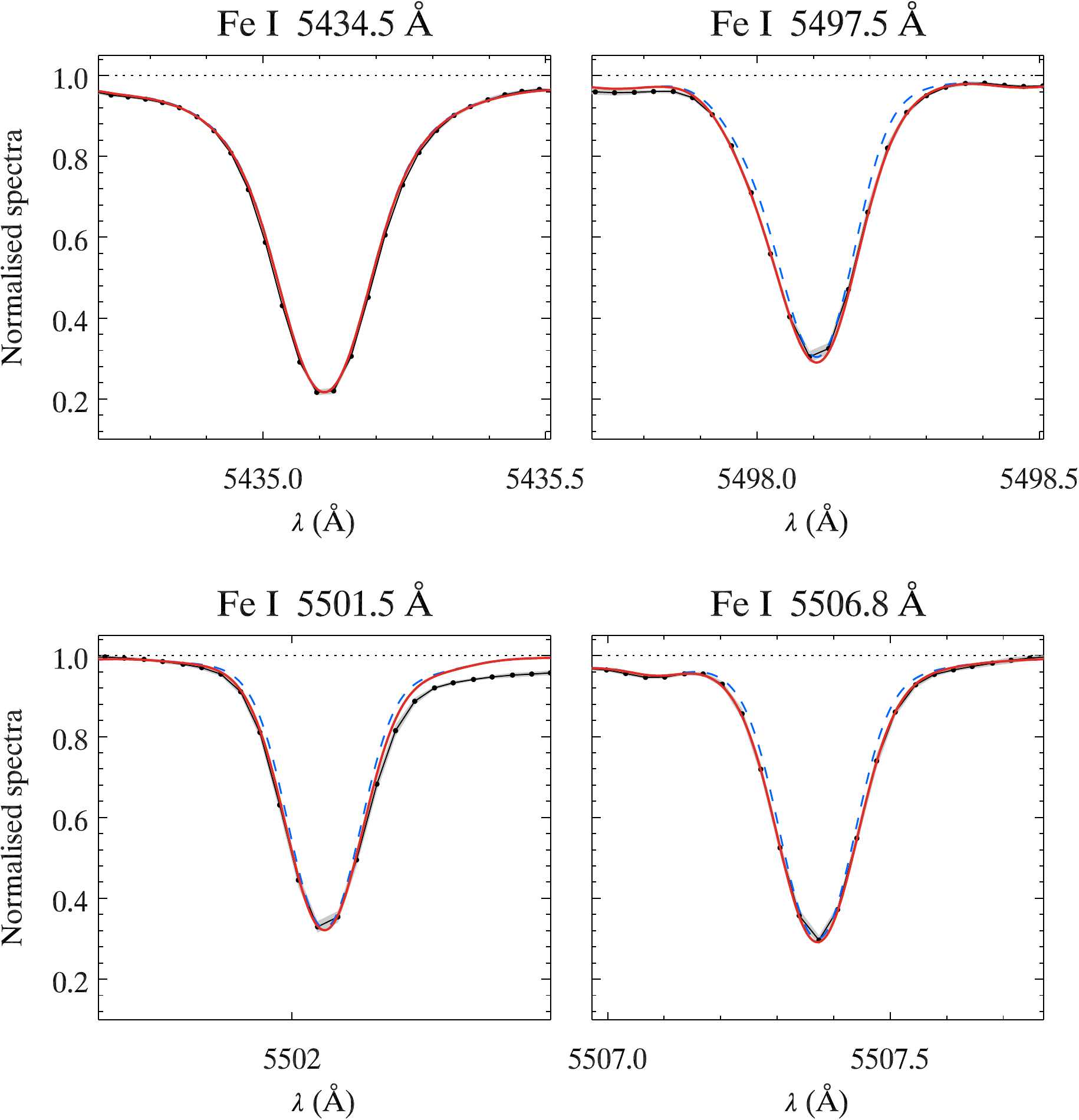}
\caption{Same as Fig.~\ref{fig:prf129333} for HD\,76151 (2009.95 epoch).}
\label{fig:prf76151}
\end{figure}

\begin{figure}[!h]
\fifps{\hsize}{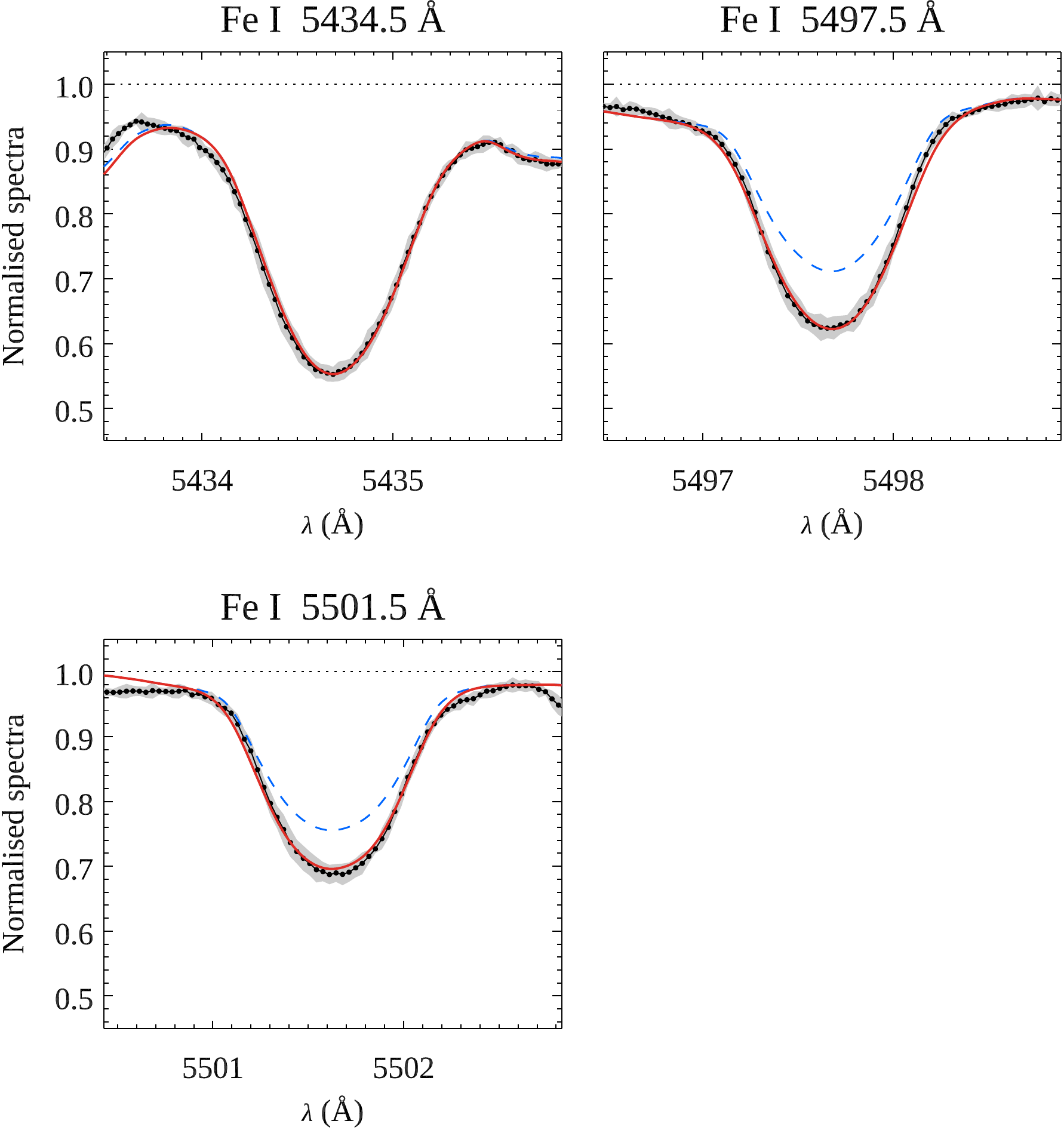}
\caption{Same as Fig.~\ref{fig:prf129333} for HD\,82558 (2016.05 epoch).}
\label{fig:prf82558}
\end{figure}

\begin{figure}[!h]
\fifps{\hsize}{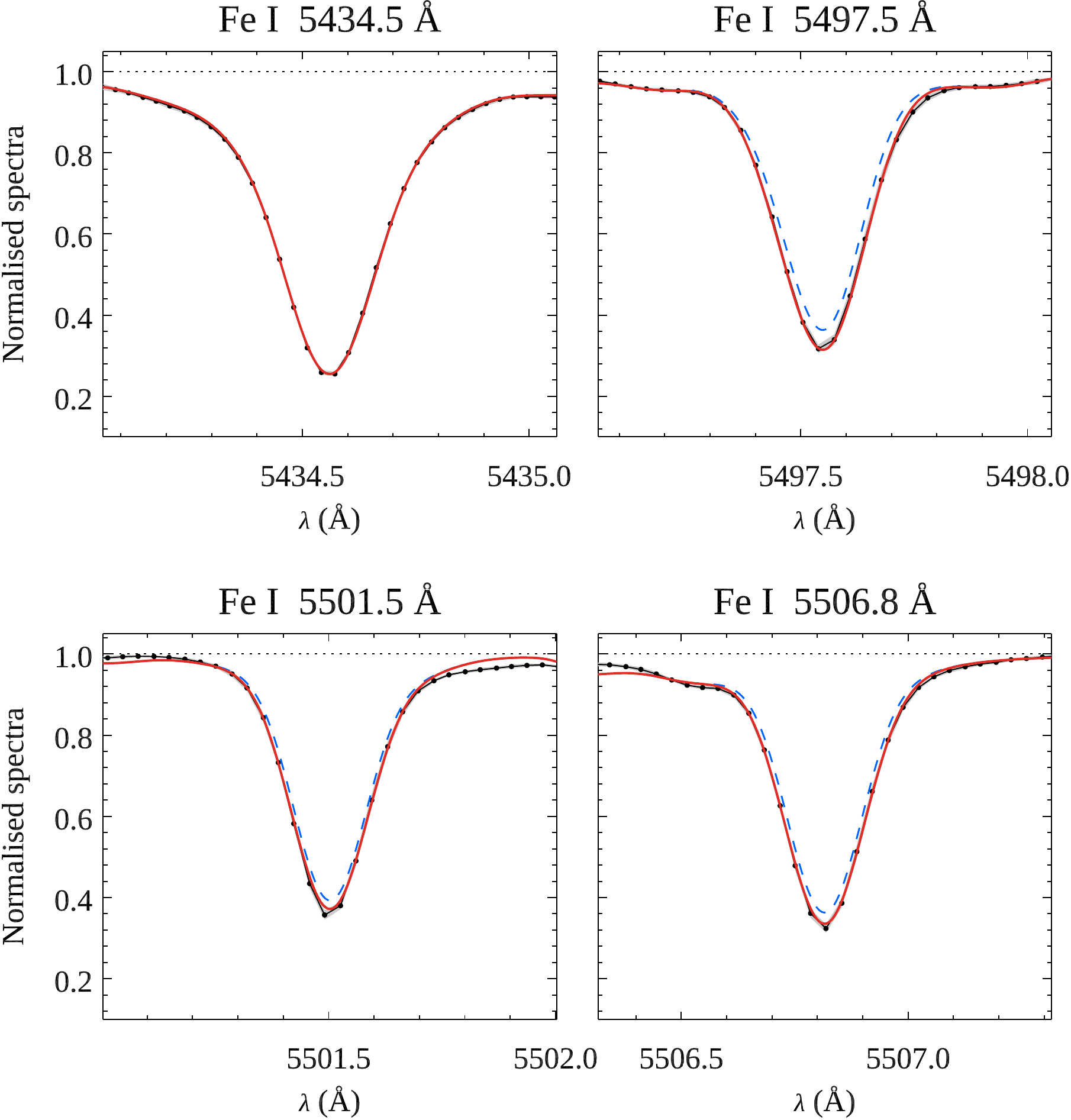}
\caption{Same as Fig.~\ref{fig:prf129333} for HD\,131156A (2013.33  epoch).}
\label{fig:prf131156A}
\end{figure}

\begin{figure}[!h]
\fifps{\hsize}{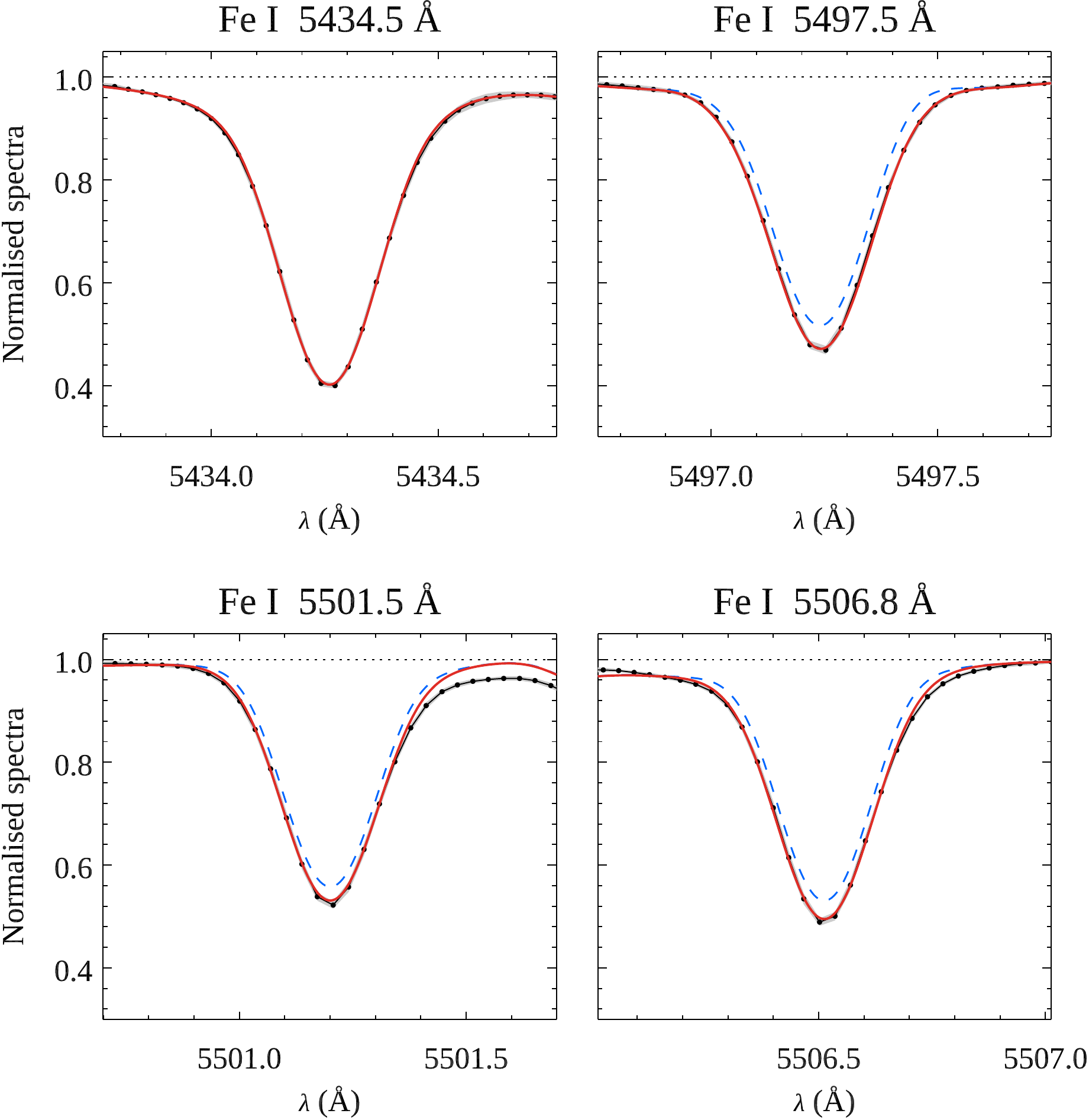}
\caption{Same as Fig.~\ref{fig:prf129333} for HD\,166435 (2010.53 epoch).}
\label{fig:prf166435}
\end{figure}

\begin{figure}[!h]
\fifps{\hsize}{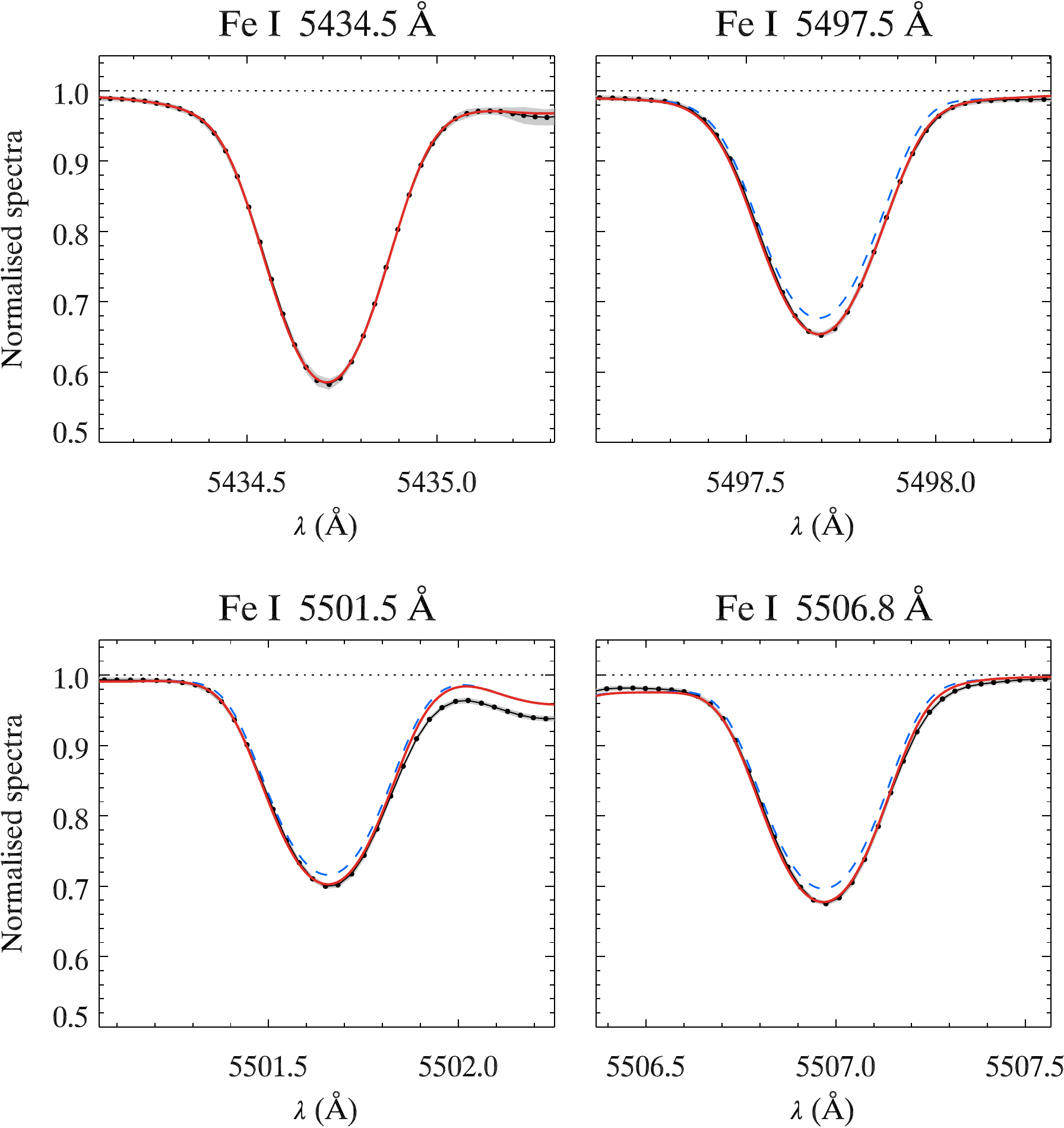}
\caption{Same as Fig.~\ref{fig:prf129333} for HD\,175726 (2008.59 epoch).}
\label{fig:prf175726}
\end{figure}

\begin{figure}[!h]
\fifps{\hsize}{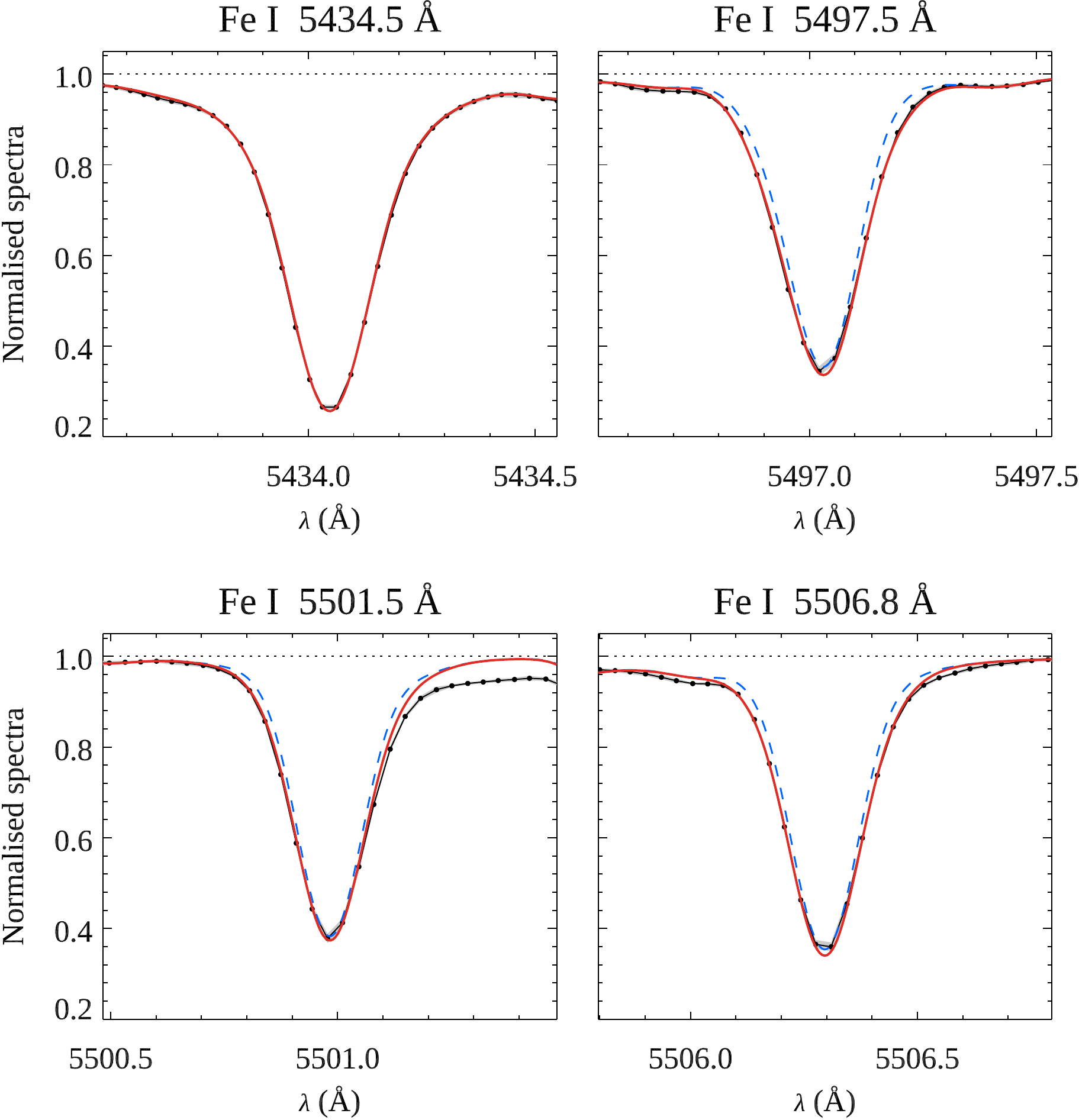}
\caption{Same as Fig.~\ref{fig:prf129333} for HD\,190771 (2010.48 epoch).}
\label{fig:prf190771}
\end{figure}

\begin{figure}[!h]
\fifps{\hsize}{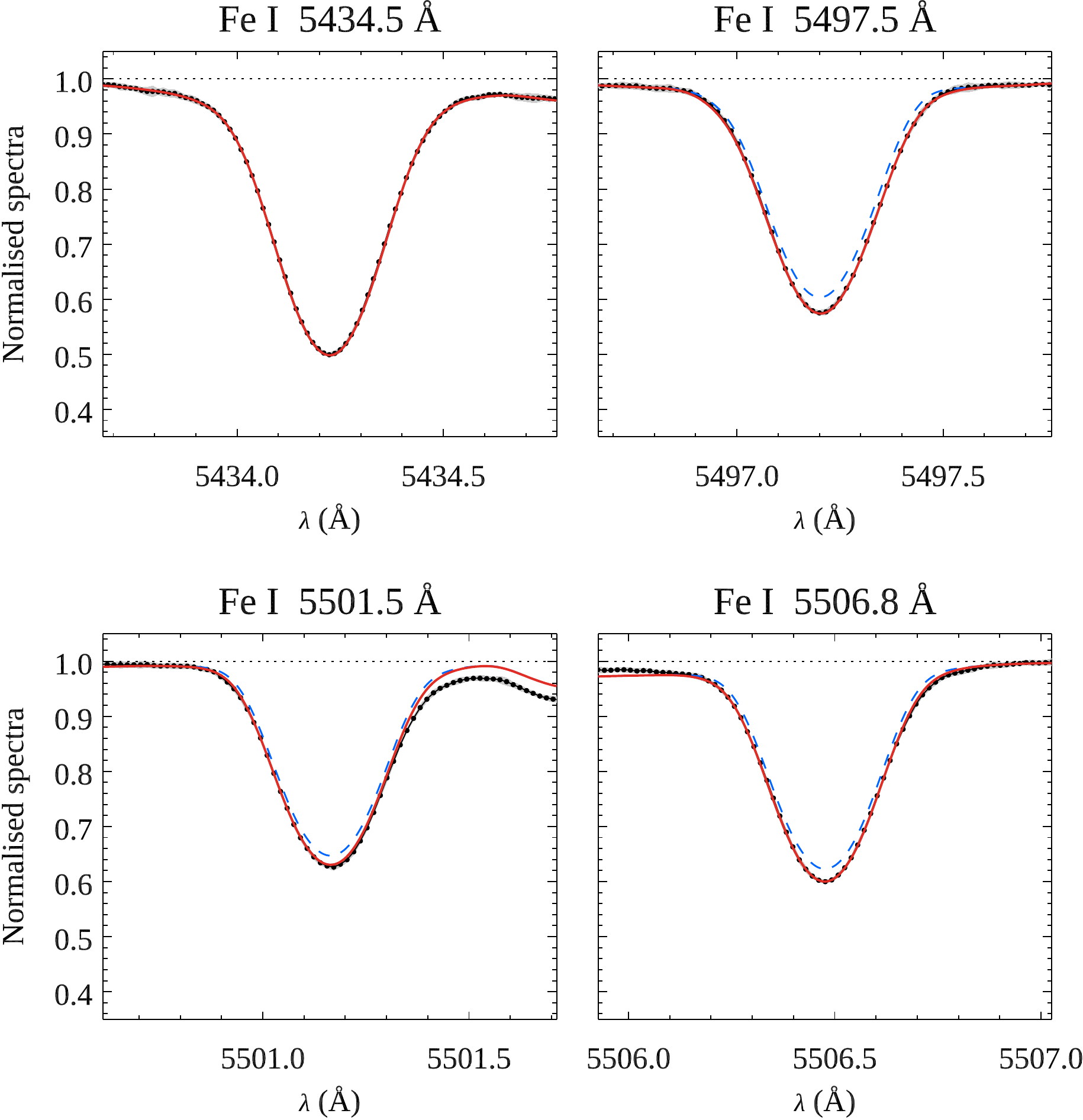}
\caption{Same as Fig.~\ref{fig:prf129333} for HD\,206860 (2013.70 epoch).}
\label{fig:prf206860}
\end{figure}

\clearpage
\section{Individual magnetic field measurements}

\longtab[1]{
\begin{longtable}{lrcccc}
\caption{Results of individual magnetic field measurements. \label{tbl:results}}\\
\hline\hline
Epoch & N & Inst. & $B$ (kG) & $f$ & \bs\ (kG) \\
\hline
\noalign{\smallskip}
\endfirsthead
\caption{Continued.}\\
\hline\hline
Epoch & N & Inst. & $B$ (kG) & $f$ & \bs\ (kG) \\
\hline
\noalign{\smallskip}
\endhead
\hline
\endfoot
\hline
\endlastfoot
\multicolumn{6}{c}{HD\,1835} \\
2013.70 & 12 & H & $3.0\substack{+0.6 \\ -0.5}$ & $0.23\substack{+0.05 \\ -0.04}$ & $0.69\substack{+0.05 \\ -0.05}$ \\[3pt]
2014.69 &  2 & E & $3.0\substack{+0.5 \\ -0.6}$ & $0.25\substack{+0.07 \\ -0.04}$ & $0.75\substack{+0.03 \\ -0.07}$ \\[3pt]
2017.95 &  7 & H & $3.2\substack{+0.6 \\ -0.7}$ & $0.19\substack{+0.06 \\ -0.03}$ & $0.61\substack{+0.06 \\ -0.05}$ \\[3pt]
\multicolumn{6}{c}{HD\,20630} \\
2012.82 & 14 & N & $3.0\substack{+1.0 \\ -0.7}$ & $0.15\substack{+0.05 \\ -0.04}$ & $0.45\substack{+0.05 \\ -0.05}$ \\[3pt]
2013.70 & 11 & H & $2.6\substack{+0.6 \\ -0.5}$ & $0.19\substack{+0.06 \\ -0.04}$ & $0.49\substack{+0.04 \\ -0.05}$ \\[3pt]
2016.84 &  7 & N & $2.9\substack{+0.6 \\ -0.7}$ & $0.18\substack{+0.06 \\ -0.03}$ & $0.52\substack{+0.05 \\ -0.06}$ \\[3pt]
2017.95 &  7 & H & $2.4\substack{+0.5 \\ -0.5}$ & $0.23\substack{+0.07 \\ -0.05}$ & $0.55\substack{+0.05 \\ -0.06}$ \\[3pt]
\multicolumn{6}{c}{HD\,29615} \\
2013.69 &  9 & H & $2.7\substack{+1.9 \\ -1.2}$ & $0.48\substack{+0.52 \\ -0.21}$ & $1.30\substack{+0.24 \\ -0.15}$ \\[3pt]
2017.95 &  7 & H & $2.7\substack{+2.2 \\ -1.1}$ & $0.51\substack{+0.49 \\ -0.23}$ & $1.38\substack{+0.23 \\ -0.16}$ \\[3pt]
\multicolumn{6}{c}{HD\,39587} \\
2007.09 &  9 & N & $3.2\substack{+1.4 \\ -1.7}$ & $0.15\substack{+0.21 \\ -0.04}$ & $0.48\substack{+0.08 \\ -0.08}$ \\[3pt]
2008.09 & 21 & N & $2.8\substack{+1.4 \\ -1.5}$ & $0.18\substack{+0.28 \\ -0.06}$ & $0.50\substack{+0.10 \\ -0.07}$ \\[3pt]
2010.76 &  7 & N & $3.1\substack{+1.6 \\ -1.5}$ & $0.15\substack{+0.17 \\ -0.05}$ & $0.46\substack{+0.10 \\ -0.08}$ \\[3pt]
2011.84 & 12 & N & $3.1\substack{+1.5 \\ -1.6}$ & $0.15\substack{+0.21 \\ -0.04}$ & $0.46\substack{+0.09 \\ -0.07}$ \\[3pt]
2015.00 & 44 & E & $3.0\substack{+1.3 \\ -1.3}$ & $0.15\substack{+0.15 \\ -0.04}$ & $0.45\substack{+0.07 \\ -0.06}$ \\[3pt]
2015.94 & 11 & N & $2.6\substack{+1.1 \\ -1.0}$ & $0.18\substack{+0.14 \\ -0.06}$ & $0.47\substack{+0.06 \\ -0.06}$ \\[3pt]
2016.84 &  5 & N & $2.5\substack{+1.3 \\ -1.3}$ & $0.18\substack{+0.27 \\ -0.06}$ & $0.45\substack{+0.19 \\ -0.07}$ \\[3pt]
2017.95 &  8 & H & $2.9\substack{+1.5 \\ -1.4}$ & $0.13\substack{+0.15 \\ -0.04}$ & $0.38\substack{+0.06 \\ -0.07}$ \\[3pt]
\multicolumn{6}{c}{HD\,56124} \\
2008.09 & 20 & N & $3.6\substack{+2.0 \\ -1.4}$ & $0.06\substack{+0.04 \\ -0.01}$ & $0.22\substack{+0.07 \\ -0.05}$ \\[3pt]
2010.84 &  4 & N & $3.1\substack{+2.1 \\ -1.1}$ & $0.07\substack{+0.03 \\ -0.02}$ & $0.22\substack{+0.09 \\ -0.05}$ \\[3pt]
2011.86 & 10 & N & $3.0\substack{+2.2 \\ -1.0}$ & $0.07\substack{+0.04 \\ -0.02}$ & $0.21\substack{+0.06 \\ -0.05}$ \\[3pt]
2012.91 &  4 & N & $3.2\substack{+2.2 \\ -1.1}$ & $0.07\substack{+0.03 \\ -0.02}$ & $0.22\substack{+0.09 \\ -0.05}$ \\[3pt]
\multicolumn{6}{c}{HD\,72905} \\
2007.08 & 12 & N & $3.3\substack{+1.1 \\ -1.2}$ & $0.18\substack{+0.11 \\ -0.04}$ & $0.59\substack{+0.06 \\ -0.07}$ \\[3pt]
2014.32 & 14 & N & $3.2\substack{+1.1 \\ -1.1}$ & $0.18\substack{+0.11 \\ -0.04}$ & $0.58\substack{+0.06 \\ -0.07}$ \\[3pt]
2015.03 & 12 & N & $3.2\substack{+0.9 \\ -1.2}$ & $0.20\substack{+0.14 \\ -0.04}$ & $0.64\substack{+0.06 \\ -0.07}$ \\[3pt]
2015.94 &  5 & N & $3.2\substack{+1.0 \\ -1.3}$ & $0.18\substack{+0.14 \\ -0.04}$ & $0.58\substack{+0.07 \\ -0.07}$ \\[3pt]
2016.88 &  5 & N & $2.9\substack{+1.1 \\ -1.1}$ & $0.20\substack{+0.15 \\ -0.05}$ & $0.58\substack{+0.07 \\ -0.07}$ \\[3pt]
\multicolumn{6}{c}{HD\,73350} \\
2007.09 &  9 & N & $2.9\substack{+0.9 \\ -0.7}$ & $0.18\substack{+0.07 \\ -0.04}$ & $0.52\substack{+0.08 \\ -0.06}$ \\[3pt]
2010.95 &  4 & N & $2.8\substack{+1.0 \\ -0.7}$ & $0.17\substack{+0.07 \\ -0.05}$ & $0.48\substack{+0.07 \\ -0.07}$ \\[3pt]
2011.05 &  7 & N & $2.9\substack{+0.9 \\ -0.7}$ & $0.18\substack{+0.07 \\ -0.04}$ & $0.52\substack{+0.07 \\ -0.07}$ \\[3pt]
2012.04 & 10 & N & $2.9\substack{+1.0 \\ -1.0}$ & $0.15\substack{+0.08 \\ -0.04}$ & $0.43\substack{+0.07 \\ -0.07}$ \\[3pt]
\multicolumn{6}{c}{HD\,76151} \\
2007.11 & 12 & N & $2.1\substack{+1.0 \\ -0.7}$ & $0.19\substack{+0.12 \\ -0.06}$ & $0.40\substack{+0.07 \\ -0.06}$ \\[3pt]
2009.95 &  7 & N & $2.0\substack{+0.8 \\ -0.7}$ & $0.22\substack{+0.15 \\ -0.07}$ & $0.44\substack{+0.06 \\ -0.07}$ \\[3pt]
2010.07 &  4 & N & $1.8\substack{+0.8 \\ -0.7}$ & $0.23\substack{+0.21 \\ -0.08}$ & $0.41\substack{+0.09 \\ -0.06}$ \\[3pt]
2012.05 & 10 & N & $1.8\substack{+0.9 \\ -0.7}$ & $0.21\substack{+0.20 \\ -0.08}$ & $0.38\substack{+0.08 \\ -0.07}$ \\[3pt]
2015.94 & 10 & N & $2.1\substack{+0.9 \\ -0.6}$ & $0.20\substack{+0.11 \\ -0.07}$ & $0.42\substack{+0.07 \\ -0.08}$ \\[3pt]
\multicolumn{6}{c}{HD\,82558} \\
2010.03 & 10 & H & $4.3\substack{+1.4 \\ -2.1}$ & $0.46\substack{+0.54 \\ -0.10}$ & $1.98\substack{+0.32 \\ -0.15}$ \\[3pt]
2011.11 & 18 & H & $4.4\substack{+1.3 \\ -2.1}$ & $0.47\substack{+0.53 \\ -0.09}$ & $2.07\substack{+0.31 \\ -0.14}$ \\[3pt]
2016.05 &  9 & E & $4.9\substack{+1.3 \\ -2.4}$ & $0.41\substack{+0.58 \\ -0.07}$ & $2.01\substack{+0.27 \\ -0.21}$ \\[3pt]
2017.96 &  9 & H & $4.3\substack{+1.4 \\ -2.1}$ & $0.46\substack{+0.54 \\ -0.10}$ & $1.98\substack{+0.32 \\ -0.14}$ \\[4pt]
\multicolumn{6}{c}{HD\,129333} \\
2006.93 &  4 & E & $3.5\substack{+1.3 \\ -1.4}$ & $0.39\substack{+0.32 \\ -0.10}$ & $1.36\substack{+0.14 \\ -0.11}$ \\[3pt]
2007.11 & 11 & N & $3.6\substack{+0.8 \\ -0.9}$ & $0.38\substack{+0.15 \\ -0.06}$ & $1.37\substack{+0.08 \\ -0.04}$ \\[3pt]
2008.23 &  9 & N & $3.8\substack{+0.9 \\ -1.0}$ & $0.37\substack{+0.15 \\ -0.06}$ & $1.41\substack{+0.09 \\ -0.06}$ \\[3pt]
2009.02 &  4 & N & $3.5\substack{+0.9 \\ -0.8}$ & $0.40\substack{+0.14 \\ -0.08}$ & $1.40\substack{+0.07 \\ -0.07}$ \\[3pt]
2012.07 & 10 & N & $4.0\substack{+0.8 \\ -1.2}$ & $0.35\substack{+0.17 \\ -0.05}$ & $1.40\substack{+0.09 \\ -0.07}$ \\[3pt]
2016.05 &  9 & E & $3.7\substack{+1.1 \\ -1.3}$ & $0.40\substack{+0.25 \\ -0.08}$ & $1.48\substack{+0.11 \\ -0.08}$ \\[3pt]
\multicolumn{6}{c}{HD\,131156A} \\
2005.50 &  8 & E & $1.1\substack{+0.7 \\ -0.2}$ & $0.71\substack{+0.29 \\ -0.36}$ & $0.78\substack{+0.16 \\ -0.18}$ \\[3pt]
2007.59 & 10 & N & $1.4\substack{+0.4 \\ -0.3}$ & $0.60\substack{+0.27 \\ -0.18}$ & $0.84\substack{+0.12 \\ -0.09}$ \\[3pt]
2008.08 & 16 & N & $1.3\substack{+0.4 \\ -0.4}$ & $0.56\substack{+0.44 \\ -0.19}$ & $0.73\substack{+0.17 \\ -0.11}$ \\[3pt]
2009.46 & 12 & N & $1.2\substack{+0.4 \\ -0.3}$ & $0.61\substack{+0.39 \\ -0.22}$ & $0.73\substack{+0.17 \\ -0.11}$ \\[3pt]
2010.07 &  7 & N & $0.9\substack{+0.5 \\ -0.1}$ & $0.79\substack{+0.21 \\ -0.40}$ & $0.71\substack{+0.09 \\ -0.16}$ \\[3pt]
2012.07 & 14 & N & $1.2\substack{+0.4 \\ -0.3}$ & $0.65\substack{+0.35 \\ -0.23}$ & $0.78\substack{+0.13 \\ -0.12}$ \\[3pt]
2013.33 & 11 & N & $1.1\substack{+0.4 \\ -0.2}$ & $0.73\substack{+0.27 \\ -0.28}$ & $0.80\substack{+0.11 \\ -0.13}$ \\[3pt]
2015.29 & 15 & N & $1.0\substack{+0.4 \\ -0.1}$ & $0.88\substack{+0.12 \\ -0.36}$ & $0.88\substack{+0.07 \\ -0.15}$ \\[3pt]
\multicolumn{6}{c}{HD\,166435} \\
2010.53 & 20 & N & $2.9\substack{+1.1 \\ -1.0}$ & $0.24\substack{+0.16 \\ -0.07}$ & $0.70\substack{+0.08 \\ -0.10}$ \\[3pt]
2011.53 &  6 & N & $2.9\substack{+1.1 \\ -1.1}$ & $0.23\substack{+0.17 \\ -0.07}$ & $0.67\substack{+0.09 \\ -0.10}$ \\[3pt]
2012.57 & 10 & N & $2.8\substack{+1.1 \\ -1.1}$ & $0.24\substack{+0.19 \\ -0.07}$ & $0.67\substack{+0.09 \\ -0.10}$ \\[3pt]
2016.35 &  7 & N & $2.9\substack{+1.1 \\ -1.0}$ & $0.24\substack{+0.14 \\ -0.07}$ & $0.70\substack{+0.09 \\ -0.09}$ \\[3pt]
\multicolumn{6}{c}{HD\,175726} \\
2008.59 & 41 & N & $3.8\substack{+2.0 \\ -2.5}$ & $0.10\substack{+0.25 \\ -0.03}$ & $0.38\substack{+0.11 \\ -0.08}$ \\[3pt]
2012.63 &  6 & N & $4.0\substack{+2.0 \\ -2.8}$ & $0.09\substack{+0.28 \\ -0.02}$ & $0.36\substack{+0.12 \\ -0.08}$ \\[3pt]
\multicolumn{6}{c}{HD\,190771} \\
2007.60 & 15 & N & $3.4\substack{+1.0 \\ -0.8}$ & $0.17\substack{+0.06 \\ -0.03}$ & $0.58\substack{+0.09 \\ -0.07}$ \\[3pt]
2008.67 & 13 & N & $3.1\substack{+0.9 \\ -0.8}$ & $0.18\substack{+0.07 \\ -0.04}$ & $0.56\substack{+0.07 \\ -0.08}$ \\[3pt]
2009.46 & 14 & N & $3.3\substack{+0.8 \\ -0.8}$ & $0.17\substack{+0.07 \\ -0.03}$ & $0.56\substack{+0.08 \\ -0.07}$ \\[3pt]
2010.48 & 10 & N & $3.1\substack{+0.8 \\ -0.7}$ & $0.19\substack{+0.07 \\ -0.04}$ & $0.59\substack{+0.08 \\ -0.07}$ \\[3pt]
2011.57 &  8 & N & $3.0\substack{+0.7 \\ -0.6}$ & $0.22\substack{+0.07 \\ -0.04}$ & $0.66\substack{+0.09 \\ -0.07}$ \\[3pt]
2012.55 & 35 & N & $3.1\substack{+0.9 \\ -0.6}$ & $0.19\substack{+0.06 \\ -0.04}$ & $0.59\substack{+0.08 \\ -0.07}$ \\[3pt]
2014.60 & 15 & N & $3.1\substack{+0.9 \\ -0.7}$ & $0.19\substack{+0.07 \\ -0.04}$ & $0.59\substack{+0.08 \\ -0.07}$ \\[3pt]
2015.45 & 16 & N & $3.1\substack{+0.8 \\ -0.7}$ & $0.19\substack{+0.06 \\ -0.04}$ & $0.59\substack{+0.08 \\ -0.07}$ \\[3pt]
2016.42 &  8 & N & $3.0\substack{+0.7 \\ -0.7}$ & $0.20\substack{+0.07 \\ -0.04}$ & $0.60\substack{+0.07 \\ -0.08}$ \\[3pt]
\multicolumn{6}{c}{HD\,206860} \\
2007.60 & 14 & N & $3.6\substack{+1.6 \\ -1.7}$ & $0.12\substack{+0.12 \\ -0.03}$ & $0.43\substack{+0.09 \\ -0.08}$ \\[3pt]
2008.63 & 11 & N & $3.6\substack{+1.5 \\ -1.9}$ & $0.11\substack{+0.14 \\ -0.03}$ & $0.40\substack{+0.08 \\ -0.08}$ \\[3pt]
2009.46 & 11 & N & $3.7\substack{+1.6 \\ -2.0}$ & $0.11\substack{+0.14 \\ -0.03}$ & $0.41\substack{+0.08 \\ -0.08}$ \\[3pt]
2010.55 & 13 & N & $3.8\substack{+1.4 \\ -1.6}$ & $0.12\substack{+0.09 \\ -0.02}$ & $0.46\substack{+0.09 \\ -0.08}$ \\[3pt]
2011.57 &  8 & N & $3.9\substack{+1.2 \\ -1.3}$ & $0.14\substack{+0.06 \\ -0.03}$ & $0.55\substack{+0.09 \\ -0.08}$ \\[3pt]
2013.70 & 14 & H & $3.0\substack{+1.6 \\ -1.5}$ & $0.14\substack{+0.18 \\ -0.04}$ & $0.42\substack{+0.07 \\ -0.07}$ \\[3pt]
2014.57 &  9 & N & $3.6\substack{+1.5 \\ -2.0}$ & $0.11\substack{+0.16 \\ -0.03}$ & $0.40\substack{+0.08 \\ -0.08}$ \\[3pt]
2015.55 & 16 & N & $3.8\substack{+1.3 \\ -1.5}$ & $0.13\substack{+0.08 \\ -0.03}$ & $0.49\substack{+0.09 \\ -0.07}$ \\[3pt]
2015.90 &  6 & N & $3.2\substack{+1.6 \\ -1.5}$ & $0.14\substack{+0.15 \\ -0.04}$ & $0.45\substack{+0.08 \\ -0.07}$ \\[3pt]
2016.51 &  3 & N & $3.5\substack{+1.3 \\ -1.6}$ & $0.14\substack{+0.13 \\ -0.03}$ & $0.49\substack{+0.07 \\ -0.07}$ \\[3pt]
\end{longtable}
\tablefoot{The table lists the epoch of each measurement, the number of individual observations combined to derive the mean spectrum, the instrument used (``E'': ESPaDOnS at CFHT, ``H'': HARPSpol at ESO 3.6-m telescope, ``N'': Narval at TBL), the field strength $B$, filling factor $f$, and the corresponding mean field strength \bs.}
}

\end{appendix}

\end{document}